\documentclass[11pt, a4paper]{article}
\pdfoutput=1
\usepackage[a4paper, total={6.4in, 10in}]{geometry}
\usepackage{graphicx}
\usepackage{color,soul}
\usepackage{xcolor}
\usepackage{alertmessage}
\usepackage{epsf}
\usepackage{amsmath}
\usepackage{array}
\usepackage[normalem]{ulem}
\usepackage{amssymb}

\usepackage{cite}
\usepackage{multirow,tabularx,tabulary}
\usepackage{multicol}
\usepackage{appendix}
\usepackage{tikz}
\usepackage{amsmath,amsfonts,amssymb,amsthm,euscript,braket,xcolor}
\usepackage{booktabs}
\usepackage{siunitx}
\usepackage{threeparttable}
\usepackage{geometry}
\usepackage{array, caption, tabularx, makecell}
\setcellgapes{3pt}
\usepackage{float}
\newcommand{\be}{\begin{equation}}
	\newcommand{\ee}{\end{equation}}

\newcommand{\Rmnum}[1]{\expandafter\@slowromancap\romannumeral #1@}
\newcommand{\bea}{\begin{eqnarray}}
	\newcommand{\eea}{\end{eqnarray}}

\numberwithin{equation}{section}

\usepackage{subfigure}
\begin{document}
	\title{\bf Quasi-normal mode of dyonic hairy black hole and its interplay with phase transitions}
	
	\author{\textbf{Supragyan Priyadarshinee}\thanks{518ph1009@nitrkl.ac.in}
		\\\\
		\textit{{\small Department of Physics and Astronomy, National Institute of Technology Rourkela, Rourkela - 769008, India}}
	}
	\date{\today}
	\maketitle
	\begin{abstract}
We study the dynamical stability of hairy dyonic black holes in the Einstein-Maxwell-scalar gravity system against the massless scalar field perturbation. We numerically obtain the corresponding quasinormal modes (QNMs) using the series solution and shooting methods for various black hole parameters. We find that the numerical values obtained from these two methods agree well with each other. The imaginary part of the QNM is always negative, indicating the stability of the dyonic hairy black hole against the scalar perturbation. We find that the decay and oscillatory modes of the scalar field perturbation increase linearly with the horizon radius for large black holes. We thoroughly investigate the behaviour of QNMs for different values of black hole parameters, including the electric charge, magnetic charge, horizon radius and hairy parameter, etc. Moreover, we also analyse the QNM near the small/large black hole phase transition and find that the nature of the QNMs is different for large and small black hole phases, suggesting QNMs as the possible probe of black hole phase transition.

	\end{abstract}
	
	
	\maketitle
	\pagestyle{plain}
	\section{Introduction}
		Black holes play a fundamental role in our understanding of the universe at a large scale. They have garnered enormous attention in the last few decades as they provide a unique and fundamental connection between thermodynamics, gravitation, and quantum theory. They are the simplest and yet the most mysterious solutions of the general theory of relativity that are believed to follow the famous ``no-hair theorem''\cite{Ruffini1971}. The theorem states that in the asymptotically flat space, the black holes are characterized by only three parameters: their mass, charge, and angular momentum. Stated differently, the black holes do not support scalar hair or other fields outside the horizon in asymptotically flat spaces. The term hair refers to additional information other than the above-mentioned parameters. The main argument favouring the no-hair theorem is based partly on the strong absorbing nature of the horizon, that tries to absorb and ingest everything around it. Despite several works supporting the original no-hair theorem \cite{Bekenstein:1972ky, Israel:1967wq, Carter:1971zc, Israel:1967za, Wald:1971iw, Robinson:1975bv,Teitelboim:1972qx, Mazur:1982db, Chase}, it is not a theorem in the rigorous mathematical sense and there are by now various counterexamples to the no-hair theorem \cite{bocharova1970exact,Bekenstein:1974sf,Bekenstein:1975ts,Bronnikov:1978mx,torii1999toward,zloshchastiev2005coexistence,berti2013numerical,herdeiro2014kerr,garfinkle1991charged,brodbeck1996instability,volkov1995number,bizon1991n,bizon1991stability,zhou1991nonlinear,bizon1990colored,Mahapatra:2022xea}.
	
	The initial attempts to study a minimally coupled scalar-gravity system were done in \cite{Bocharova, Bekenstein:1974sf, Bekenstein:1975ts} for the asymptotic flat spaces. However, the constructed hairy geometry turned out to be un-physical as the scalar field diverged on the horizon, and stability analysis further confirmed its unstable nature \cite{Bronnikov:1978mx}. A straightforward proof of the no-scalar hair theorem for the spherically symmetric asymptotic flat spacetime was provided in \cite{Bekenstein:1971hc,Bekenstein:1995un,Sudarsky:1995zg}, see also \cite{Heusler:1992ss}. For a recent discussion and review on the issue of scalar hair in asymptotic flat spaces, see \cite{Herdeiro:2015waa,Hertog:2006rr}. It was soon realised that the introduction of the cosmological constant or other scales may pave the way for a regular scalar field configuration near the horizon, as the cosmological constant may act as an effective potential outside the horizon which may stabilise the scalar field. Hairy black hole solution for the minimally coupled scalar-gravity system in the asymptotic dS space was obtained in \cite{Zloshchastiev:2004ny}, however, it turned out to be unstable \cite{Torii:1998ir}. In the case of asymptotically AdS space, stable hairy black hole solutions with various horizon topologies were obtained and analysed in \cite{Torii:1998ir,Torii:2001pg,Winstanley:2002jt,Martinez:2004nb,Martinez:2005di,Martinez:2006an,Hertog:2004dr,Henneaux:2004zi,Henneaux:2006hk,Amsel:2006uf,Mahapatra:2020wym}. In recent years, countless works addressing various physical aspects of the hairy black holes in different asymptotic spaces have appeared, for a necessarily biased selection see \cite{Dias:2011at,Dias:2011tj,Bhattacharyya:2010yg,Basu:2010uz,Anabalon:2012ta,Anabalon:2012ih,Anabalon:2012tu,Kleihaus:2013tba,Kolyvaris:2013zfa,Gonzalez:2013aca,Anabalon:2009qt,Charmousis:2009cm,Guo:2023mda,Guo:2021zed}.
	
	 Recent progress in string theory, particularly in gauge/gravity duality \cite{Maldacena:1997re}, has generated much interest in hairy black hole geometries in AdS spaces. The idea that the gauge/gravity duality can play a vital role in understanding strongly coupled field theory at finite chemical potential and temperature non-perturbatively have provided an unique and interesting field of application for these geometries. For example, the hairy AdS black holes play central role in the area of holographic superconductors \cite{Gubser:2008px,Hartnoll:2008vx,Hartnoll:2008kx,Dey:2014voa,Dey:2014xxa,Mahapatra:2013vta}. Similarly, they are an integral part of holographic QCD model building \cite{Gubser:2008ny}. Moreover, black holes in AdS space exhibit rich phase structure and are thermodynamically stable as opposed to its asymptotic flat counter parts.  For instance, there can be various types phase transitions in AdS black holes, such as the Hawking/Page phase transition \cite{Hawking:1982dh} or the Van der Waals type small/large black hole phase transitions \cite{Chamblin:1999tk,Chamblin:1999hg,Sahay:2010wi,Dey:2015ytd}.  These different phases and transitions between them, moreover, have interesting physical interpretation in the dual boundary theory \cite{Witten:1998zw,Dudal:2017max}, making the black holes in AdS space a fruitful arena of research. Therefore, one can expect that introduction of the scalar hair might modify the thermodynamic properties of the AdS black holes in a nontrivial way and are worth investigating.
	
	In a similar context, dyonic black holes, which possess electric and magnetic charge, have also appeared extensively in the gauge/gravity duality literature. Because of electromagnetic duality, it is possible to construct a black hole that carries both electric and magnetic charges in four dimensions. In the dual boundary description, these dyonic black holes correspond to a field theory in ($2+1$) dimensions with a $U(1)$ conserved charge ($q_e$) and in a constant magnetic field background ($q_M$). The presence of magnetic charge not only enriches the thermodynamic phase diagram of AdS black holes in the gravity side but also allows us to introduce a background magnetic field in the dual boundary side, thereby providing an approach to probe physics related to the Hall effect \cite{Hartnoll:2007ai}, ferromagnetism \cite{Dutta:2013dca},  magnetohydrodynamics \cite{Caldarelli:2008ze}, the Nernst effect \cite{Hartnoll:2007ih} etc. See \cite{Goldstein:2010aw,Kundu:2012jn,Caldarelli:2016nni,Donos:2015bxe,Sadeghi:2016dvc,Kruglov:2020aqm,Kruglov:2019ybs,Panahiyan:2018fpb,Hajkhalili:2018thm,Hendi:2016uni,Chaturvedi:2014vpa,Kim:2015wba,
	Amoretti:2020mkp,Bhatnagar:2017twr,Lindgren:2015lia,Zhou:2015dha,Khimphun:2017mqb,Bai:2020ezy,Li:2016nll,BravoGaete:2019rci} for other related discussions on dyonic black holes and their holographic applications to field theory.
	
		However, despite their profound importance in the context of gauge/gravity duality, the analysis of hairy dyonic black holes is rather limited \cite{Caldarelli:2016nni}. This is mainly due to the difficulty in constructing such solutions analytically, as it requires a simultaneous solution of the Einstein-Maxwell-Scalar system with a well-behaved profile for the scalar and gauge fields. Accordingly, most hairy dyonic black hole solutions have been discussed numerically \cite{Cadoni:2011kv}. Moreover, analytic expressions of conserved charges and thermodynamics observables are often difficult to obtain in these systems. Recently, in \cite{Priyadarshinee:2021rch}  some of these issues were addressed and a number of exact analytic dyonic black hole solutions were found which were simultaneously supported by a regular and well behaved scalar field. The constructed dyonic hairy AdS black hole solutions were obtained by analytically solving the Einstein-Maxwell-Scalar gravity system using potential reconstruction method\cite{Mahapatra:2018gig,BOHRA2020135184,Bohra:2020qom,Dudal:2021jav,dudal2018interplay,mahapatra2019interplay,he2013phase,Arefeva:2018hyo,Arefeva:2020vae,Arefeva:2022avn,Arefeva:2021mag}. The solution rely on an arbitrary function $A(z)$ that adds a new parameter $a$, corresponding to the strength of the scalar hair. Moreover, the hairy dyonic solutions were obtained not only for planar, as is more relevant for the gauge/gravity duality application point of view,  but for spherical and hyperbolic horizon topologies as well. The thermodynamic analysis of these black holes further suggested that the hairy black holes are not only thermodynamically stable but also thermodynamically favored. In particular, the free energies of the hairy black holes are lower than the nonhairy black holes at low temperatures. For the spherical horizon, like their nonhairy counterpart, the Hawking/Page and small/large Van der Waals type phase transitions were found. Interestingly, with scalar hair, unlike their nonhairy counterpart, the small/large black hole phase transition was found to appear in the grand-canonical ensemble as well. Since these dyonic hairy black holes were found to be thermodynamically stable, it is natural to ask whether these black holes are dynamically stable or not. In this work, one of our main aims is to undertake this study and investigate the dynamically stability by analysing the scalar field perturbation and the corresponding QNM of these black holes.
	
	QNMs characterize the perturbations in the surrounding geometry of a black hole. These are basically the solution of linearized wave equation about the black hole background, subject to the ingoing wave boundary condition at the horizon \cite{regge1957stability,zerilli1970gravitational,vishveshwara1970scattering}. The spectrum of QNMs is generally complex because of the absorbing nature of the horizon, with real and imaginary parts of it describing the oscillation and damping time of the perturbation respectively. Since the QNMs are uniquely characterized by the black hole parameters, they can provide valuable insight into the structure of the black hole. In astrophysical and theoretical contexts as well, QNMs have a significant impact. In the former, they predominate the ring down dynamics of the final black hole created by the fusing of compact objects, and a direct measurement of the mode with the lowest damping assists in determining the final mass and angular momentum of the object \cite{PhysRevD.76.064034}. From the gauge/gravity point of view, QNMs provide the thermalization timescales in the strongly coupled CFT \cite{horowitz2000quasinormal}. Because of their utmost importance in the gravitational physics, the discussion of QNMs have undergone extensive computations, be it from theory or experiments \cite{Pretorius:2005gq,Campanelli:2005dd,Creighton:1999pm,Tsunesada:2005fe}. Various advanced and effective methods have been implemented, both analytically and numerically, for its computation. For review and more details on this topic see \cite{Berti:2009kk,Berti:2003ud,Kokkotas:1999bd,Ferrari:1996vx,Chan:1999sc,Cardoso:2001hn}.
	
	From the above discussion, it is also natural to ask whether the dynamical perturbation in the black hole background can capture the imprints of the black hole thermal phase transition. Recently promising research have started concerning the probing of black hole phase transitions using QNMs. This notion was first discussed by \cite{koutsoumbas2006quasi}, where it was noticed that QNMs of electromagnetic perturbation show distinct behaviour in the MTZ black hole compared to the vacuum topological black hole phase \cite{martinez2004exact}. The QNMs of scalar field perturbation on phase transition was further discussed in \cite{shen2007phase}. The QNMs of electromagnetic and gravitational perturbations similarly showed contrasting behaviour in different black hole phases \cite{koutsoumbas2008phase}. In \cite{liu2014signature}, scalar QNMs in RN-AdS black hole background were found to follow different pattern in the small and the large black hole phases — both for the isobaric and isothermal phase transition processes near the first order transition line. This discussion was then generalised to higher derivative gravity theories \cite{Mahapatra:2016dae}, where again scalar QNMs were found to follow different pattern with horizon radii in small-large black hole phases. For more details on these directions see \cite{Wei:2018aqm,Liang:2017ceh,Li:2017kkj,Zou:2017juz,Chabab:2017knz,Chabab:2016cem,Zhang:2020khz}.
	
	Our main aim in this work is twofold. First, we investigate the dynamical stability of the charged dyonic hairy black holes under the scalar field perturbation and find the corresponding QNMs. For this purpose, we employ two methods (i) the power series method \cite{horowitz2000quasinormal} and (ii) the shooting method \cite{liu2014signature, Mahapatra:2016dae} to compute the scalar QNMs and analyze the stability of the dyonic hairy black hole for different parameters. We find that the results from both these methods agree well with each other. We thoroughly investigate the behaviour of QNMs for different values of black hole parameters, including the electric charge, magnetic charge, horizon radius, and hairy parameter $a$, and analyze the dependence of real and imaginary parts of the QNMs on these parameters. For all parameter values, the imaginary part of the QNMs is found to be negative, suggesting the dynamical stability of the dyonic hairy black holes. Moreover, we also analytically show that the imaginary part of the QNMs is always negative in the hairy case. Second, we then compute scalar QNMs near the small/large black hole phase transition and try to analyze whether they carry any imprints of the said phase transition. We find the affirmative answer and that QNMs follow different pattern with horizon radii in the small and large hairy black hole phases. This is true both in the canonical as well as in the grand canonical ensembles.
	
	This paper is organized as follows. In section \ref{section2}, we briefly discuss the Einstein-Maxwell-scalar gravity system and present the analytic solution of the hairy dyonic black hole. In section \ref{section3}, we analyse the thermodynamics and phase transition of the hairy dyonic black holes. In section \ref{section4}, we will discuss the scalar field perturbation and present our numerical results of the QNMs for various values of the black hole parameters. In section \ref{section5}, we discuss the behaviour of QNMs near the small/large black hole transition line. Finally, section \ref{section6} contains a summary of our results and outlook for future work.

	\section{Hairy dyonic black hole solution}\label{section2}
	To construct the dyonic hairy black hole solution, we consider the following Einstein-Maxwell-scalar gravity system with action,
	\begin{equation}\label{actionEF}
			S_{EMS}=\frac{1}{16 \pi G_4}\int_M d^4 x \sqrt{-g} [R-\frac{f(\phi)}{4} F_{\mu \nu} F^{\mu \nu} -\frac{1}{2}\partial_\mu\phi\partial^\mu\phi-V(\phi)] \,,
	\end{equation}
	where $G_4$ is the $4$-dimensional Newton's constant, $F_{\mu\nu}$ is the field strength tensor of $U(1)$ Gauge field $A_\mu$, $\phi$ is the scalar field, and $f(\phi)$ is the coupling between the $U(1)$ gauge and scalar field. Here, we concentrate on the simplest case $f(\phi)=1$, akin to no direct coupling between the scalar and gauge field. It is also possible to take other
useful couplings, such as the nonminimal linear $f(\phi) \propto \phi$ and exponential $f(\phi) \propto e^{-\phi}$ couplings, which have been considered extensively in the literature in recent years, to obtain analogous dyonic hairy black holes.  The electric and magnetic charge, hence the dyonic properties of the gravity system, lie within the structure of the electromagnetic field strength tensor. $V(\phi)$ is the potential of the scalar field $\phi$.

Since we are interested in constructing the dyonic hairy black hole solution with spherical horizon topology, we consider the following Ans$\ddot{a}$tze for the metric $g_{\mu\nu}$, field strength tensor $F_{\mu\nu}$, and scalar field $\phi$
	\begin{eqnarray}\label{metric}
		ds^2&=&\frac{L^2}{x^2}\left[-g(x) dt^2+\frac{e^{2 A(x)}}{g(x)} dx^2+d y_1^{2}+\sin{y_1}^2 d y_2^{2}\right] \,, \nonumber\\
		\phi&=&\phi(x)\,,\nonumber\\
		A_\mu&=&A_t(x){\delta^t}_{\mu}+q_M \cos(x_1) \,,
	\end{eqnarray}
	where $L$ is the AdS length scale and $A_t$ and $q_M$ contain the information about the electric and magnetic charges. The radial coordinate $x=1/r$ runs from $x=0$ (asymptotic boundary) to $x=h$ (horizon radius), or to $x=\infty$ for thermal AdS (without horizon). The variation of above action~(\ref{actionEF}) gives the following Einstein, Maxwell, and scalar equation of motion, respectively
	\begin{eqnarray}	
		R_{\mu\nu}-\frac{1}{2}g_{\mu\nu}R+\frac{f(\phi)}{4}(\frac{g_{\mu\nu}}{2}F^2-2F_{\mu\rho}{F_\nu}^\rho) \nonumber&\\ +\frac{1}{2}\left(\frac{g{\mu\nu}}{2}\partial_\rho\phi\partial^\rho\phi-\partial_\mu\phi\partial_\nu\phi+g_{\mu\nu}V(\phi)\right)&=&0 \,, \nonumber\\
		\frac{1}{\sqrt{-g}}\partial_\mu[\sqrt-gf(\phi)F_{\mu\nu}]&=&0\label{2} \,, \nonumber\\
		\frac{1}{\sqrt{-g}}\partial_\mu[\sqrt{-g}\partial^\mu\phi]-\frac{F^2}{4}\frac{\partial f(\phi)}{\partial\phi}-\frac{\partial V(\phi)}{\partial\phi}&=&0 \,,
		\label{1}
	\end{eqnarray}
Substituting Eq.~(\ref{metric}) into the above equation, we get the following three Einstein equations of motion,

	\begin{eqnarray}
		& & tt: \  \frac{A'(x)}{x} - \frac{g'(x)}{2xg(x)} + \frac{\phi'(x)^2}{8}  + \frac{3}{2x^2}  + \frac{z^2f(z)B_{t}'(x)^2}{8L^2g(x)}  \nonumber \\
		& & +  \frac{q_{M}^{2} e^{2A(x)} x^2 f(x)}{8L^2g(x)}  +\frac{e^{2A(x)}L^2V(x)}{4z^2g(x)} - \frac{e^{2A(x)}}{2g(x)} =0\,,
		\label{Einsteintt}
	\end{eqnarray}
	\begin{eqnarray}
		& & zz: \ - \frac{g'(x)}{x} + g(z) \left(\frac{3}{x^2} - \frac{\phi'(x)^2}{4} \right)  + \frac{z^2f(x)B_{t}'(x)^2}{4L^2}   \nonumber \\
		& & +  \frac{q_{M}^{2} e^{2A(x)} x^2 f(x)}{4L^2}  +\frac{e^{2A(x)}L^2V(x)}{2x^2} - e^{2A(x)}=0 \,,
		\label{Einsteinzz}
	\end{eqnarray}
	\begin{eqnarray}
		& &  y_i y_i: \ g''(x) -  g'(x) \left(A'(x)+ \frac{4}{x}\right) + g(x)\left(\frac{6}{x^2} + \frac{4A'(x)}{x} + \frac{\phi'(x)^2}{2}\right)  \nonumber \\
		& &  - \frac{z^2f(x)B_{t}'(x)^2}{2L^2} - \frac{q_{M}^{2} e^{2A(x)} x^2 f(x)}{2L^2} + \frac{e^{2A(x)}L^2V(x)}{x^2} =0 \,.
		\label{Einsteinyiyi}
	\end{eqnarray}

the above Eqs. (\ref{Einsteintt}), (\ref{Einsteinzz}), and (\ref{Einsteinyiyi}) combiningly can be further rearranged as
\begin{eqnarray}\label{eq5}
		& &	\frac{2 A'(x)}{x}+\frac{1}{2} \phi '(x)^2=0 \,, \nonumber\\
		& &	-\frac{A'(x) g'(x)}{2 g(x)}-\frac{q_M^2 x^2 e^{2 A(x)} f(x)}{2 L^2 g(x)}+\frac{e^{2 A(x)}}{g(x)}-\frac{x^2 f(x) A_t'(x)^2}{2 L^2 g(x)}+\frac{g''(x)}{2 g(x)}-\frac{g'(x)}{x g(x)}=0 \,, \nonumber\\
		& &	-\frac{A'(x) g'(x)}{2 g(x)}+\frac{2 A'(x)}{x}+\frac{L^2 e^{2 A(x)} V(x)}{x^2 g(x)}-\frac{e^{2 A(x)}}{g(x)}+\frac{g''(x)}{2 g(x)}-\frac{3 g'(x)}{x g(x)}+\frac{6}{x^2}=0 \,.
		\end{eqnarray}
Similarly, one gets the following equation of motion for the scalar field,
	\begin{equation}\label{scalar}
			 \phi ''(x)+\left(A'(x)-\frac{4}{x}\right) \phi '(x) -\frac{x^4 e^{-2 A(x)} f'(x) \left(q_M^2 e^{2 A(x)}-A_t'(x)^2\right)}{2 L^4}-\frac{\partial V(\phi)}{\partial\phi}=0\,,
	\end{equation}
and for the gauge field,
	\begin{equation}
		A_t'(x) \left(\frac{A'(x)}{L^4}-\frac{f'(x)}{L^4 f(x)}\right)-\frac{A_t''(x)}{L^4}=0\,.
	\end{equation}
Therefore, there are a total of five equations in the Einstein-Maxwell-scalar system. However, only four of them are independent. Here, we consider the scalar equation (\ref{scalar}) as a constrained equation and take
	the remaining equations as independent. To solve these equations, we impose the following boundary conditions,
	\begin{eqnarray}
		g(0)&=&1, \; \;\, \text{and}\;\quad  g(h)=0 \,, \\ \nonumber
		A_t(0)&=&\mu_e, \; \text{and} \; \quad A_t(h)=0\,, \\ \nonumber
		A(0)&=&1\,.
		\label{boundarycdtn}
	\end{eqnarray}
	These boundary conditions are chosen to ensure that the spacetime asymptotes to AdS at the boundary $x\rightarrow0$. The parameter $\mu_e$ is the leading term of the near boundary expansion of the time component of the $U(1)$ gauge field $A_t(x)$, and corresponds to the chemical potential of the dual boundary theory. Using Gauss's theorem, one can also find a relation between $\mu_e$ and the electric charge of the black hole. Apart from these boundary conditions, we further demand that the scalar field $\phi$ remains real everywhere in the bulk and goes to zero at the asymptotic boundary $\phi(0)=0$. Using these boundary conditions, the Einstein-Maxwell-scalar field equations can be solved in analytic forms in terms of a single functions $A(x)$. The different forms of $A(x)$, however, correspond to different $V(x)$, i.e., different $A(x)$ attributes to different dyonic hairy black hole solutions. Therefore, one can construct a large family of physically allowed dyonic hairy black hole solutions for the Einstein-Maxwell-scalar gravity system of Eq. (\ref{actionEF}) by choosing different forms of $A(x)$.
	
In \cite{Priyadarshinee:2021rch}, two different forms of $A(x)$: (i) $A(x)=-\log(1+a x)$ and (ii) $A(x)=-a x$ were considered. These forms of $A(x)$ were chosen not just for their simplicity but also to have better control over the integrals that appear in  field equations. Here, we will mainly concentrate on the first form $A(x)=-\log(1+a x)$, as analogous investigation for the second form can be straightforwardly done. For $A(x)=-\log(1+a x)$, the gravity solution is
\begin{equation}
			\begin{split}
				g(x) =& 1 + \frac{a x (a x-2)+2 \log (a x+1)}{a^2 \left(a h \left(a h-2\right)+2 \log \left(a
					h+1\right)\right)} \Bigl\{ \log \left(a h+1\right) \left(\log \left(a h+1\right)-2 \log \left(h\right)-1\right)  \nonumber \\&
				-a^2 + a h \left(2 \log \left(\frac{h}{a h+1}\right)+a h \log \left(\frac{1+a h}{h}\right)-1\right)-2
				\text{Li}_2\left(-a h\right) \Bigr\}  \nonumber \\&
				-  \frac{\left(q_e^2 +q_M^2  \right) \left(a x (a x-2)+2 \log (a x+1) \right)}{4 a^4 \left(a h \left(a h-2\right)+2 \log \left(a
					h+1\right)\right)} \Bigl\{ 2 \left(\log \left(a h+1\right)-3\right) \log \left(a h+1\right) \nonumber \\&
				+ a h \left(-4 \log \left(a h+1\right)+a h \left(2 \log \left(a h+1\right)-1\right)+6\right) \Bigr\} \nonumber \\&
				+ \left(q_e^2 +q_M^2  \right) \Bigl[ \frac{2 \left(a^2 x^2-2 a x-3\right) \log (a x+1)+a x (6-a x)+2 \log ^2(a x+1)}{4 a^4} \Bigl]  \nonumber \\&
				+ \frac{4 \text{Li}_2(-a x)+a x (-a x+2 (a x-2) \log (x)+4)+4 \log (x) \log (a x+1)}{2 a^2} \nonumber \\&
				+ \frac{a x (a x-2)-2 \log (a x+1) (a x (a x-2)+\log (a x+1)-1)}{2 a^2}  \,,
				\label{sphericalcase}
			\end{split}
		\end{equation}
where $\text{Li}_2$ is the Polylogarithm function. One can easily check that this expression of $g(x)$ reduces to the standard dyonic RN-AdS expression in the limit $a\rightarrow0$, indicating the consistency of the obtained solution. Similarly, the solution for the scalar and gauge fields are,
	\begin{eqnarray}
		\phi(x) &=& 4\sinh{\sqrt{ax}} \,, \nonumber \\
		A_t(x) &=& \mu_e \left(1-\frac{\log{(1+a x)}}{\log{(1+a h)}}   \right) \,.
	\end{eqnarray}
	We can similarly write down the analytic expression of the potential $V(x)$. However, it is too lengthy and, at the same time, not very illuminating; therefore, we skip to write it down here for brevity. Notice that in the limit $a\rightarrow 0$, the scalar field goes to zero, and all other expressions reduce to the standard nonhairy dyonic expressions. The Ricci and Kretschmann scalar are finite everywhere outside the horizon, indicating the nonsingular nature of the constructed bulk spacetime. In particular, there are no additional singularities in the dyonic hairy case than those already present in the nonhairy case. Moreover, the scalar field goes to zero only at the asymptotic boundary $x=0$, implying the existence of
	a well-behaved dyonic hairy black hole solution. Similarly, $V$ is also regular everywhere outside the horizon, and it asymptotes to $V(x=0)=-6$ at the AdS boundary for all $a$. Additionally, provided that the chemical potential $\mu_e$ and magnetic charge $q_M$ are not too large, $V$ is also bounded from above by its boundary value,\textit{ i.e.}, $V(0)\geq V(z)$, thereby satisfying the Gubser criterion to have a well-defined dual boundary theory \cite{Gubser:2000nd}. However, for higher values of $\mu_e \gtrsim 2$ and $q_M \gtrsim 2$, the Gubser criterion can be violated.
	
	The black hole solution has the temperature and entropy,
	\begin{eqnarray}
		T&=&-\frac{g'(h)}{e^{A(h)}4 \pi} \,, \nonumber \\
		S&=&\frac{L^2 \pi}{G_4 h^2} \,,
	\end{eqnarray}
	and the relation between the chemical potential and the black hole charge is
	\begin{eqnarray}
		Q_e &=&\frac{q_e \omega_2}{16 \pi G_4} \,, \nonumber \\
		q_e &=& \frac{a \mu_e}{\log{(1+a h)}}\,,
\label{BHchargerelation}
	\end{eqnarray}
	where $\omega_2$ is the area of the unit two-sphere. Similarly, using the holographic renormalization procedure, we can computet the mass of the black hole. In the constant chemical potential grand canonical ensemble, it is given by
\begin{equation}\label{massGibbsphcase1}
			\begin{split}
				M_{HR}^G =& \frac{a^2 \left(-a \left(4 a^2+5\right) h^2+8 \left(a^2+5\right) h-60 h \left(a h-2\right)\left(\coth ^{-1}\left(2 a h+1\right)-\log (4)\right)+30 a\right)}{360 \pi  G_4 \left(a h \left(a h-2\right)+2 \log \left(a h+1\right)\right)} \nonumber \\&
				+ \frac{4 a \left(-2 a^2+15 \log \left(a h\right)+5+60 \log (2)\right) \log \left(a h+1\right)+60 a
					\text{Li}_2\left(-a h\right)-30 a \log ^2\left(a h+1\right)}{360 \pi  G_4 \left(a h \left(a
					h-2\right)+2 \log \left(a h+1\right)\right)} \nonumber \\&
				+ \frac{a \mu_e^2 \left(2 \left(a^2 h^2-2 a h-3\right) \log \left(a h+1\right)+a z_h
					\left(6-a h\right)+2 \log ^2\left(a h+1\right)\right)}{48 \pi  G_4 \log ^2\left(a h+1\right)
					\left(a h \left(a h-2\right)+2 \log \left(a h+1\right)\right)} \nonumber \\&
				+\frac{ q_M^2 \left(2 \left(a^2 h^2-2 a h-3\right) \log \left(a h+1\right)+a h \left(6-a
					h\right)+2 \log ^2\left(a h+1\right)\right)}{48 a \pi  G_4 \left(a h \left(a h-2\right)+2 \log
					\left(a h+1\right)\right)} \,,
			\end{split}
		\end{equation}
whereas for the fixed charge canonical ensemble, the mass expression is
	\begin{equation}
		M_{HR}^C=M_{HR}^G-\frac{q_e^2 \log(a h+1)}{16 \pi a G_4}\,.
	\end{equation}
Using these thermodynamic expressions, we can similarly compute the free energy expression for both the canonical and grand canonical ensembles using the standard thermodynamic techniques. The free energy expressions will be useful to analyse the phase structure of the hairy dyonic black holes.

At this point, we like to emphasize that the above constructed hairy solution corresponds to the primary hair. In order to show that the scalar hair is of primary nature, we need to show that the conserved charges depend only on the respective independent integration constants. We have already established in Eq.~(\ref{BHchargerelation}) that the black hole electric charge $Q_e$ depends only on the integration constants $q_e$. Similarly, it can also be easily shown from Eq.~(\ref{massGibbsphcase1}) that the mass of the hairy black hole also depends on the independent integration constant. Moreover, since the hairy solution smoothly reduces to the dyonic black hole expression in the limit $a\rightarrow 0$, it further establishes the primary nature of the scalar hair.

Before ending this section, let us also mention that apart from the above obtained dyonic hairy black hole solution, there also exists a second solution corresponding to no horizon. This no-horizon solution corresponds to a thermal-AdS solution, and it can be obtained by taking the limit $h\rightarrow \infty$ in the black hole solution given above \footnote{Here, we are referring this without horizon solution as thermal-AdS for simplicity even though the curvature is not constant throughout the spacetime.}. In the hairy case, the thermal-AdS solution can have a non-trivial structure in the bulk spacetime, however, it will always go to AdS asymptotically. Interestingly, just like in the RN-AdS case, there can be a Hawking/Page type thermal-AdS/black hole phase transition between these two solutions.
	
	\section{Black hole phase transition}\label{section3}
\begin{figure}[h!]
		\centering
		\subfigure[\label{zhvsTvsaMu0qM0sphcase1}Hawking temperature $T$ as a function of horizon radius $z_h$ for various values of $a$. Here $\mu_e=0$ and $q_M=0$ are used.]{\includegraphics[width=0.45\linewidth]{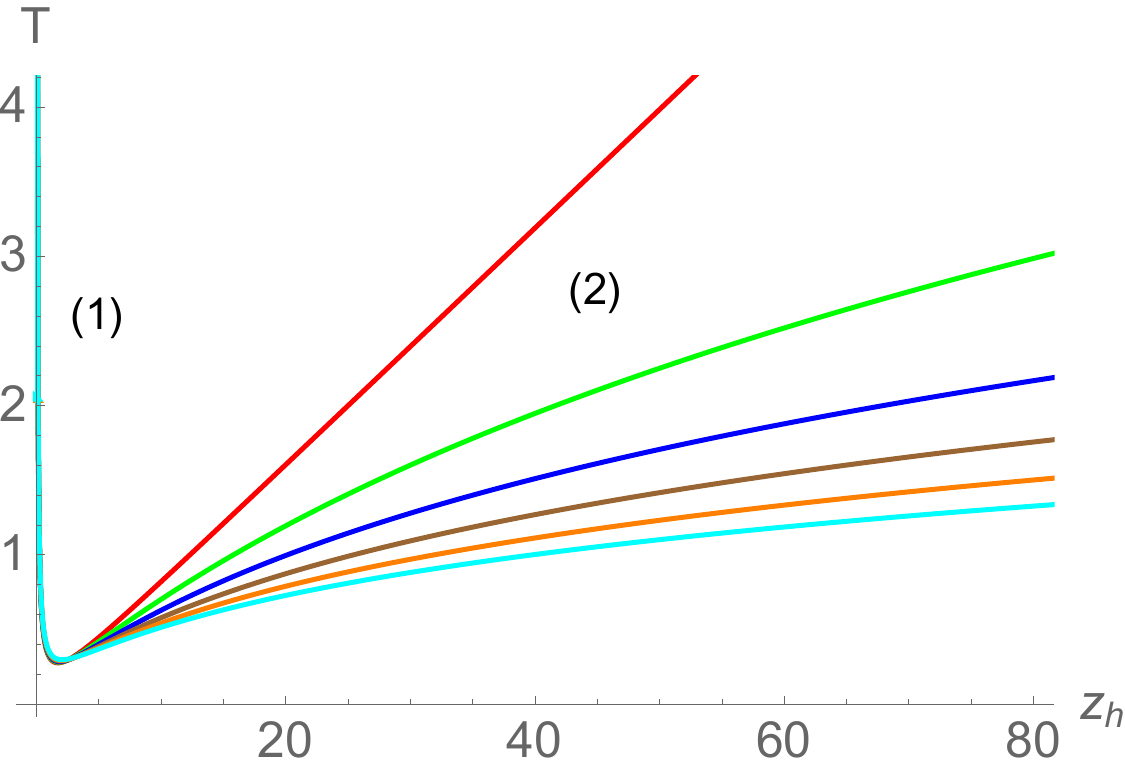}}	
		\hfill
		\subfigure[	\label{TvsdeltaGvsaMu0qM0sphcase1}Gibbs free energy difference $\Delta G$ as a function of Hawking temperature $T$ for various values of $a$. Here $\mu_e=0$ and $q_M=0$ are used.]{	\includegraphics[width=0.45\linewidth]{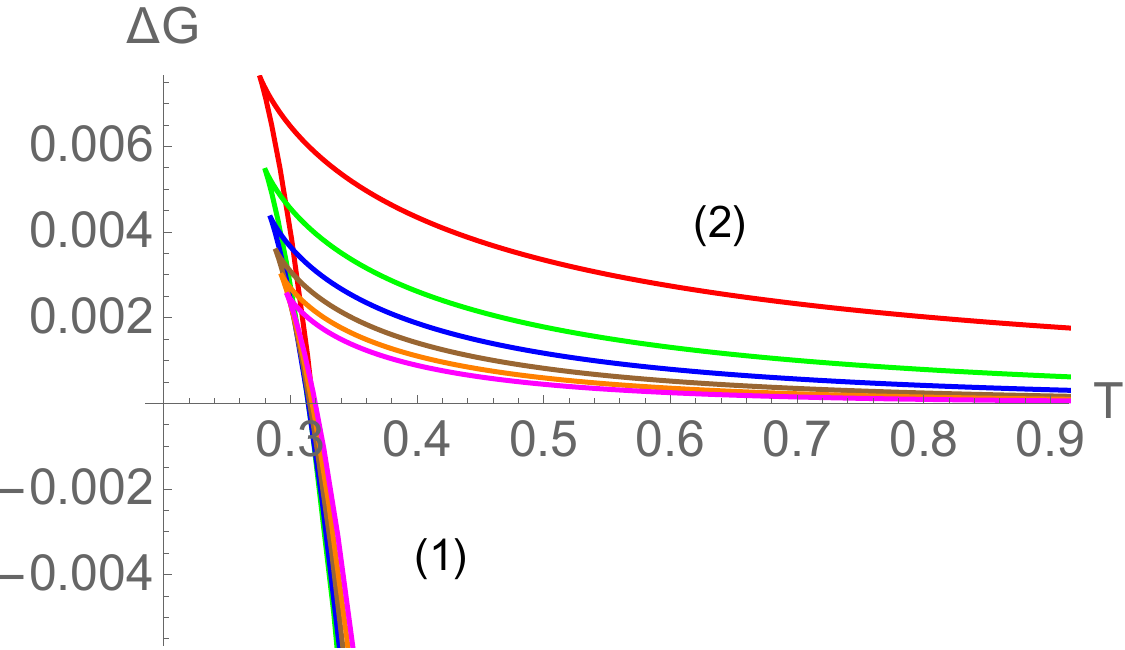}}
		\caption{Red, green, blue, brown, orange, and cyan curves correspond to $a=0$, $0.05$, $0.10$, $0.15$, $0.20$, and $0.25$ respectively.}\label{figg}
	\end{figure}

		In this section we discuss the thermodynamical behaviour of the dyonic hairy black holes discussed above. The variation of Hawking temperature with horizon radius for $\mu_e=0$ and $q_M=0$ is shown in Fig.~\ref{zhvsTvsaMu0qM0sphcase1}. The usual nonhairy Schwarzschild-AdS black hole exists only above a certain minimum temperature $T_{min}$, and below this minimum temperature the black hole does not exist, thereby exhibiting an interesting Hawking/Page transition
	between the Schwarzschild-AdS black hole and thermal-AdS. We find that this interesting behaviour and phase transition continue to hold in the presence of scalar hair as well. In particular, there again appear two black hole branches, large and small, at all temperatures $T\geq T_{min}$. The large black hole branch [indicated by (1) in Fig.~\ref{zhvsTvsaMu0qM0sphcase1}] has a positive specific heat and is stable, whereas the small black hole branch [indicated by (2) in Fig.~\ref{zhvsTvsaMu0qM0sphcase1}] has a negative specific heat and is unstable. Moreover, the free energy of the large black hole branch is always smaller than the small black hole branch. However, the free energy of the large black hole can become higher than the thermal-AdS at lower temperatures, implying the Hawking/page phase transition between them. This is shown in Fig.~\ref{TvsdeltaGvsaMu0qM0sphcase1}, where the free energy difference between hairy black hole and thermal-AdS is plotted for various values of $a$.
	
	The large black hole branch is also locally stable. This can be seen from the specific heat ($C=T(\frac{dS}{dT})$) behaviour. In particular, local stability measures the response of the equilibrium system under a small fluctuation in thermodynamical variables and can be quantified by the specific heat. We find that the specific heat is always positive for the large black hole branch. This can also be seen from the fact that the entropy $S \propto \frac{1}{h^2}$, and therefore, the slope of the $S-T$ curve is always positive for the large black hole branch.
	
	\begin{figure}[h!]
		\centering
		\subfigure[\label{one} Hawking temperature $T$ as a function of horizon radius $z_h$ for various values of $a$. Here $\mu_e=0.3$ and $q_M=0$ are used.]{	\includegraphics[width=0.45\linewidth]{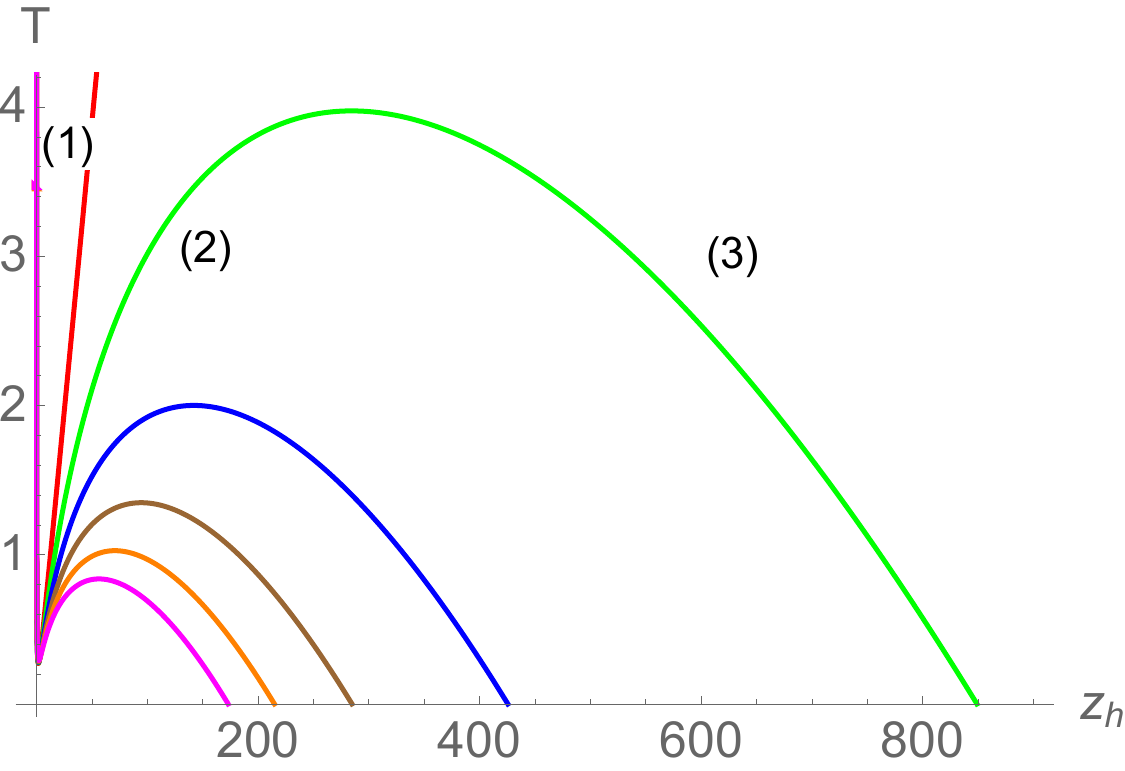}}	
		\hfill
		\subfigure[	\label{two} Gibbs free energy $\Delta G$ as a function of Hawking temperature $T$ for various values of $a$. Here $\mu_e=0.3$ and $q_M=0$ are used.]{				\includegraphics[width=0.45\linewidth]{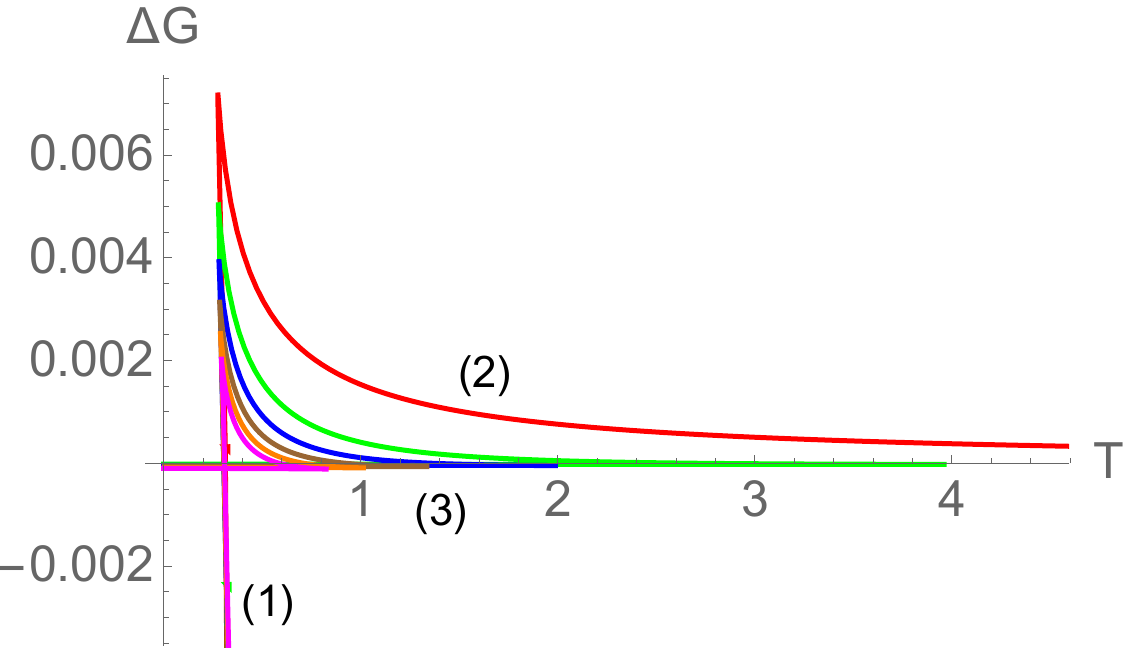}}
		\caption{Red, green, blue, brown, orange, and magenta curves correspond to $a=0$, $0.05$, $0.10$, $0.15$, $0.20$, and $0.25$ respectively.}\label{fig}
	\end{figure}
	
	
		The thermodynamic structure becomes even more interesting for the small but finite $\mu_E$. For a small $\mu_E$, still in the case $q_M=0$, the Hawking/page phase transition continues to exist, with a large stable black hole solution dominating the thermodynamics at a higher temperature, whereas the thermal-AdS solution dominating at a lower temperature. Interestingly, with scalar hair, a new small black hole branch can appear, which remains stable at low temperatures [indicated by (3) in Fig.~\ref{one}]. In particular, the Hawking temperature now has local minima and maxima and vanishes at a finite radius $h^{ext}$, \textit{i.e.}, at least one stable black hole branch always exists at each temperature. Moreover, the magnitude of this $h^{ext}$ decreases with $a$. The free energy behaviour, shown in Fig.~\ref{two}, further suggests a first-order phase transition between the large black hole branch (1) and small black hole branch (3) as the temperature is lowered. This is the famous small/large black hole phase in the context of charged AdS black hole \cite{Chamblin:1999tk,Chamblin:1999hg}. Notice that the free energy of the unstable second branch (2) is always higher than the stable first and third branches and therefore is always thermodynamically disfavoured.

At this point, it is important to remember that the usual spherical RN-AdS black hole exhibits only the Hawking/Page phase transition in the grand canonical ensemble, and the small/large black hole phase transition appears only in the canonical ensemble. Here, we see that in the presence of scalar hair the small/large black hole phase transition can take place in the grand-canonical ensemble as well.

For higher values of $a$, only one stable black hole branch appears that remains thermodynamically preferred at all temperatures. In particular, the size of the second unstable branch decreases with $a$ and then completely disappears. This leads to the merging of small and large black hole branches to form a single black hole branch that remains stable at all temperatures $T\geq 0$. Therefore, the small/large phase transition ceases to exist at higher values of $a$. A similar scenario persists for larger values of chemical potential as well. Overall, this thermodynamic behaviour in the grand canonical ensemble is analogous to the famous Van der Waals type phase transition, where a first-order critical line stops at a second-order critical point.
	
	\begin{figure}[h!]\label{energy}
		\centering
		\subfigure[\label{three} Hawking temperature $T$ as a function of horizon radius $z_h$ for various values of $a$. Here $q_e=0.1$ and $q_m=0$ are used.]{	 \includegraphics[width=0.45\linewidth]{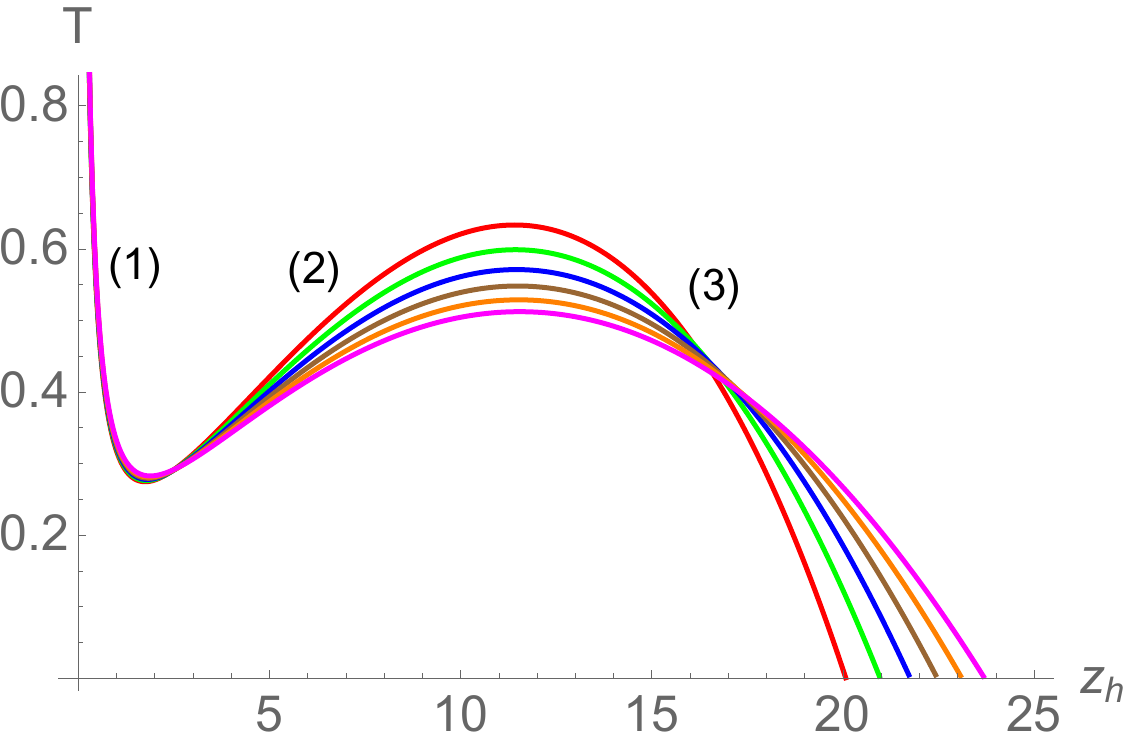}}	
		\hfill
		\subfigure[\label{four} Helmholtz free energy $F$ as a function of Hawking temperature $T$ for various values of $a$. Here $q_e=0.1$ and $q_m=0$ are used. ]{\includegraphics[width=0.45\linewidth]{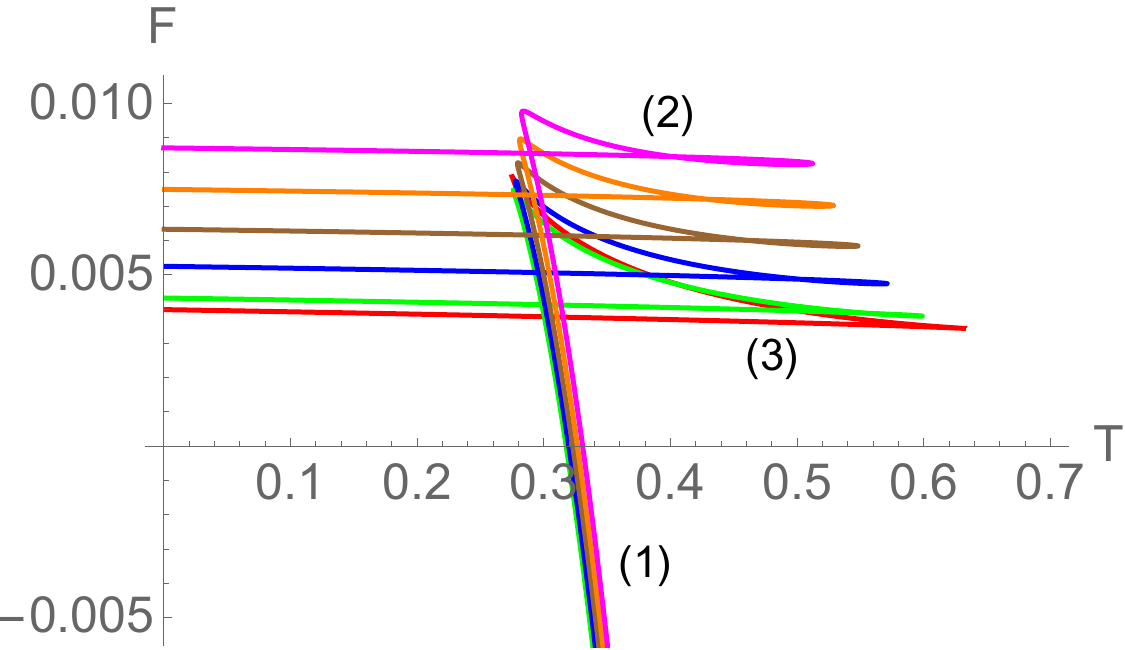}}
		\caption{Red, green, blue, brown, orange, and magenta curves correspond to $a=0.0$, $0.02$, $0.04$, $0.06$, $0.08$, and $0.1$ respectively.}
	\end{figure}\label{free}
	
\begin{figure}[h!]
		\centering
		\subfigure[\label{five} Hawking temperature $T$ as a function of horizon radius $z_h$ for various values of $a$. Here $q_M=0$ and $a=0.02$ are used.]{		\includegraphics[width=0.45\linewidth]{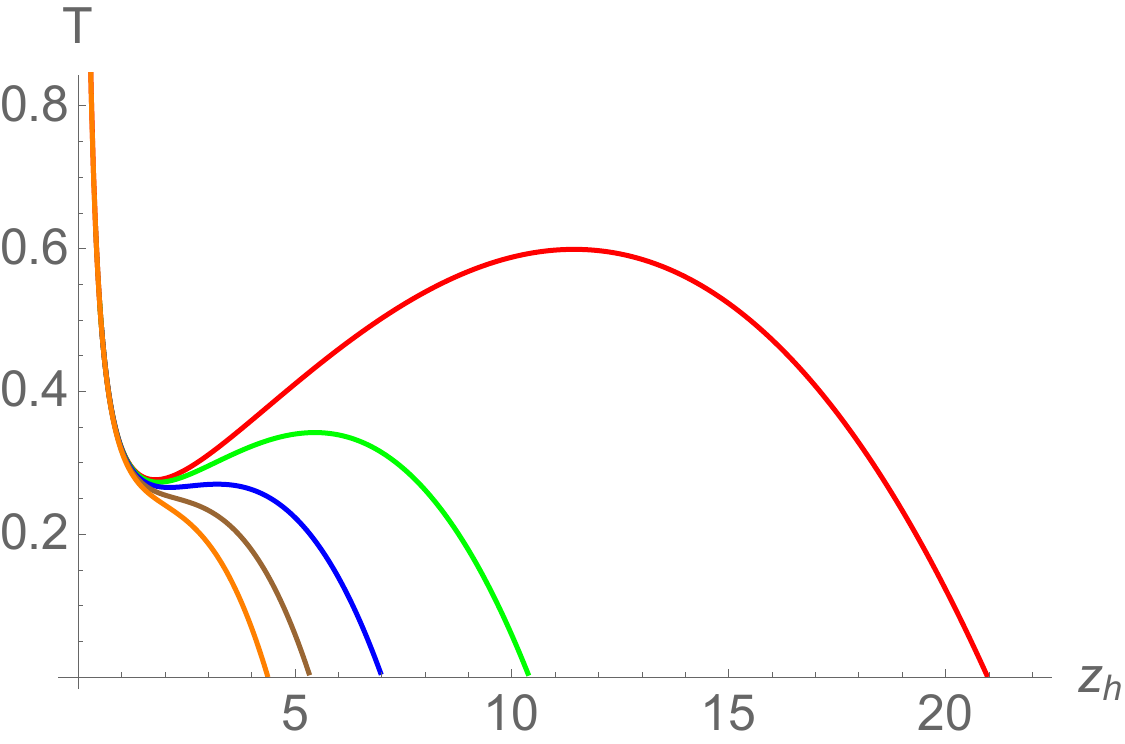}}	
		\hfill
		\subfigure[	\label{six} Helmholtz free energy $F$ as a function of Hawking temperature $T$ for various values of $a$. Here $q_M=0$ and $a=0.02$ are used.]{		\includegraphics[width=0.45\linewidth]{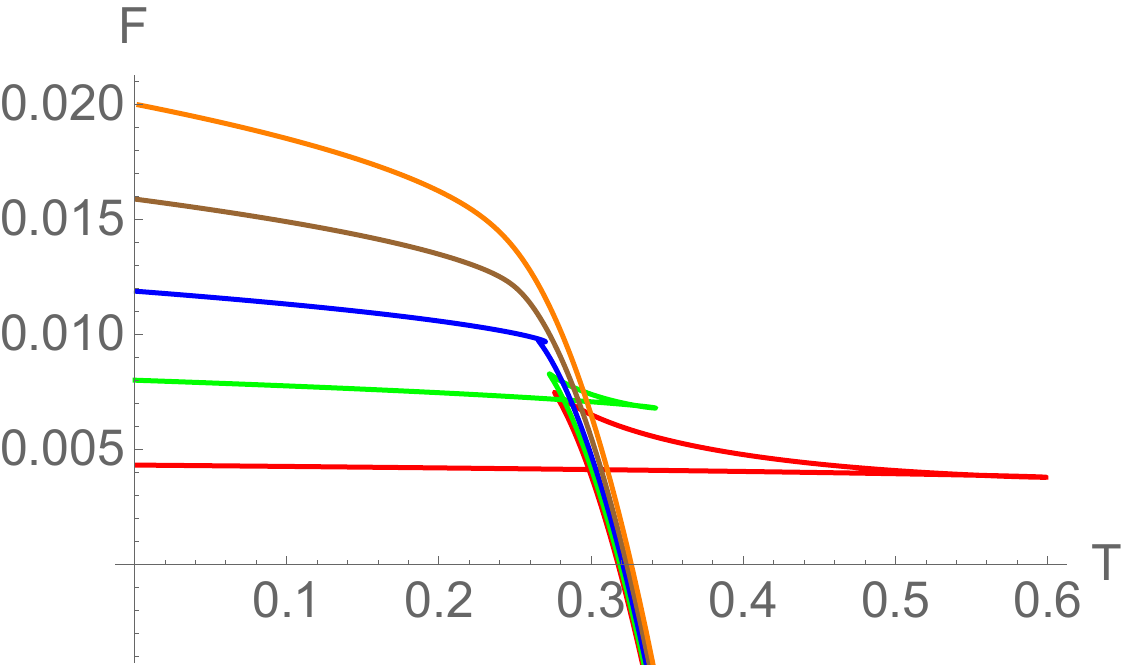}}
		\caption{Red, green, blue, brown, and orange curves correspond to $q_e=0.1$, $0.2$, $0.3$, $0.4$, and $0.5$ respectively.}
	\end{figure}

	
		Similar results appear in the fixed charge canonical ensemble. The thermodynamic behaviour of the $q_e=0$ case is exactly similar to the $\mu_e=0$ case discussed above. The thermodynamic behaviour for a small but finite fixed charge $q_E$ is shown in Figs.~\ref{three} and \ref{four}. It is well known that the usual charged RN-AdS spherical black hole exhibits a swallow-tail like structure in the free energy and undergoes a small/large black hole phase transition as the temperature is altered in the canonical ensemble \cite{Chamblin:1999tk}. We find that similar results persist for the charged hairy cases as well. The difference arises in the magnitude of  the small/large black hole transition temperature, which increases as the parameter $a$ increases. Similarly, most of the results for higher $q_e$ values persist as well. In particular, there appears a critical charge $q_e^{crit}$ above which the unstable branch [indicated by (2) in Figs.~\ref{three} and \ref{four}] disappears and we have a single black hole branch, which is stable at all temperatures, \textit{i.e.}, the small/large black hole phase transition ceases to exist above $q_e^{crit}$. This is shown in Figs.~\ref{five} and \ref{six}. The $q_e^{crit}$, therefore, defines a second-order critical point on which the first-order small/large black hole phase transition line stops. As usual, the magnitude of $q_e^{crit}$ can be found by analysing the inflexion point of temperature.
	
	Let us also mention that the thermodynamic behaviour of the hairy black hole remains qualitatively the same for different values of $q_M$ in the canonical as well as grand canonical ensemble. In the next section we will discuss about scalar field perturbation around these hairy black holes.
	
	\section{Scalar field perturbation and QNMs}\label{section4}
	We now study the dynamics of massless scalar field perturbation in the background geometry of the constructed hairy dyonic black holes. The main aims of this section are to first analyse the QNM spectrum of scalar field perturbation and then to see whether the signature of thermodynamical black hole phase transition can be reflected in the QNM of massless scalar field. To compute the QNM, we employ two different numerical methods.
	
	The massless scalar field perturbation in a curved background is governed by the Klein-Gordon equation, \footnote{This scalar field $\Phi$ here should not be confused with the hair scalar field $\phi$ used in the previous sections to construct hairy dyonic black holes.}
	\begin{equation}\label{laplace}
		\nabla_{\mu}\nabla^{\mu}\Phi(x)=\frac{1}{\sqrt{-g}}\partial_{\mu}(\sqrt{-g}g^{\mu\nu}\partial_\nu\Phi)=0\,.
	\end{equation}
We are mainly interested in the positive frequency mode solution of Eq.~(\ref{laplace}). We consider the following factorization
	\begin{equation}\label{spherical}
		\Phi=\frac{\psi_{\omega l}(r)}{r} Y_{lm}(\theta, \phi) e^{-i\omega t}, ~~~ (\omega>0)
	\end{equation}
	where $Y_{lm}(\theta,\phi)$ represents spherical harmonics. Using eqs.~(\ref{metric}) and (\ref{spherical}), the Klein-Gordon equation reduces to the following ordinary differential equation,
	\begin{equation}\label{ode}
		\left(-\frac{k(r)}{e^{A(r)}}\frac{d}{dr}\left(\frac{k(r)}{e^{A(r)}}\frac{d}{dr}\right)+\left[\omega ^2-V_{eff}(r) \right] \right)\psi (r)=0\,,
	\end{equation}
	where the effective potential is given by
	\begin{equation}
		V_{eff}(r)=\frac{k(r)}{e^{A(r)}}\left[\frac{l(l+1) e^{A(r)}}{r^2}+\frac{g(r)}{e^{A(r)}}+\frac{d}{dr}\left(\frac{g(r) r}{ e^{A(r)}}\right)\right] \,,
	\end{equation}
with
	\begin{equation}
		k(r)=g(r) r^2\,.
	\end{equation}
Using Tortoise coordinate $r*=\int dr\frac{e^{A(r)}}{k(r)}$, we can also recast the above wave equation in to the Schr\"{o}dinger form
	\begin{equation}
		\left(\frac{\partial^2}{\partial r_*^2}+\left[\omega^2-V_{eff}(r)\right]\right)\psi(r)=0\,.
\label{waveeqscalar}
	\end{equation}

	\begin{figure}[hbt!]
		\centering
		\subfigure[\label{ff}$r$ vs $V_{eff}(r)$ behaviour for  $q_M=0.3$ and $q_e=0.2$ for different values of $a$. Here $z_h=1$ is used.]{\includegraphics[width=0.41\linewidth]{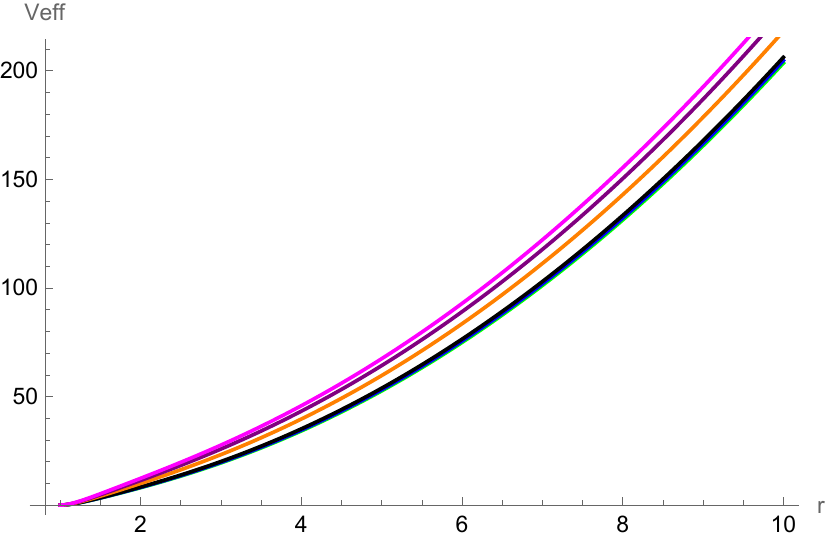}}
		\hfill
		\subfigure[\label{ff1}$r$ vs $V_{eff}(r)$ behaviour for $q_M=0.3$ and $\mu_e=0.2$ for different values of $a$. Here $z_h=1$ is used.]{\includegraphics[width=0.41\linewidth]{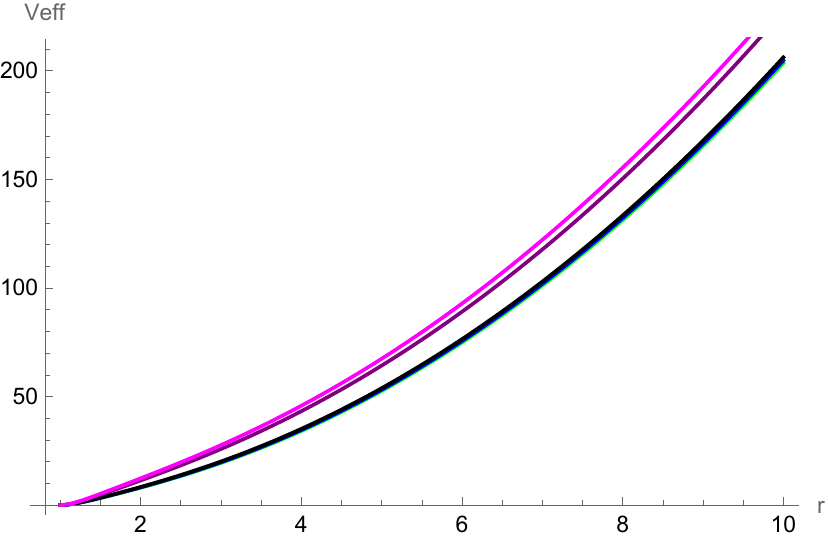}}
		\caption{ Here green, blue, black, orange, purple, and magenta curves correspond to $a=0.01, 0.05, 0.1, 0.5, 0.8$, and $1.0$ respectively.}\label{effectivepotential}
	\end{figure}

The behaviour of the aforementioned effective potential is shown in Figs.~\ref{ff} and \ref{ff1}. Notice that the effective potential is positive outside the horizon, and it vanishes at the horizon. Also, the effective potential increases with an increase in hairy parameters. This suggests that the hairy parameter might modify QNM values nontrivially. The dynamics of the scalar field can be analysed in a without-horizon background (such as the thermal-AdS). This leads to ordinary normal modes with a discrete set of real $\omega$ in AdS spacetime. However, in the presence of a black hole, the field can now fall into the black hole and decay. In this case, a complex set of frequencies  generally describe the decay of perturbations outside the horizon. Quasinormal modes are defined as solutions which are purely outgoing $\Phi\sim e^{-i \omega(t-r_*)}$ near the boundary and purely ingoing $\Phi\sim e^{-i \omega(t+r_*)}$ near the horizon. The QNMs in the black hole background therefore consist of a discrete set of complex frequencies $\omega$.  Below, we will compute the QNMs values numerically for various values of the hairy parameter and analyse the dynamical stability of the hairy black holes. For the numerical computation, we used two different techniques: (i) the power series method; and (ii) the shooting method. Below, we will first discuss these techniques in detail and then subsequently discuss the numerical results.

	\subsection{QNMs using the power series method}
	We first adopt the Horowitz-Hubney power series method to calculate the QNMs \cite{horowitz2000quasinormal}. Here, we first take the following form of the perturbative scalar field
	\begin{equation}\label{eq20}
		\psi(r)=e^{-i \omega r*} \zeta(r)\,,
	\end{equation}
and recast the wave equation (\ref{waveeqscalar}) into the following form
	\begin{equation}
		\Lambda(r)\frac{d^2 \zeta(r)}{d^2 r}+\left[(\Lambda'(r)-2 i\omega)\right]\zeta(r)-\mathcal{V}(r)\zeta(r)=0 \,,
\label{waveequation}
	\end{equation}
	where $\Lambda(r)=\frac{k(r)}{e^{A(r)}}$ and $\mathcal{V}(r)=\frac{V_{eff}}{\Lambda(r)}$. To compute the QNMs, one then expand the solution in a power series about the black hole horizon and impose the boundary condition that the solution vanish at the asymptotic boundary. For numerical purposes, it would be better to consider the coordinate $x=\frac{1}{r}$, in terms of which the whole spacetime maps into a finite range $0\leq x \leq h$. In terms of $x$, we have the wave equation
	\begin{equation}\label{19}
		(x-h)s(x)\frac{d^2}{dx^2}\zeta(x)+t(x)\frac{d\zeta(x)}{dx}+\frac{u(x)}{(x-h)}\zeta(x)=0 \,,
	\end{equation}
	where the various coefficients are $s(x)=\frac{-x^4 \Lambda(x)}{(x-h)}$, $t(x)=-(2 \Lambda(x)x^3+\Lambda'(x)x^2+2i\omega x^2)$, and $u(x)=\frac{\mathcal{V}(x) (x-h)}{\Lambda(x)}$. Next, we consider the
	 Forbenius solution near the horizon
	\begin{equation}\label{20}
		\zeta(x)=(x-h)^\alpha\displaystyle\sum_{n=0}^{\infty}a_n (x-h)^n \,.
	\end{equation}
By substituting Eq.~(\ref{20}) to Eq.~(\ref{19}), we get the indicial equation $\alpha(\alpha-1)s_0+\alpha t_0+u_0=0$. From this, we get $\alpha=0$ and $\alpha=\frac{i\omega}{\Lambda'(r_h)}$ along with $u_0=0$. As we are interested only in ingoing mode at the horizon, we take $\alpha=0$. The near horizon solution for $\zeta$ then takes the form
	\begin{equation}
		\zeta(x)=\displaystyle\sum_{n=0}^{\infty}a_n(\omega)(x-h)^n \,.
	\end{equation}
Similarly, the polynomials $s$, $t$ and $u$ can be expended near the horizon
	\begin{eqnarray}
		s(x)=\displaystyle\sum_{n=0}^{\infty}s_n (x-h)^n,~~ t(x)=\displaystyle\sum_{n=0}^{\infty}t_n (x-h)^n, ~~u(x)=\displaystyle\sum_{n=0}^{\infty}u_n (x-h)^n \,,
	\end{eqnarray}
with $s_0=2h^2\kappa$, $t_0=2h^2(\kappa-i \omega)$, and $u_0=0$. Here $\kappa=\Lambda'(h)/2$ is the surface gravity. Comparing the power of $(x-h)^{n-2}$, one similarly gets the recurrence relation
	\begin{equation}
		a_n=-\frac{\displaystyle\sum_{k=0}^{n-1}
			\left[k(k-1)S_{n-k}+kt_{n-k}+u_{n-k}\right]a_k}{n(n-1) s_0+n t_0} \,.
	\end{equation}
We next impose the boundary condition at the asymptotic boundary
	\begin{equation}
		\zeta(x=0)=\displaystyle\sum_{n=0}^{\infty}a_n (-h)^n=0\,.
		\label{23}
	\end{equation}
For a fixed value of black hole parameters, the QNM values are then given by the roots of Eq.~(\ref{23}). Note that various coefficients $a_n$ are $\omega$ dependent. In the expression (\ref{23}) the full sum can not be implemented in numerical computations. Therefore, we take the partial sum from $0$ to some finite $N$ and find $\omega$ values from the resulting polynomial expression. In our case, we find that for $N=100$, the QNM value converges.

Before we explicitly discuss our numerical results from the power series method, it is also possible to make some general remarks about the nature of the QNMs. For instance, it can be shown that there can be no solutions with $i\omega$ pure real. If $i\omega$ were real, then the whole wave equation (\ref{waveequation}) would be real and the solutions $\zeta$ would have to be real. This, in turn, implies that if there
were a local extremum at some point, let's say $\tilde{r}$, then $\zeta'(\tilde{r})=0$ and $\zeta''(\tilde{r})$ would have the same
sign as $\zeta(\tilde{r})$. So if $\zeta$ were positive (negative) at $\tilde{r}$, it would have to increase (decrease) as $r$ increased. Therefore, $\zeta$ could never approach zero asymptotically as opposed to the requirement of the boundary condition (\ref{23}).

We can also make other concrete comments about the nature of the QNMs. In particular, it can be analytically shown that the imaginary part of the QNM would always be negative in the hairy black hole background. To show this, it is convenient to work with the ingoing Eddington coordinates $v=t+r_*$. In terms of this coordinate, the metric (\ref{metric}) becomes
	\begin{eqnarray}\label{metric1}
		ds^2=
		-k(r) dv^2+2 e^{A(r)} dv dr +r^2\left(d y_1^2+\sin^2{y_1} d y_2^2\right) \,.	
	\end{eqnarray}
The tortoise coordinate $r_*$ near horizon $r=r_h$ then takes the form
	\begin{equation}
		r_*=\int dr \frac{e^{A(r)}}{k(r)}\approx\frac{e^{A(r)}}{k'(r_h)} \ln(r-r_h) \,.
\label{tortoise}
	\end{equation}
Now multiplying Eq.~(\ref{waveequation}) by $\bar{\zeta}$ and integrating from $r_h$ to $\infty$, we get
	\begin{eqnarray}\label{eqn30}
		\int_{r_h}^\infty \left[\bar{\zeta}(r)\Lambda(r)\frac{d^2 \zeta(r)}{d^2 r}
		+(\Lambda'(r)-2 i\omega)\zeta'(r)\bar{\zeta}(r)-\mathcal{V}(r)\zeta(r)\bar{\zeta}(r)\right] dr=0\,,
	\end{eqnarray}
which is equivalent to
	\begin{eqnarray}\label{89}
		\int_{r_h}^\infty \left[\bar{\zeta}(r)\frac{d}{dr}\left(\Lambda(r)\frac{d\zeta(r)}{dr}\right)
		-2 i\omega\zeta'(r)\bar{\zeta}(r)-\mathcal{V}(r)\zeta\bar{\zeta}(r)\right] dr=0\,.
	\end{eqnarray}
Using the fact that $\Lambda(r_h)=0$ and $\zeta(\infty)=0$, the above equation can be further simplified to
\begin{eqnarray}
		\int_{r_h}^\infty \left[ \Lambda(r)|\zeta'(r)|^2 + 2i\omega \bar{\zeta}\zeta'(r) + \mathcal{V}(r) |\zeta(r)|d^2   \right] dr=0 \,.
\label{waveeqimgpart}
	\end{eqnarray}
Notice that the first and third terms in the above equation are always real. The imaginary part of (\ref{waveeqimgpart}) then yields
	\begin{equation}
		\int_{r_h}^{\infty}\left[\omega \bar{\zeta}(r)\zeta'(r)+\bar{\omega}\bar{\zeta}'(r)\zeta(r)\right] dr=0 \,,
	\end{equation}
by integrating the second term by parts, we get
	\begin{equation}
		(\omega-\bar{\omega})\int_{r_h}^{\infty}\zeta(r)\zeta'(r) dr=\bar{\omega}|\zeta(r_h)|^2 \,.
	\end{equation}
Substituting this back into the Eq.~(\ref{waveeqimgpart}), gives us the relation
	\begin{equation}
		\int_{r_h}^{\infty}\left[\Lambda(r)|\zeta(r)'|^2+\mathcal{V}(r)|\zeta(r)|^2\right] dr=-\frac{|\omega^2| |\zeta(r_h)|^2}{\text{Im}\omega}\,.
	\end{equation}
Since $\Lambda(r)$ and $\mathcal{V}(r)$ are both positive, the only solutions that can exist for $\zeta$ is when $\text{Im}\omega<0$, i.e., there are no solutions with $\text{Im}\omega>0$. The positive values of $\text{Im}\omega$ indicate instabilities in the background geometry as the mode would then grow with time. The above analysis therefore suggests that the hairy dyonic hair black holes are stable against the malssless scalar field perturbation.

\subsection{QNMs using the shooting method}
We now illustrate the second method to compute the QNM numerically. In the shooting method, one sets up the initial value problem at the horizon and then integrate the equation of motion from the horizon to the asymptotic boundary. The QNMs are then determined by demanding the perturbation $\psi$ to vanish at the boundary \cite{liu2014signature,Mahapatra:2016dae}.

To illustrate this procedure, let us take the wave equation (\ref{ode}) and impose the ingoing wave boundary condition at the horizon. This is mathematically equivalent to requiring $\psi \propto (r-r_h)^{-i\omega/4 \pi T}$ at the horizon. To impose the ingoing boundary condition, we use
\begin{equation}
\psi(r)= \zeta(r) e^{-i \omega r*}  = \zeta(r) e^{ -i \omega \int dr \frac{e^{A(r)}}{k(r)}} \,,
\end{equation}
where, using Eq.~(\ref{tortoise}), one can easily show that $\psi$ approaches $(r-r_h)^{-i\omega/4 \pi T}$ at the horizon. In terms of $\zeta$, the wave equation reduces to
\begin{equation}
	\zeta''(r)+\zeta'(r)\left[\frac{k'(r)}{k(r)}-A'(r)-\frac{2 i \omega e^{A(r)}}{k(r)}\right]+\zeta(r)\left[\frac{l(l+1) e^{A(r)}}{r^2 k(r)}+\frac{A'(r)}{r}-\frac{k'(r)}{r k(r)}\right]=0\,.
\label{waveeqshooting}
\end{equation}
Now we setup the initial value problem for $\zeta$ at the horizon. For this purpose, we use a different coordinate $\mathcal{Y}=r_h/r$ and take a series solution for $\zeta$ as
\begin{equation}
	\zeta(z)= \alpha_0 + \alpha_1 (\mathcal{Y}-1)+ \alpha_2 (\mathcal{Y}-1)^2+\ldots \,,
	\end{equation}
where the coefficients $\alpha_i$ are determined by substituting the above series expansion back into the wave equation (\ref{waveeqshooting}) and equating the coefficient of each power of $(\mathcal{Y}-1)$ to zero.\footnote{Using the linearity of the wave equation, we can set $\alpha_0=1$} Once the profile of $\zeta$ (and hence $\zeta'$) at the horizon is set, we can numerically integrate the wave equation from the horizon to the boundary. The QNMs are then again determined by imposing the condition $\zeta (\mathcal{Y}\rightarrow 0) = 0$ at the asymptotic boundary, i.e., the boundary condition $\zeta (0) = 0$ is satisfied only for a particular set of frequencies $\omega$.

\subsection{Numerical results}\label{NMR}
We now discuss our numerical results of QNMs. Among the various modes, the one with the lowest frequency corresponds to the primary mode, which plays a dominant role in characterizing the stability and dynamical behaviour of the black hole under perturbation. Here we concentrate only on the primary mode. Similarly, we mainly discuss results with $l=0$ case here as results with $l\neq 0$ can be analogously obtained.

\begin{table}[ht!]
	\setlength\tabcolsep{2pt}
	\footnotesize
	\renewcommand{\arraystretch}{1}
	\centering
	\begin{threeparttable}
		\begin{minipage}{0.5\linewidth}
			\centering
			\begin{tabular}{@{}
					r S[table-format=-2.3] S[table-format=-2.3]
					@{\hspace{12pt}} !{\vrule width 0.2pt} @{\hspace{12pt}}
					r S[table-format=-2.3] S[table-format=-2.3]
					@{}}
				\toprule
				\multicolumn{1}{c}{$r_h$} & $\omega_{Re}$ & \multicolumn{1}{c}{$\omega_{Im}$} &
				\multicolumn{1}{c}{$r_h$} & $\omega_{Re}$ & $\omega_{Im}$ \\
				\midrule[0.08cm]
				\multicolumn{6}{c}{\textbf{\textbf{$a=0.1$}}}\\
			~	1  & 2.947 & -2.737  & 5  &9.690 & -13.398 \\
			10  & 18.830 & -26.714   & 25  & 46.505 & -66.671 \\
			\multicolumn{6}{c}{\textbf{\textbf{$a=0.2$}}} \\
			~	1 & 3.101 & -2.802   & 5   & 9.907 & -13.470 \\
			10  & 19.052 & -26.788  & 25  & 46.729 & -66.744 \\
			\multicolumn{6}{c}{\textbf{\textbf{$a = 0.3$}}} \\
			~	1 & 3.101 & -2.802   & 5   & 10.122 &-13.542 \\
			10  & 19.272 & -26.860  & 25  & 46.952 & -66.817 \\
			\bottomrule
			\end{tabular}
		\caption{QNMs obtained using the shooting method with $q_M=0$, $q_e=0.1$.}\label{shootingtable}
		\end{minipage}
		\hfil
		\begin{minipage}{0.5\linewidth}
			\centering
			\begin{tabular}{@{}
					r S[table-format=-2.3] S[table-format=-2.3]
					@{\hspace{12pt}} !{\vrule width 0.2pt} @{\hspace{12pt}}
					r S[table-format=-2.3] S[table-format=-2.3]
					@{}}
				\toprule
				\multicolumn{1}{c}{$r_h$} & $\omega_{Re}$ & \multicolumn{1}{c}{$\omega_{Im}$} &
				\multicolumn{1}{c}{$r_h$} & $\omega_{Re}$ & $\omega_{Im}$ \\
				\midrule[0.08cm]
				\multicolumn{6}{c}{\textbf{\textbf{$a=0.1$}}}\\
				~	1  & 2.947 & -2.737  & 5  &9.690 & -13.398 \\
			10  & 18.830 & -26.715   & 25  & 46.505 & -66.671 \\
			\multicolumn{6}{c}{\textbf{\textbf{$a = 0.2$}}} \\
			~	1 & 3.101 & -2.802   & 5   & 9.907 & -13.470 \\
			10  & 19.052 & -26.788  & 25  & 46.729 & -66.744 \\
				\multicolumn{6}{c}{\textbf{\textbf{$a = 0.3$}}} \\
				~	1 & 3.101 & -2.802   & 5   & 10.122 &-13.542 \\
				10  & 19.272 & -26.860  & 25  & 46.952 & -66.817 \\
				\bottomrule
			\end{tabular}
			\caption{QNMs obtained using the power series method with $q_M=0$, $q_e=0.1$.}\label{powers}
		\end{minipage}
	\end{threeparttable}
\end{table}

In Tables \ref{shootingtable} and \ref{powers}, we have compared the QNM values obtained using the shooting and power series methods for different values of the hairy parameter and horizon radius. We find that the results from both these methods agree up to the fourth decimal place. Here we have decomposed the QNM into real and imaginary parts $\omega=\omega_{Re} + i \omega_{Im}$. We have checked for other values of $a$ as well, and found that the QNMs values from both these methods match reasonably well.

\begin{table}[htb]
	\setlength\tabcolsep{2pt}
	\footnotesize
	\renewcommand{\arraystretch}{1}
	\centering
	\begin{threeparttable}
		\begin{minipage}{0.5\linewidth}
			\centering
			\begin{tabular}{@{}
					r S[table-format=-2.3] S[table-format=-2.3]
					@{\hspace{12pt}} !{\vrule width 0.2pt} @{\hspace{12pt}}
					r S[table-format=-2.3] S[table-format=-2.3]
					@{}}
				\toprule
				\multicolumn{1}{c}{$r_h$} & $\omega_{Re}$ & \multicolumn{1}{c}{$\omega_{Im}$} &
				\multicolumn{1}{c}{$r_h$} & $\omega_{Re}$ & $\omega_{Im}$ \\
				\midrule[0.08cm]
				\multicolumn{6}{c}{\textbf{\textbf{$a=0.01$}}}\\
				~	1 &2.811 &-2.678  & 5 & 9.493 & -13.332 \\
				10  & 18.629 & -26.649 & 25 & 46.303 & -66.605 \\
				50 & 92.516 & -133.201  & 100 & 184.981 & -266.401 \\
				125 &231.100 & -332.833 & 200 & 368.978 & -531.432 \\
				\multicolumn{6}{c}{\textbf{$ a=0.1$}} \\
				~	1  & 2.947 & -2.737  & 5  &9.690 & -13.398 \\
				10  & 18.830 & -26.715   & 25  & 46.505 & -66.671 \\
				50  & 92.718 & -133.266  & 100 & 185.178 & -266.459  \\
				125 & 231.411 & -333.055  & 200 &370.115 & -532.844 \\
				\multicolumn{6}{c}{\textbf{\textbf{$a = 0.2$}}} \\
				~	1 & 3.101 & -2.802   & 5   & 9.907 & -13.470 \\
				10  & 19.052 & -26.788  & 25  & 46.729 & -66.744 \\
				50  & 92.942 & -133.340  & 100 & 185.403 & -266.532 \\
				125 & 231.636 & -333.128  & 200 & 370.339 & -532.918 \\
				\bottomrule
			\end{tabular}
			\caption{$q_M=0$, $q_e=0.1$}\label{tabqM0qe0pt1}
		\end{minipage}
		\hfil
		\begin{minipage}{0.5\linewidth}
			\centering
			
			\begin{tabular}{@{}
					r S[table-format=-2.3] S[table-format=-2.3]
					@{\hspace{12pt}} !{\vrule width 0.2pt} @{\hspace{12pt}}
					r S[table-format=-2.3] S[table-format=-2.3]
					@{}}
				\toprule
				\multicolumn{1}{c}{$r_h$} & $\omega_{Re}$ & \multicolumn{1}{c}{$\omega_{Im}$} &
				\multicolumn{1}{c}{$r_h$} & $\omega_{Re}$ & $\omega_{Im}$ \\
				\midrule[0.08cm]
				\multicolumn{6}{c}{\textbf{\textbf{$a=0.01$}}}\\
				~		1 & 2.799 & -2.687 & 5 & 9.492 & -13.332 \\
				10 & 18.629 & -26.649 & 25 & 46.303 & -66.596 \\
				50    & 92.516 &-133.201  & 100   & 184.982	&-266.401 \\
				125   & 231.100	&-332.833  & 200   & 368.978&-531.432 \\
				\multicolumn{6}{c}{\textbf{$ a=0.1$}} \\
				~	1     &2.936	&-2.745   & 5     & 9.689	&-13.398 \\
				10    & 18.830	&-26.715  & 25    &46.505	&-66.671 \\
				50  & 92.718 & -133.266  & 100 & 185.178 & -266.459  \\
				125   & 231.411	&-333.055  & 200   & 370.115	&-532.844 \\
				\multicolumn{6}{c}{\textbf{\textbf{$a = 0.2$}}} \\
				~	1     &3.091	&-2.809  & 5     & 9.906	&-13.470 \\
				10    & 19.052  &-26.788  & 25    & 46.729	&-66.744  \\
				50    & 92.942 &-133.340  & 100   & 185.402	&-266.532 \\
				125   &231.636	&-333.128  & 200   & 370.339	&-532.917  \\
				\bottomrule
			\end{tabular}
			\caption{$q_M=0$, $q_e=0.3$}\label{tabqM0qe0pt3}
		\end{minipage}
		\hfil
		\begin{minipage}{0.5\linewidth}
			\centering
			\begin{tabular}{@{}
					r S[table-format=-2.3] S[table-format=-2.3]
					@{\hspace{12pt}} !{\vrule width 0.2pt} @{\hspace{12pt}}
					r S[table-format=-2.3] S[table-format=-2.3]
					@{}}
				\toprule
				\multicolumn{1}{c}{$r_h$} & $\omega_{Re}$ & \multicolumn{1}{c}{$\omega_{Im}$} &
				\multicolumn{1}{c}{$r_h$} & $\omega_{Re}$ & $\omega_{Im}$ \\
				\midrule[0.08cm]
				\multicolumn{6}{c}{\textbf{\textbf{$a=0.01$}}}\\
				~		1 &2.813 & -2.677 & 5 & 9.493 &-13.332 \\
				10 &18.629 & -26.649 & 25 &46.303 &-66.605 \\
				50 &92.516 &-133.201 & 100 & 184.982 & -266.401 \\
				125 &231.100 &-332.833 & 200 & 368.978 &-531.432 \\
				\multicolumn{6}{c}{\textbf{$ a=0.1 $}} \\
				~		1 &2.949 &-2.736 & 5 & 9.690 & -13.398 \\
				10 & 18.830 & -26.715 & 25 & 46.505 & -66.671\\
				50 &92.718 & -133.266 & 100 & 185.178 & -266.459 \\
				125 & 231.411 & -333.055 & 200 & 370.115&-532.844 \\
				\multicolumn{6}{c}{\textbf{\textbf{$a = 0.2$}}} \\
				~		1 &3.102 & -2.801 & 5 & 9.907 &-13.471 \\
				10 & 19.052 & -26.788 & 25 &46.729 &-66.744 \\
				50 &92.943 & -133.339 & 100 & 185.403 & -266.532 \\
				125 & 231.636 &-333.129 & 200 & 370.339 &-532.917 \\
				\bottomrule
			\end{tabular}
			\caption{$q_M=0$, $q_e=0$}\label{tabqM0qe0}
		\end{minipage}
		\hfil
		\begin{minipage}{0.5\linewidth}
			\centering
			
			\begin{tabular}{@{}
					r S[table-format=-2.3] S[table-format=-2.3]
					@{\hspace{12pt}} !{\vrule width 0.2pt} @{\hspace{12pt}}
					r S[table-format=-2.3] S[table-format=-2.3]
					@{}}
				\toprule
				\multicolumn{1}{c}{$r_h$} & $\omega_{Re}$ & \multicolumn{1}{c}{$\omega_{Im}$} &
				\multicolumn{1}{c}{$r_h$} & $\omega_{Re}$ & $\omega_{Im}$ \\
				\midrule[0.08cm]
				\multicolumn{6}{c}{\textbf{\textbf{$a=0.01$}}}\\
				1     & 2.801	&-2.686  & 5     &9.492	&-13.332 \\
				10 & 18.629 & -26.649 & 25    & 46.303	&-66.605 \\
				50    &92.516	&-133.201  & 100   & 184.982	&-266.401 \\
				125   &231.100	&-332.833   & 200   & 368.977	&-531.432 \\
				\multicolumn{6}{c}{\textbf{$ a=0.1 $}} \\
				1     & 2.938	&-2.744   & 5     & 9.689	&-13.398 \\
				10    & 18.830	&-26.715  & 25    & 46.505	&-66.671 \\
				50    & 92.718 &-133.266  & 100   & 185.178	&-266.459\\
				125   & 231.411	&-333.055  & 200   & 370.115	&-532.844 \\
				\multicolumn{6}{c}{\textbf{\textbf{$a = 0.2$}}} \\
				1     &3.092	&-2.808   & 5     & 9.906	&-13.470 \\
				10    & 19.052  &-26.788   & 25    & 46.729	&-66.744  \\
				50    &92.942	&-133.340 & 100   & 185.403	&-266.532 \\
				125   &231.636	&-333.128   & 200   & 370.339 &-532.917 \\
				\bottomrule
			\end{tabular}
			\caption{$q_M=0.2$, $q_e=0.2$}\label{tabqM0pt2qe0pt2}
		\end{minipage}
	\end{threeparttable}
\end{table}

\begin{figure}[htbp!]
	\centering
	\subfigure[\label{f3}$a$ vs $\omega_{Re}$]{\includegraphics[width=0.41\linewidth]{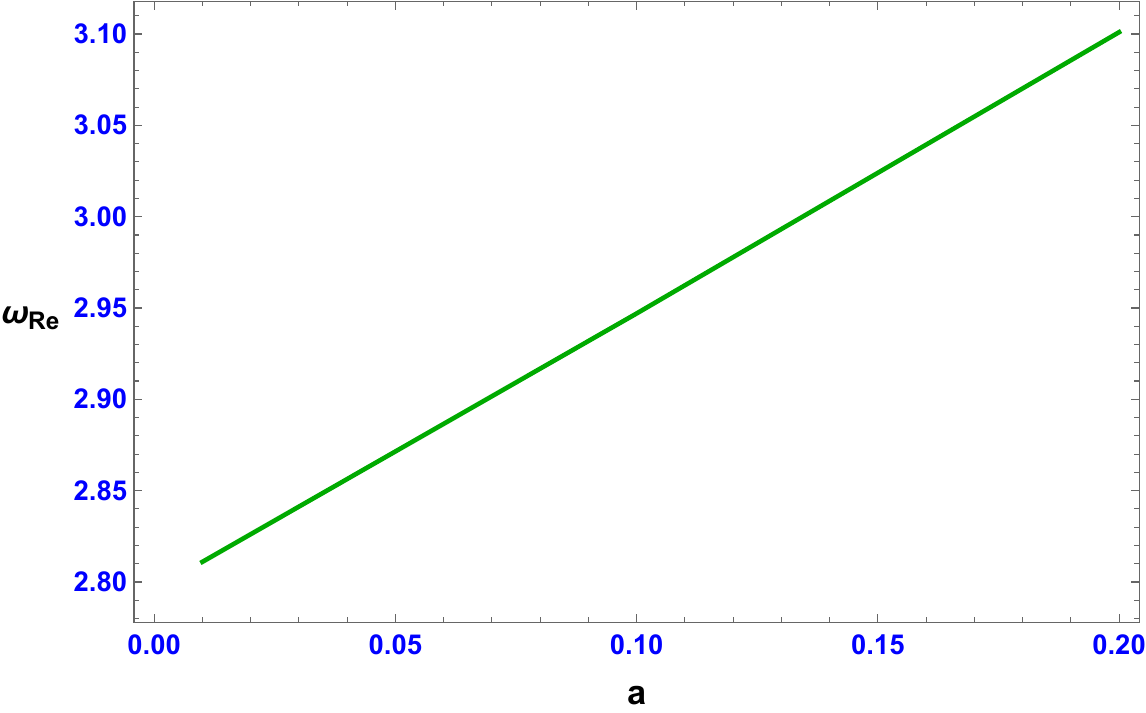}}
	\hfill
	\subfigure[\label{f4}$a$ vs $\omega_{Im}$]{\includegraphics[width=0.41\linewidth]{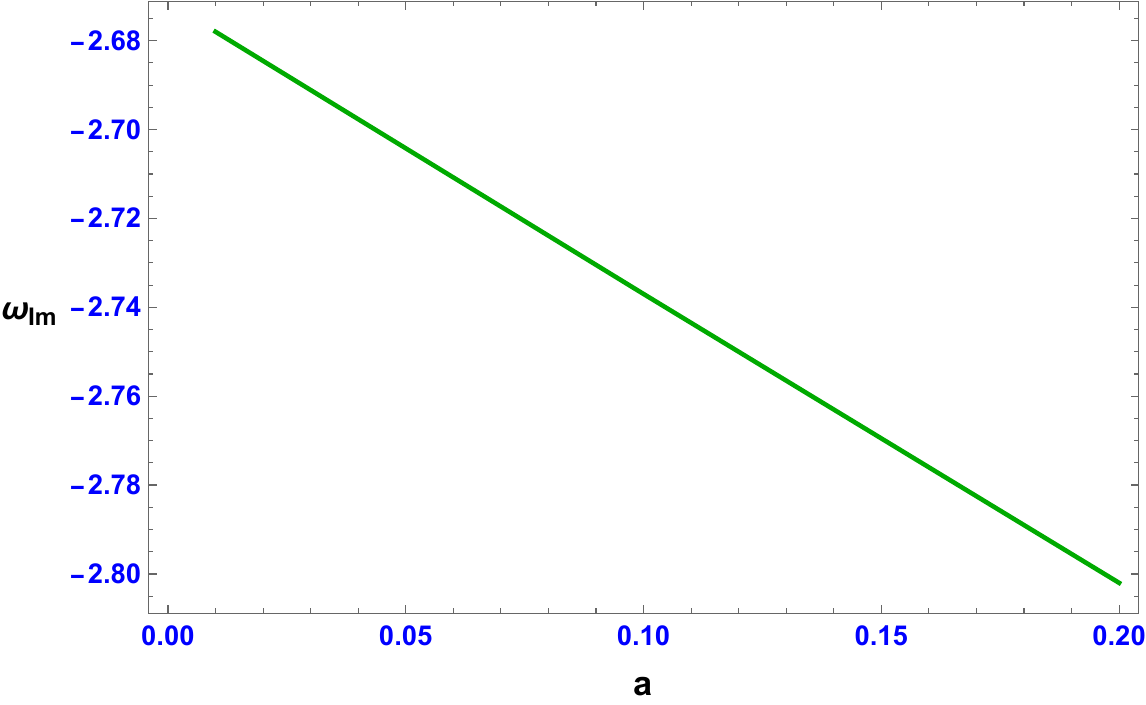}}
	\hfill
	\caption{QNM behavior as a function of $a$ for $q_e=0.1, q_M=0, r_h=1$.}\label{avsomega}
\end{figure}

We perform a through analysis of the scalar QNM with various black hole parameters in the hairy background. Our numerical results are shown in Tables \ref{tabqM0qe0pt1}, \ref{tabqM0qe0pt3}, \ref{tabqM0qe0}, and \ref{tabqM0pt2qe0pt2}. The results are mentioned up to three decimal places. One of our main results is that the imaginary part of QNM is always negative for all parameters of the hairy black hole. This indicates the stability of hairy black holes against the scalar field perturbations. This result also agrees well with the analytic argument presented above, where we showed that $\text{Im}\omega$ should always be negative in the hairy black hole background.
\\
\\
A few observations are in order from the numerical results:

$\bullet$For a fixed values of $\{r_h$, $q_e$, $q_M\}$, the QNM values depend marginally on the hairy parameter $a$. In particular, the variation in the imaginary part generally occurs at decimal places. This can be explicitly observed from Fig.~\ref{avsomega}. Also, note that both the oscillating and damping modes of the QNM depend linearly on $a$. In particular, the magnitude of both $\omega_{Re}$ and $\omega_{Im}$ increases with $a$. In fact, by fitting the curve, we find that $\omega_{Re} \simeq 1.526a$ and $\omega_{Im} \simeq - 0.653 a$. This suggests that the hairy black holes return to the equilibrium configuration faster for larger values of $a$.

\begin{figure}[hbtp!]
		\centering
		\subfigure[\label{f5}$r_h$ vs $\omega_{Im}$]{\includegraphics[width=0.31\linewidth]{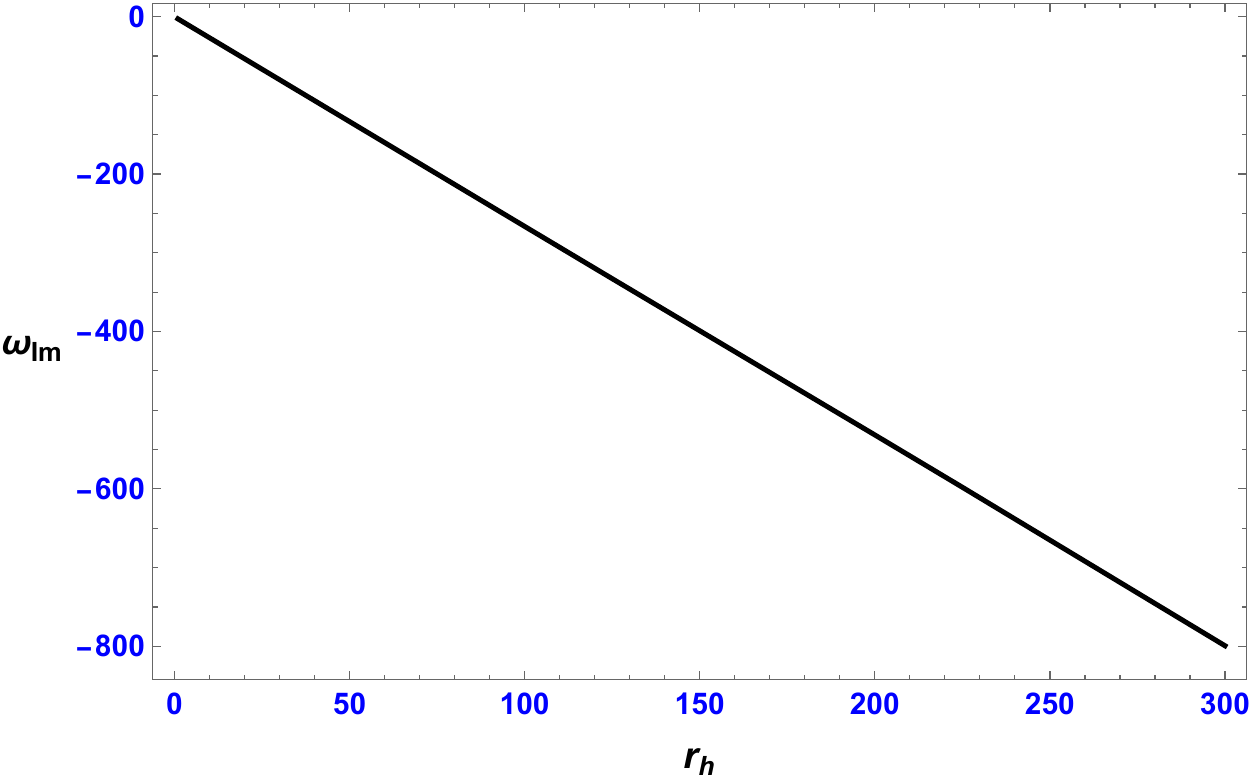}}
		\hfill
		\subfigure[\label{f6}$r_h$ vs $\omega_{Re}$]{\includegraphics[width=0.31\linewidth]{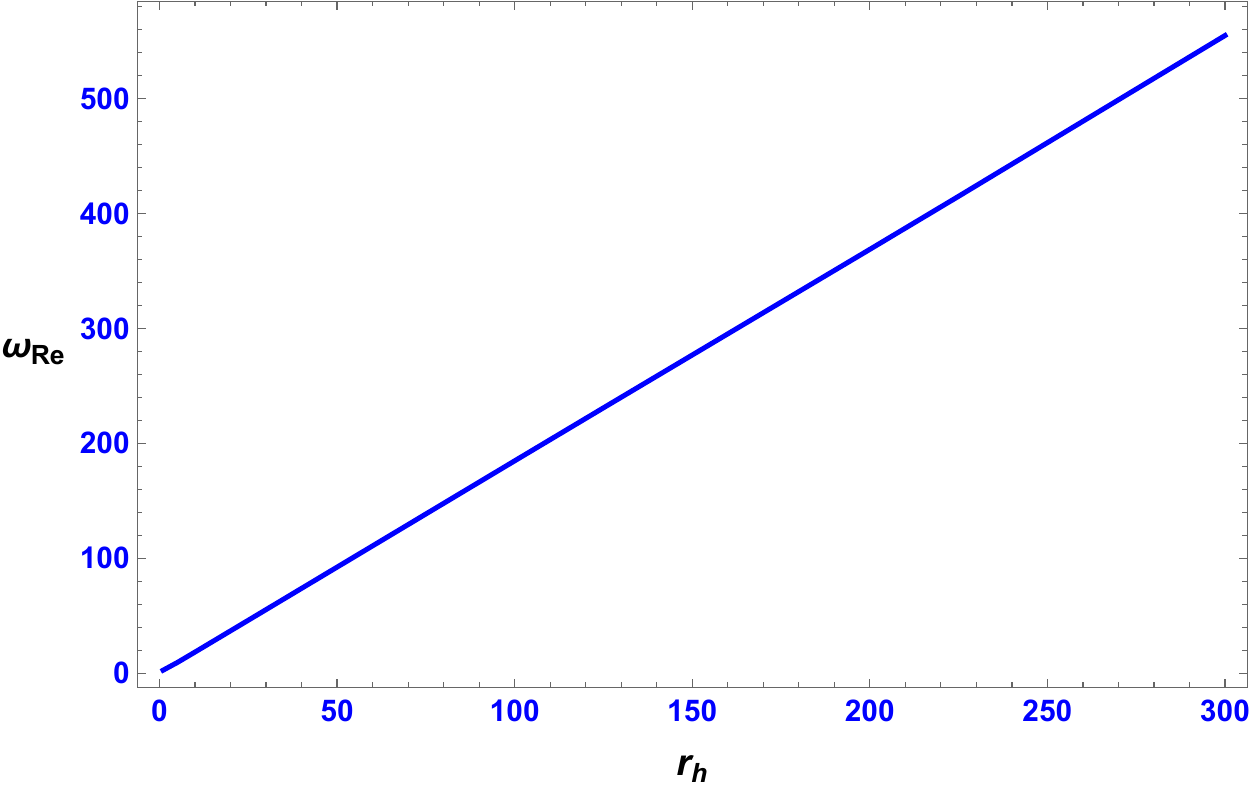}}
		\hfill
		\subfigure[\label{f7}$\omega_{Re}$ vs $\omega_{Im}$]{\includegraphics[width=0.31\linewidth]{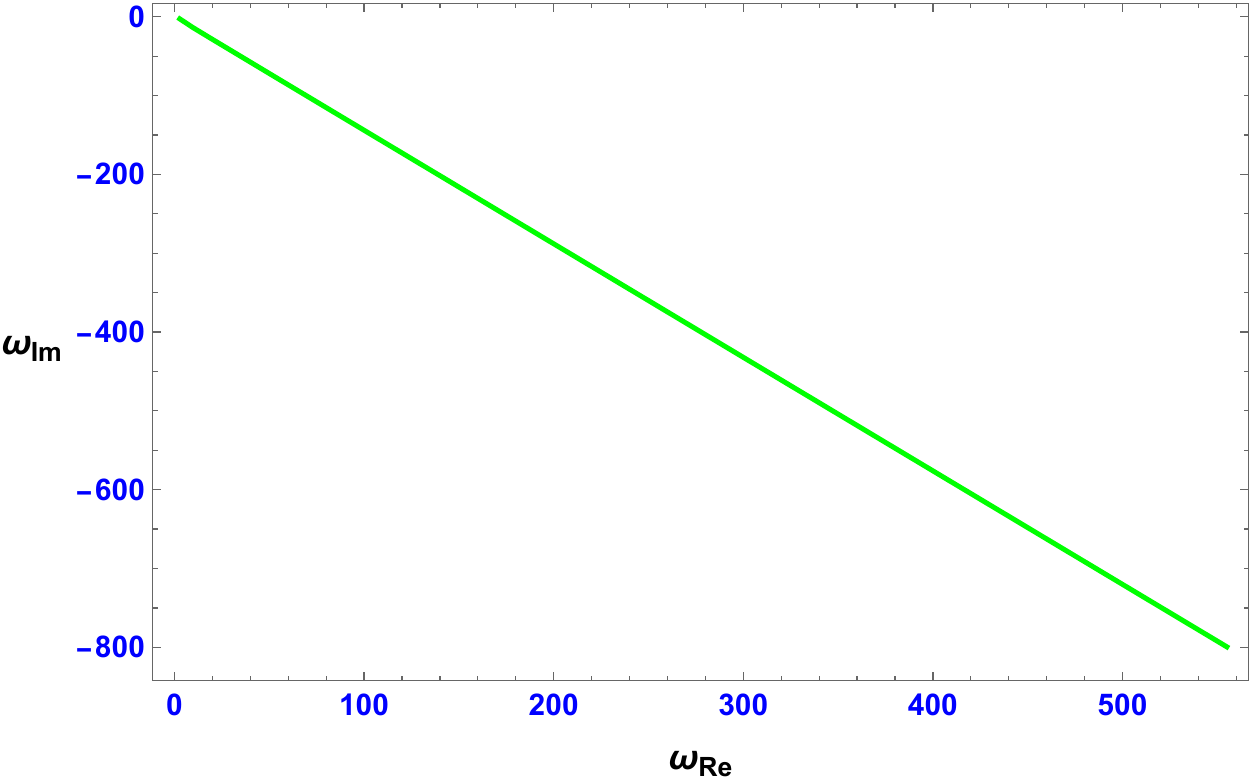}}
		
		\caption{QNM behavior as a function of $r_h$ for $a=0.01,q_e=0$, $q_M=0.3$.}\label{fig6}
\end{figure}

\begin{figure}[htbp!]
	\centering
	\subfigure[\label{f}$r_h$ vs $\omega_{Im}$]{\includegraphics[width=0.31\linewidth]{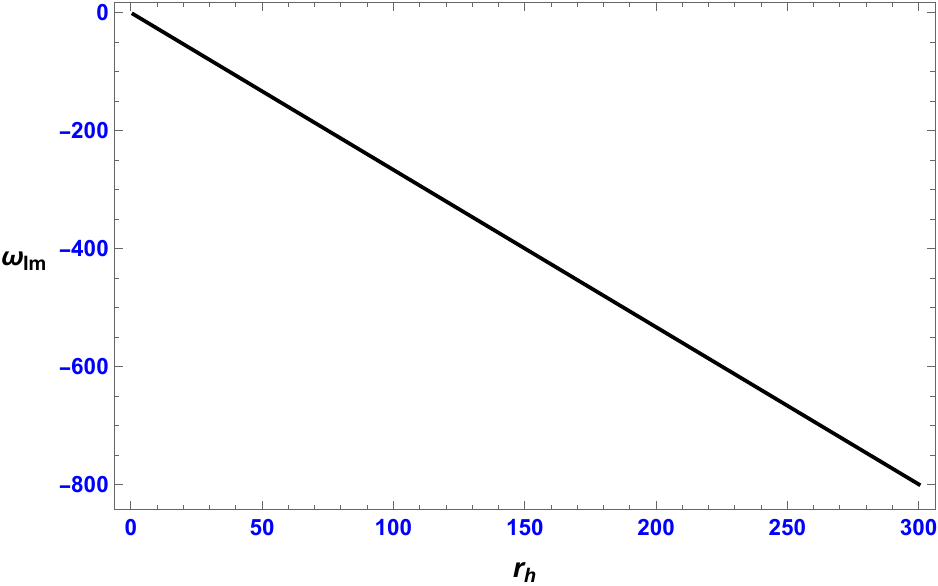}}
	\hfill
	\subfigure[\label{f1}$r_h$ vs $\omega_{Re}$]{\includegraphics[width=0.31\linewidth]{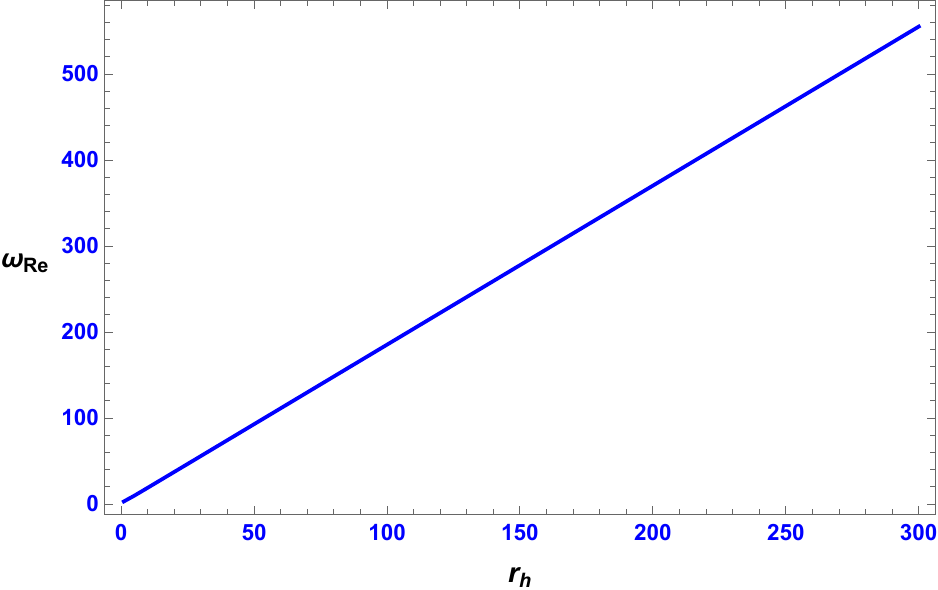}}
	\hfill
	\subfigure[\label{f2}$\omega_{Re}$ vs $\omega_{Im}$]{\includegraphics[width=0.31\linewidth]{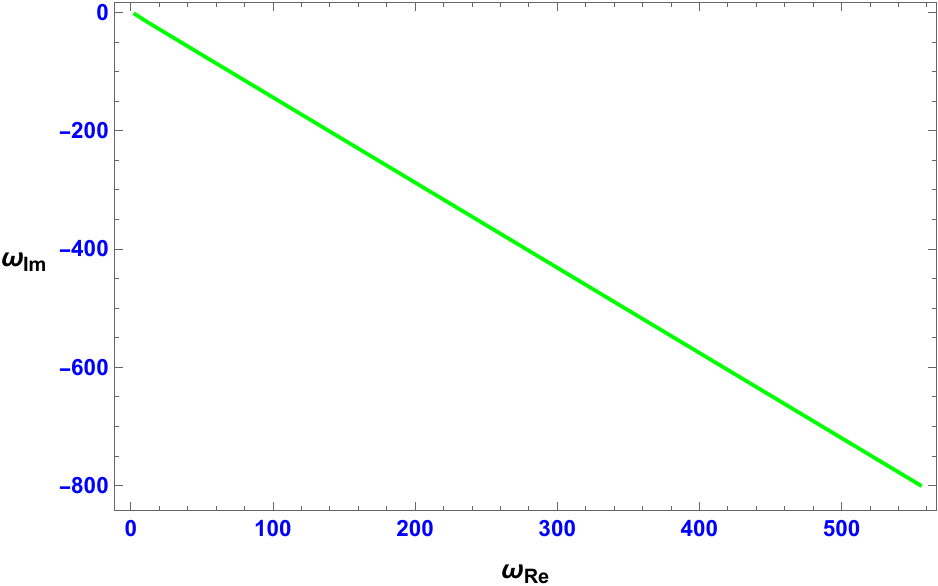}}
	
	\caption{QNM behavior as a function of $r_h$ for $a=0.1, q_e=0$, $q_M=0$.}\label{fig7}
\end{figure}

\begin{figure}[hbtp!]
		\centering
		\subfigure[\label{f8}$r_h$ vs $\omega_{Im}$]{\includegraphics[width=0.31\linewidth]{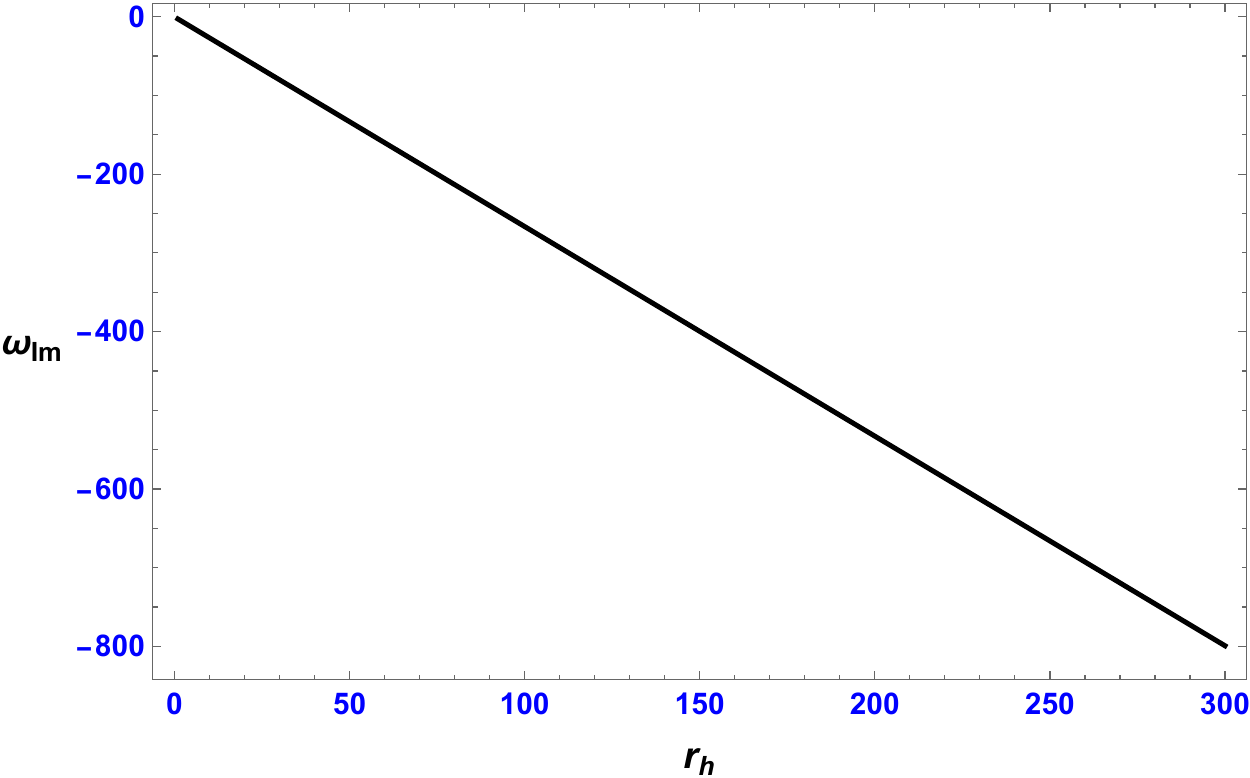}}
		\hfill
		\subfigure[\label{f9}$r_h$ vs $\omega_{Re}$]{\includegraphics[width=0.31\linewidth]{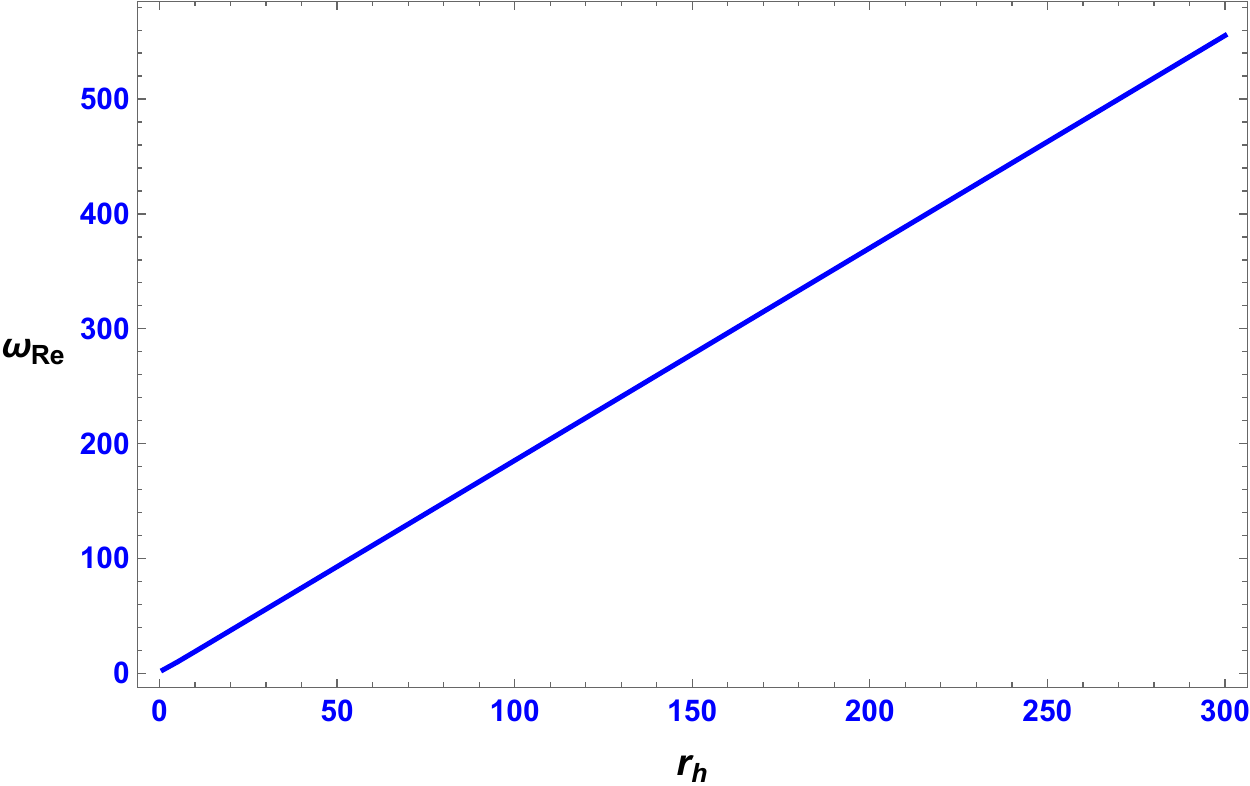}}
		\hfill
		\subfigure[\label{f10}$\omega_{Re}$ vs $\omega_{Im}$]{\includegraphics[width=0.31\linewidth]{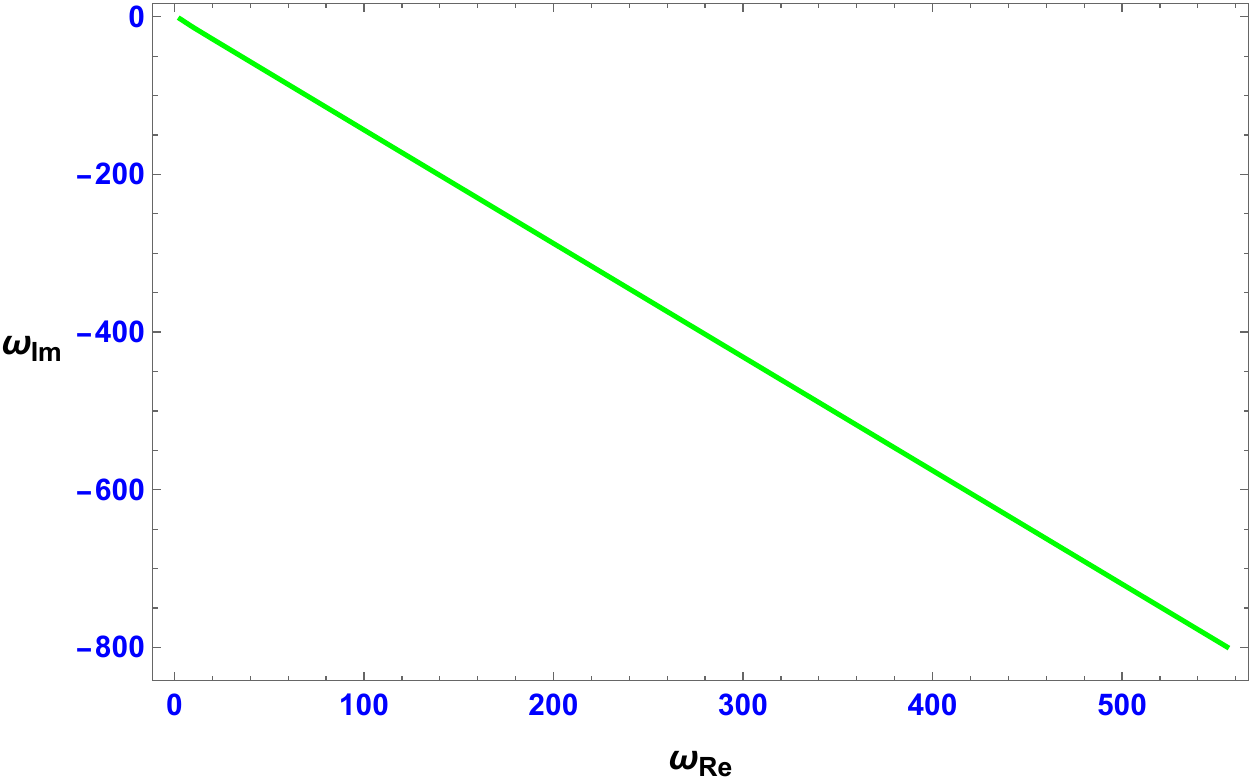}}
		
		\caption{QNM behavior as a function of $r_h$ for $a=0.2, q_e=0.2$, $q_M=0.2$.}\label{fig8}
\end{figure}

$\bullet$ Similar to the Schwarzschild-AdS black hole, the QNM frequencies increase linearly with the horizon radius for large size black holes. This is illustrated in Figs.~\ref{fig6}, \ref{fig7}, and \ref{fig8}. Notably, the oscillating mode can be approximated by the relation $\omega_{Re} \simeq 1.84 r_h$ (Figs. \ref{f5}, \ref{f}, and \ref{f8}), whereas the damping mode can be described by $\omega_{Im} \simeq -2.66 r_h$ (Figs. \ref{f6}, \ref{f1}, and \ref{f9}). Additionally, the $\omega_{Re}$ vs $\omega_{Im}$ curves (Figs. \ref{f7}, \ref{f2}, and \ref{f10}) display a negative slope for large black holes, indicating the presence of a ringdown geometry for the system.

$\bullet$ We further find that the electric charges $q_e$ have a limited effect on the QNMs. This result is consistent with the results of \cite{Wang:2000gsa}, where it was shown that the QNM values do not vary significantly with electric charge in the RN-AdS black hole background. Essentially, the different profiles of $r_h$ vs $\omega_{Re}$ (or $\omega_{Im}$) for various values of the black hole charge nearly overlap with each other. We are seeing the same structure in the hairy charged black hole case as well.

$\bullet$ As we vary $q_M$, the QNM structure remains similar to what we have seen for varying $q_e$. Here as well, the QNMs vary mildly with $q_M$. The remembrance of QNMs for fixed $q_e$ and $q_M$ is expected considering that the metric is symmetric in electric and magnetic charges. Hence, it is expected that the pure constant magnetic charge system behaves analogously to the pure constant electric charge system. Essentially, the QNM values are determined by the square of charges $q_{e}^2+q_{M}^2$.

\begin{figure}[htb]
	\centering
	\subfigure[\label{ft}  $a=0.1, q_e=0.1$, $q_M=0$]{\includegraphics[width=0.4\linewidth]{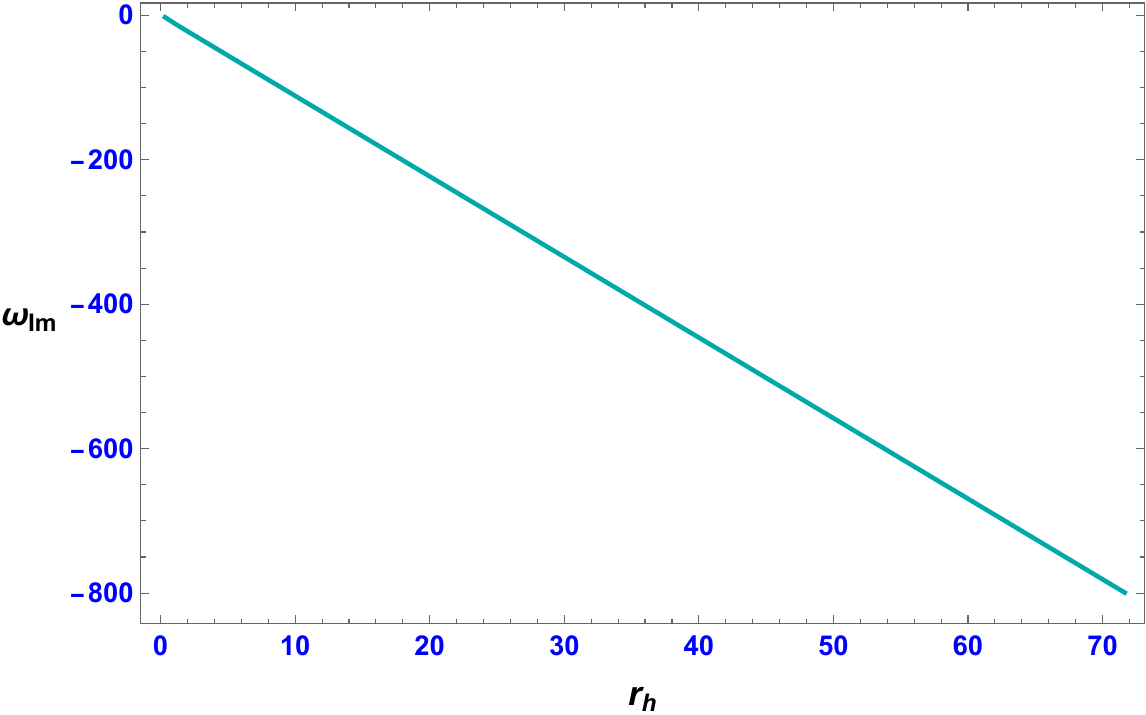}}
	\hfill
	\subfigure[\label{ft1} $a=0.2, q_e=0.1$, $q_M=0$]{\includegraphics[width=0.4\linewidth]{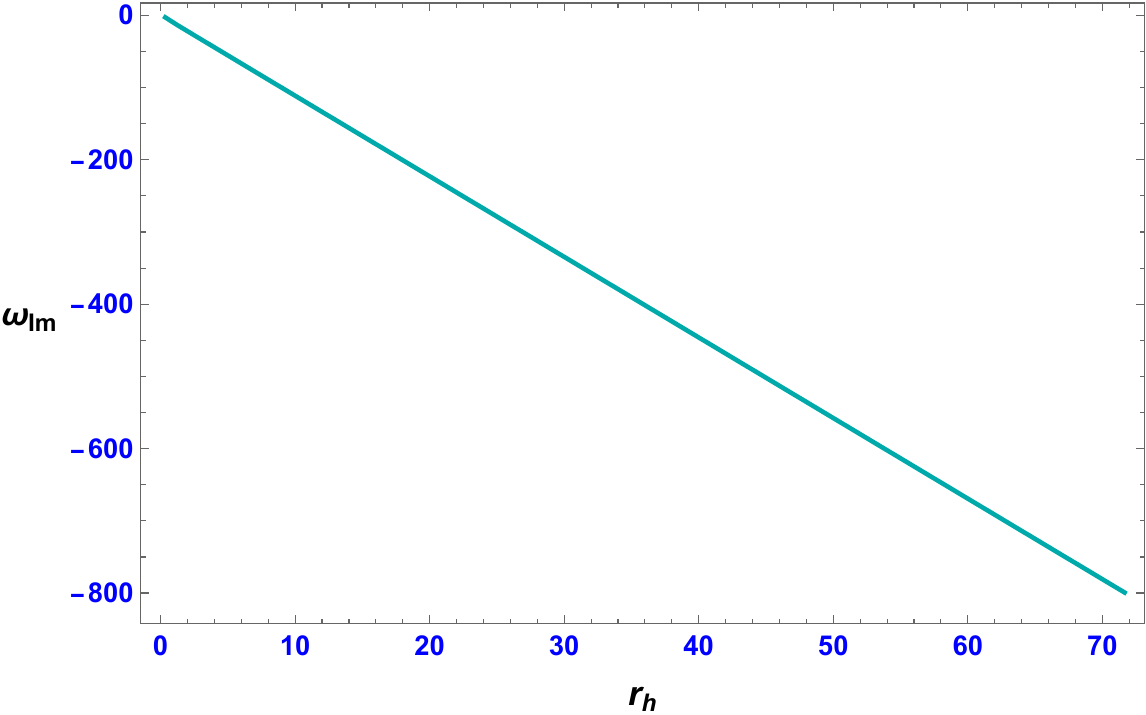}}
	\caption{$T$ vs $\omega_{Im}$.}\label{figTvsomegaim}
	\subfigure[\label{ft2}  $a=0.1, q_e=0.1$, $q_M=0$]{\includegraphics[width=0.4\linewidth]{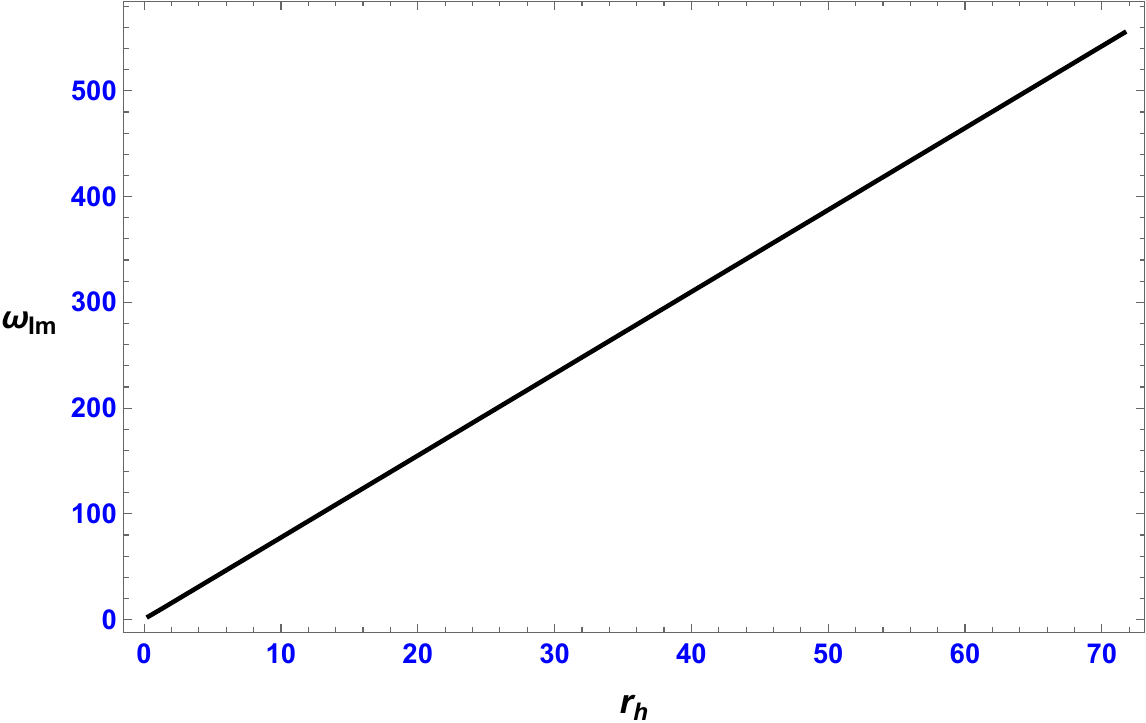}}
	\hfill
	\subfigure[\label{ft3} $a=0.2, q_e=0.1$, $q_M=0$]{\includegraphics[width=0.4\linewidth]{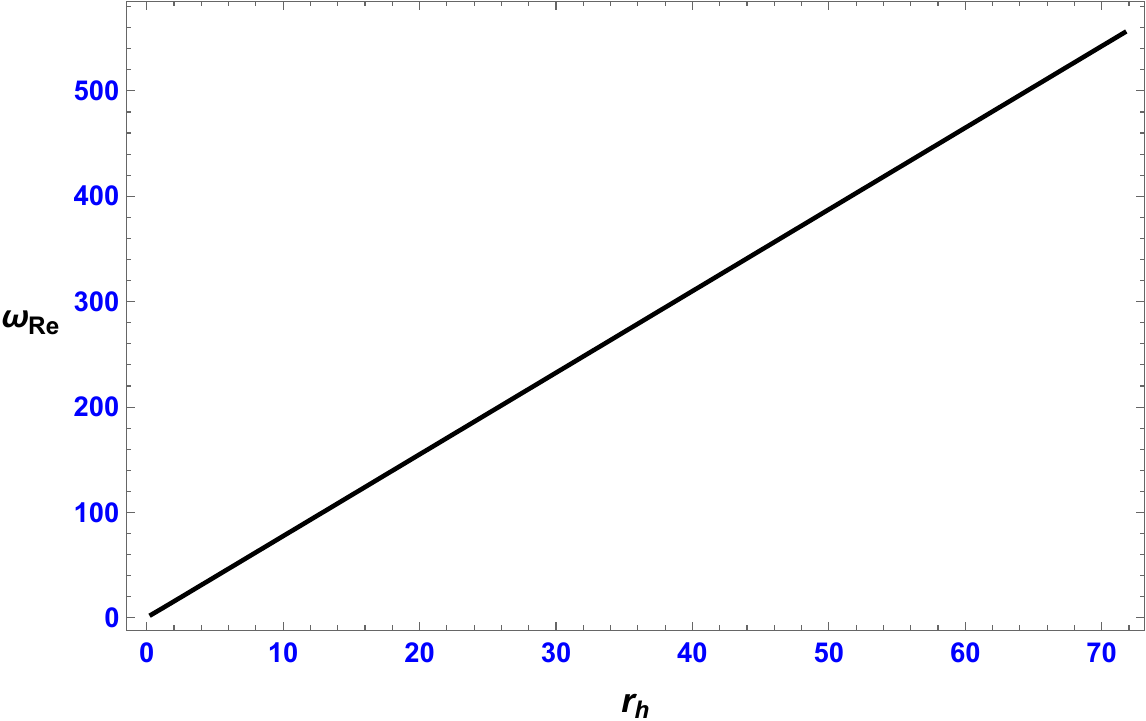}}
	\caption{$T$ vs $\omega_{Re}$.}\label{figTvsomegare}
\end{figure}

$\bullet$ we also observe a linear correlation between the imaginary and real parts of the frequency and temperature for different values of $a$ for large black holes. This correlation is approximately given by $\omega_{Im}\simeq -11.176 T$ (Fig. \ref{figTvsomegaim}) and $\omega_{Re}\simeq 7.740 T$ (Fig.\ref{figTvsomegare}), respectively, which resembles the analogous findings discussed in \cite{horowitz2000quasinormal} for the four-dimensional case. According to the AdS/CFT correspondence, the damping mode and oscillating mode can be linked to the damping time scale ($\tau_d=1/\omega_{Im}$) and oscillating time scale ($\tau_o=1/\omega_{Re}$) of thermal equilibrium, enabling the evaluation of the scalar field's decay and oscillation rate.

\begin{figure}[htbp!]
	\centering
	\subfigure[\label{f11}$r_h$ vs $\omega_{Im}$]{\includegraphics[width=0.31\linewidth]{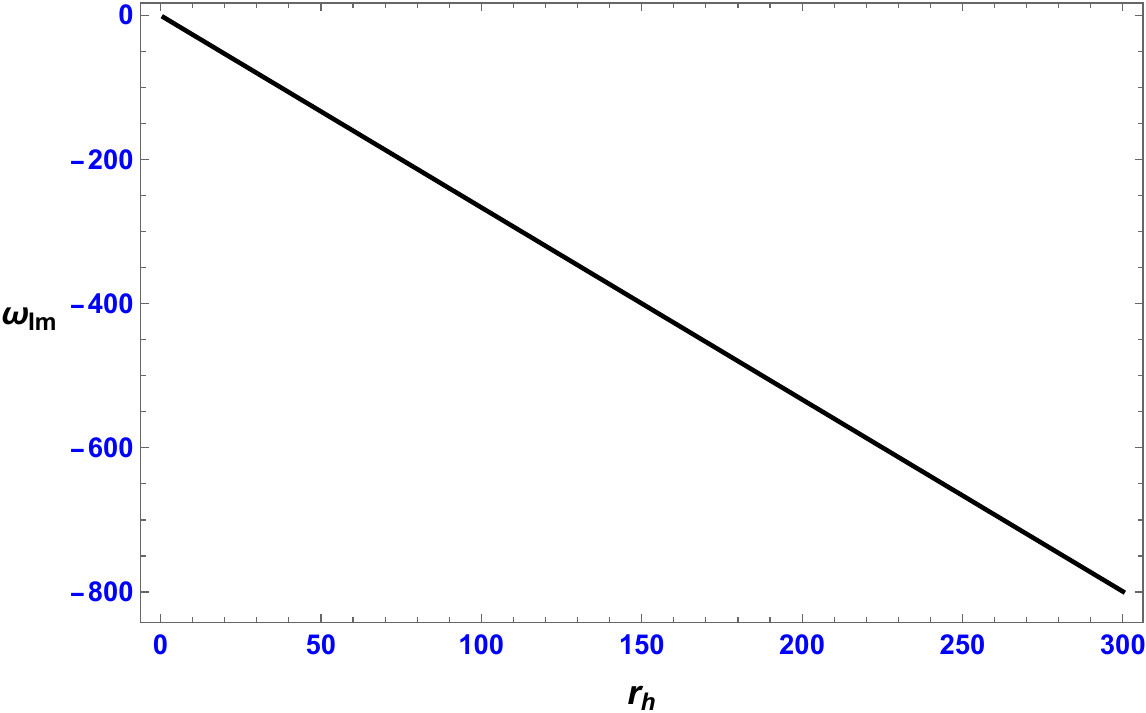}}
	\hfill
	\subfigure[\label{f12}$r_h$ vs $\omega_{Re}$]{\includegraphics[width=0.31\linewidth]{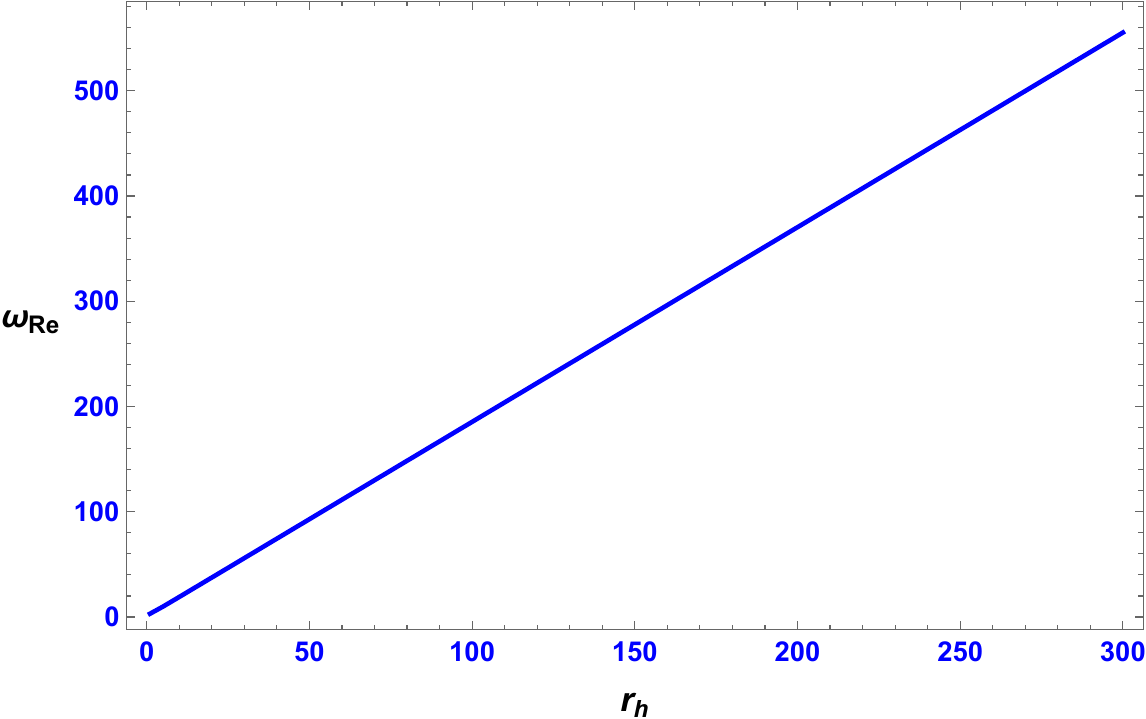}}
	\hfill
	\subfigure[\label{f13}$\omega_{Re}$ vs $\omega_{Im}$]{\includegraphics[width=0.31\linewidth]{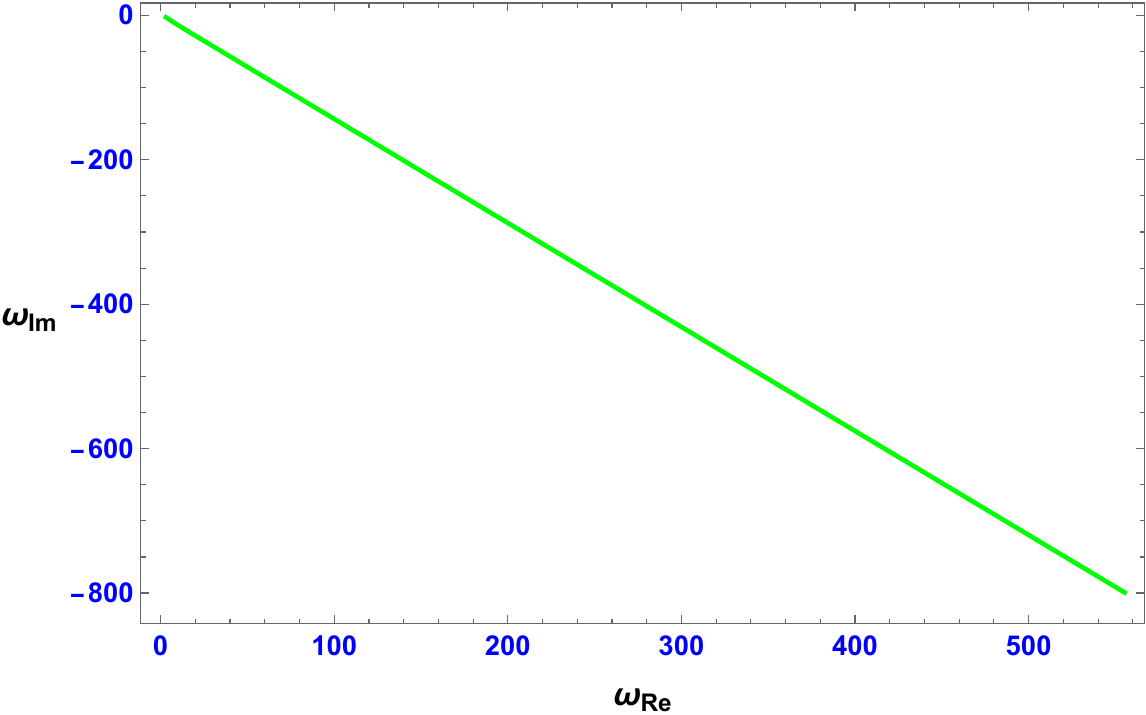}}
	
	\caption{$q_M=0$, $a=0.2, \mu_e=0.1$.}\label{qm0a0pt2mue0pt1}
\end{figure}

\begin{figure}[htbp!]
	\centering
	\subfigure[\label{f16}$r_h$ vs $\omega_{Im}$]{\includegraphics[width=0.31\linewidth]{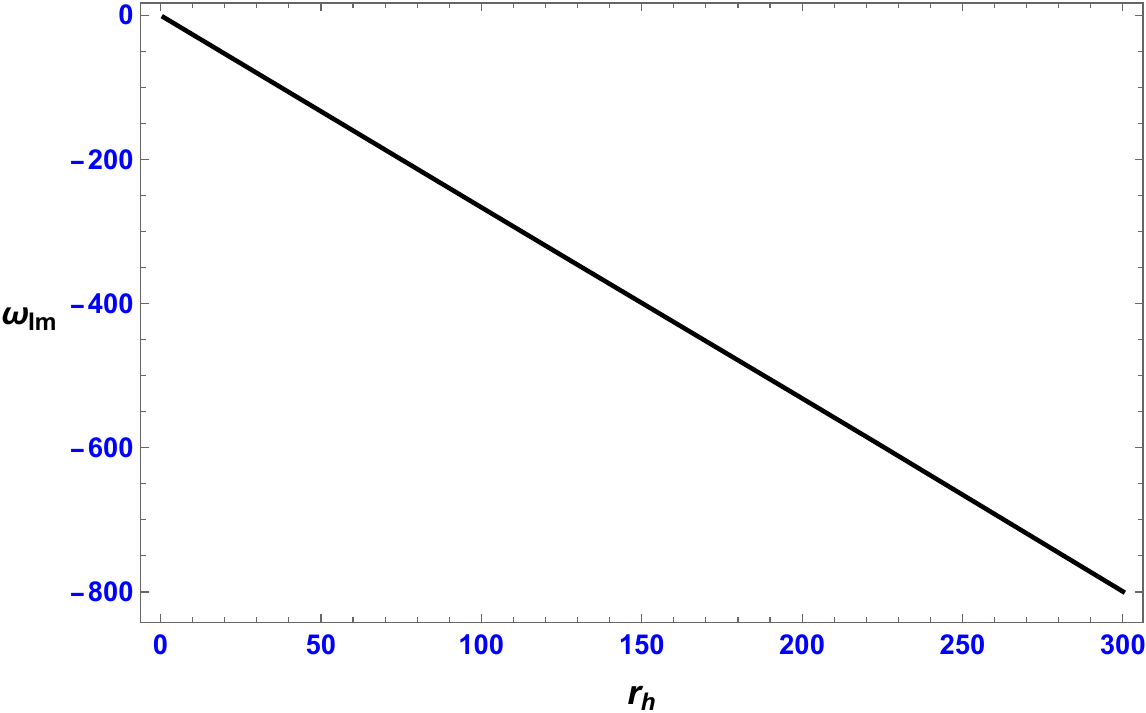}}
	\hfill
	\subfigure[\label{f17}$r_h$ vs $\omega_{Re}$]{\includegraphics[width=0.31\linewidth]{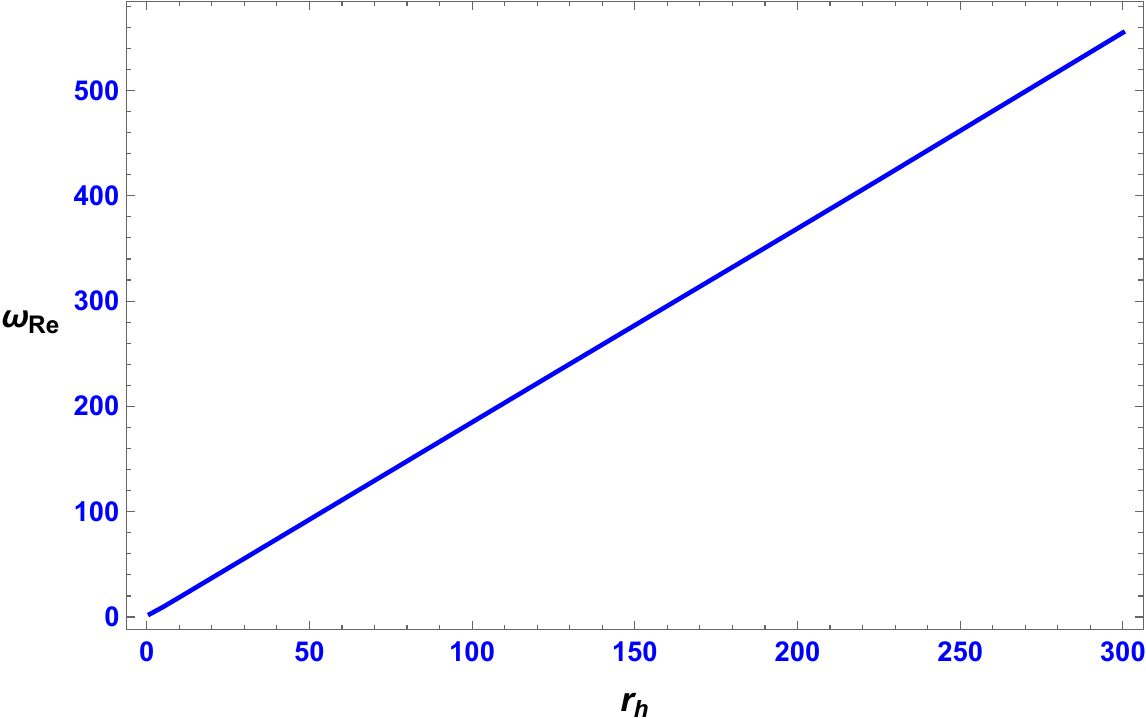}}
	\hfill
	\subfigure[\label{f18}$\omega_{Re}$ vs $\omega_{Im}$]{\includegraphics[width=0.31\linewidth]{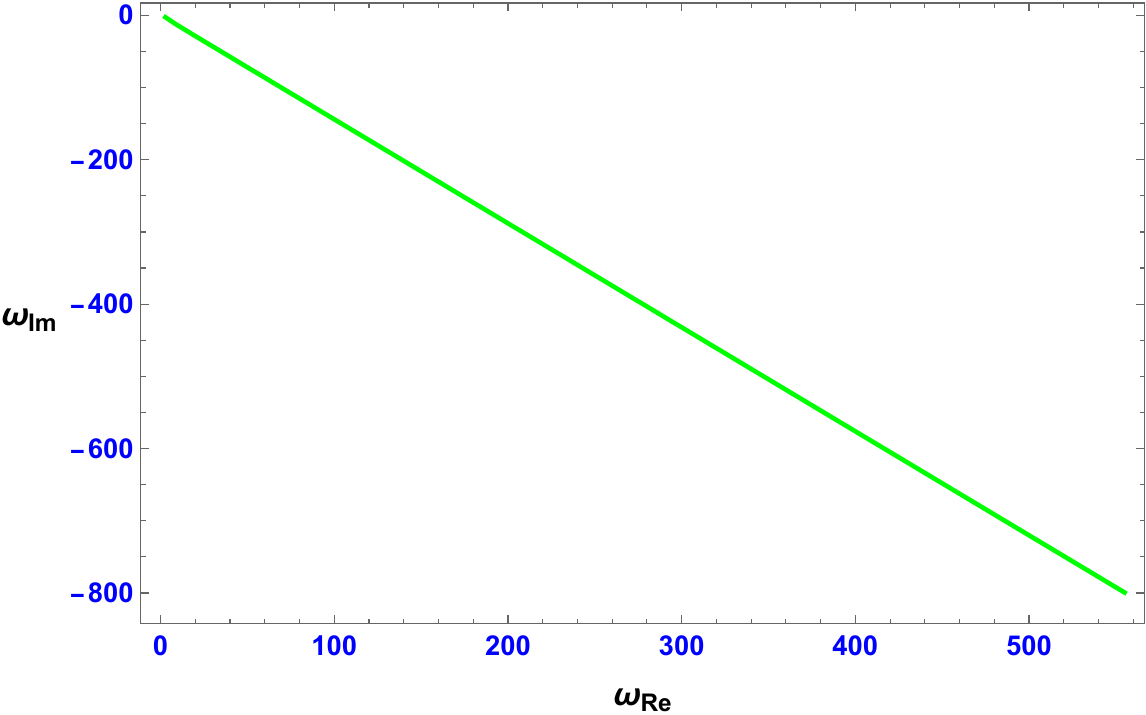}}
	
	\caption{$q_M=0$, $a=0.01, \mu_e=0.3$.}\label{qm0a0pt01mue0pt3}
\end{figure}

$\bullet$ To make our analysis more complete, we have also analysed how the QNM changes with the chemical potential. This will be useful in the next section when we will try to probe the black hole phase transition using the QNM structure. Our results for different chemical potential is shown in Tables~\ref{tab6} and \ref{tab7}. The QNMs behave the same way as in case of fixed charge -- both the oscillation and decaying modes increase with increase in the horizon radius for large black holes and the slope of the $\omega_{Re}$ vs $\omega_{Im}$ curve is always negative. Importantly, $\omega_{Im}$ is again always negative, indicating the overall dynamical stability of the hairy black hole in the fixed chemical potential ensemble. With horizon radius, the real and imaginary parts of QNM linearly scale as $\omega_{Re} \simeq 1.84 r_h$ and $\omega_{Im} \simeq -2.66 r_h$. The dependence of QNMs on $r_h$ for fixed values of $a$, $\mu_e$, and $q_M$ is shown in Figs.~\ref{qm0a0pt2mue0pt1} and \ref{qm0a0pt01mue0pt3}.
\begin{table}[htpb!]
	\setlength\tabcolsep{2pt}
	\footnotesize
	\renewcommand{\arraystretch}{1}
	\centering
	\begin{threeparttable}
		\begin{minipage}{0.5\linewidth}
			\centering
			\begin{tabular}{@{}
					r S[table-format=-2.3] S[table-format=-2.3]
					@{\hspace{12pt}} !{\vrule width 0.2pt} @{\hspace{12pt}}
					r S[table-format=-2.3] S[table-format=-2.3]
					@{}}
				\toprule
				\multicolumn{1}{c}{$r_h$} & $\omega_{Re}$ & \multicolumn{1}{c}{$\omega_{Im}$} &
				\multicolumn{1}{c}{$r_h$} & $\omega_{Re}$ & $\omega_{Im}$ \\
				\midrule[0.08cm]
				\multicolumn{6}{c}{\textbf{\textbf{$a=0.01$}}}\\
				~	1 &2.812 &-2.679  & 5 & 9.493 & -13.333 \\
				10  & 18.629 & -26.649 & 25 & 46.303 & -66.605 \\
				50 & 92.517 & -133.202  & 100 & 184.982 & -266.402 \\
				125 &231.100 & -332.833 & 200 & 368.976 & -531.492 \\
				\multicolumn{6}{c}{\textbf{$ a=0.1 $}} \\
				~	1  & 2.948 & -2.738  & 5  &9.690 & -13.399 \\
				10  & 18.830 & -26.715   & 25  & 46.505 & -66.671 \\
				50  & 92.718 & -133.267  & 100 & 185.179 & -266.459  \\
				125 & 231.421 & -333.055  & 200 &370.115 & -532.845 \\
				\multicolumn{6}{c}{\textbf{\textbf{$a = 0.2$}}} \\
				~	1 & 3.101 & -2.803   & 5   & 9.907 & -13.471 \\
				10  & 19.052 & -26.788  & 25  & 46.729 & -66.744 \\
				50  & 92.943 & -133.340  & 100 & 185.403 & -266.533 \\
				125 & 231.636 & -333.129  & 200 & 370.340 & -532.918 \\
				\bottomrule
			\end{tabular}
			\caption{$q_M=0$, $\mu_e=0.1$.}\label{tab6}
		\end{minipage}
		\hfil
		\begin{minipage}{0.5\linewidth}
			\centering
			
			\begin{tabular}{@{}
					r S[table-format=-2.3] S[table-format=-2.3]
					@{\hspace{12pt}} !{\vrule width 0.2pt} @{\hspace{12pt}}
					r S[table-format=-2.3] S[table-format=-2.3]
					@{}}
				\toprule
				\multicolumn{1}{c}{$r_h$} & $\omega_{Re}$ & \multicolumn{1}{c}{$\omega_{Im}$} &
				\multicolumn{1}{c}{$r_h$} & $\omega_{Re}$ & $\omega_{Im}$ \\
				\midrule[0.08cm]
				\multicolumn{6}{c}{\textbf{\textbf{$a=0.01$}}}\\
				~	1 & 2.799 & -2.687 & 5 & 9.488 & -13.335 \\
				10 & 18.627 & -26.650 & 25 & 46.302 & -66.605 \\
				50    & 92.516 &-133.202  & 100   & 184.982	&-266.402 \\
				125   & 231.096	&-332.827  & 200   & 368.956 &-531.402 \\
				\multicolumn{6}{c}{\textbf{$ a=0.1 $}} \\
				~	1     &2.936	&-2.747   & 5     & 9.686 &-13.401 \\
				10    & 18.830	&-26.715  & 25    &46.505	&-66.671 \\
				50  & 92.718 & -133.267  & 100 & 185.178 & -266.459  \\
				125   & 231.412	&-333.056  & 200   & 370.115&-532.845 \\
				\multicolumn{6}{c}{\textbf{\textbf{$a = 0.2$}}} \\
				~	1     &3.088	&-2.811  & 5     & 9.903	&-13.473 \\
				10    & 19.049  &-26.789  & 25    & 46.728	&-66.745  \\
				50    & 92.942 &-133.340  & 100   & 185.402	&-266.532 \\
				125   &231.636	&-333.129  & 200   & 370.339	&-532.918  \\
				\bottomrule
			\end{tabular}
			\caption{$q_M=0$, $\mu_e=0.3$.}\label{tab7}
		\end{minipage}
	\end{threeparttable}
\end{table}

\begin{figure}[htpb!]
	\centering
	\subfigure[\label{f14}$a$ vs $\omega_{Re}$]{\includegraphics[width=0.41\linewidth]{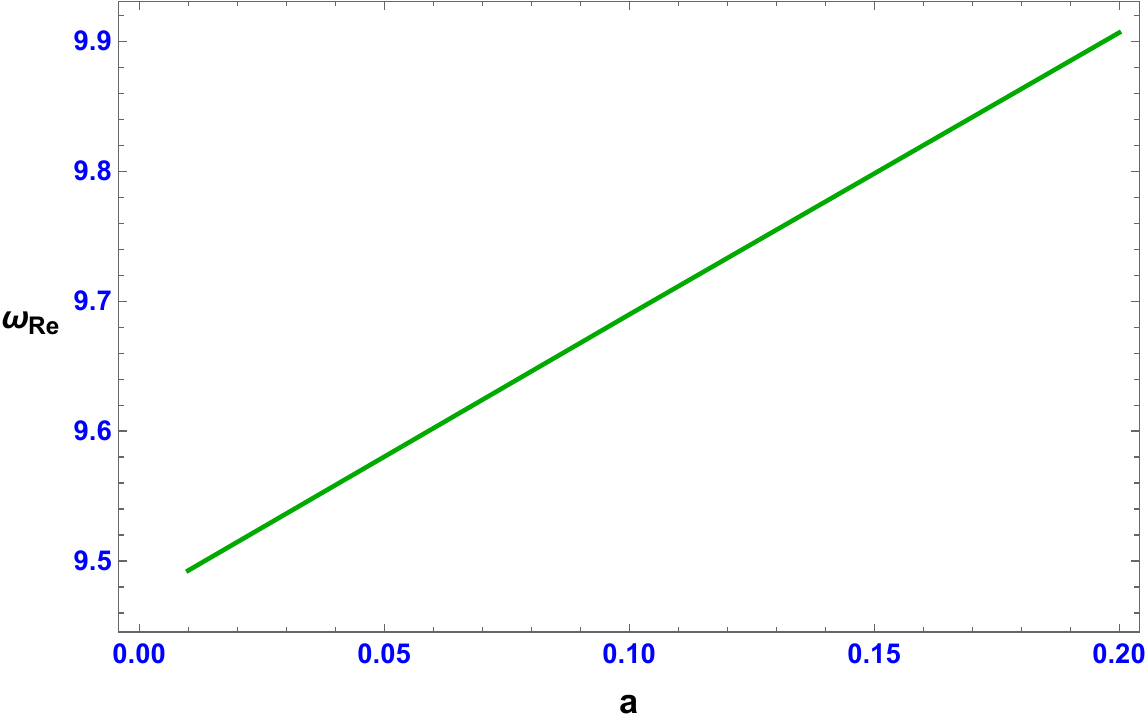}}
	\hfill
	\subfigure[\label{f15}$a$ vs $\omega_{Im}$]{\includegraphics[width=0.41\linewidth]{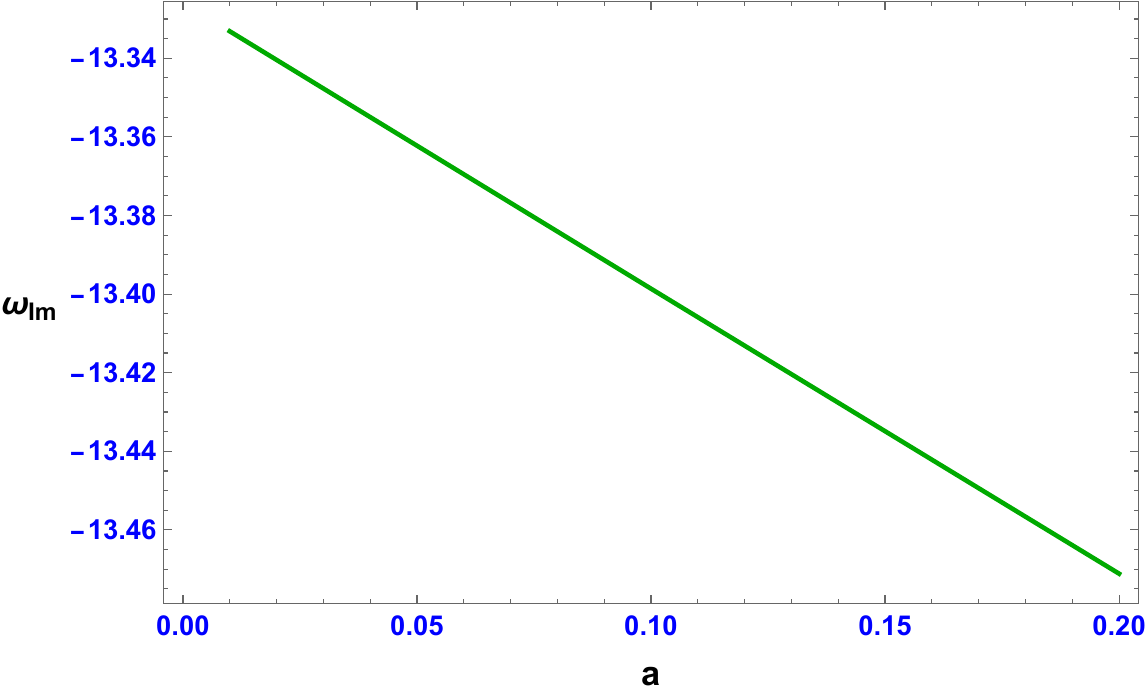}}
	\caption{$q_M=0,\mu_e=0.1,r_h=5$.}\label{avsomega1}
\end{figure}

We further analyze the dependence of QNM on the hairy parameter. With $a$, the absolute value of both the oscillating and the damping modes increase. This dependence is shown Fig.~\ref{avsomega1}. Again, like in the fixed charge case, the variation is slightly more for $\omega_{Re}$ than for $\omega_{Im}$.

Here we have presented QNM results for hairy black holes for specific values of the hair parameter $a=0.05, 0.1, 0.2$ while keeping $q_e=0.1$ and $q_M=0$ fixed. However, we like to mention that analogous results also occur for other values of $a$ as well.


	Our overall analysis, therefore, suggests that the hairy black holes are dynamically stable against the massless scalar field perturbation in both fixed charge and potential configurations. Since the imaginary part of QNM is inversely related to the time scale of thermalization $\tau=1/\omega_{Im}$, our analysis also suggests that the hairy black holes come to thermal equilibrium at a faster rate than their nonhairy counterparts.
	
\section{QNM and black hole phase transition}\label{section5}
In this section, we investigate whether the signature of Van der Waals like small/large phase transition of hairy charged black holes can be reflected in the QNM structure of the massless scalar perturbation. As discussed in the last section, the behaviour of scalar QNMs in the hairy black hole background resembles closely to that of Schwarzschild and RN-AdS black holes for larger horizon radius. However, during the phase transition, specifically between the small and large black hole phases, the QNM can exhibit significant differences.
	\begin{figure}[hbtp!]
	\centering
	\subfigure[$a=0.05$, $q_e=0.1, q_M=0$.]{\includegraphics[width=0.3\linewidth]{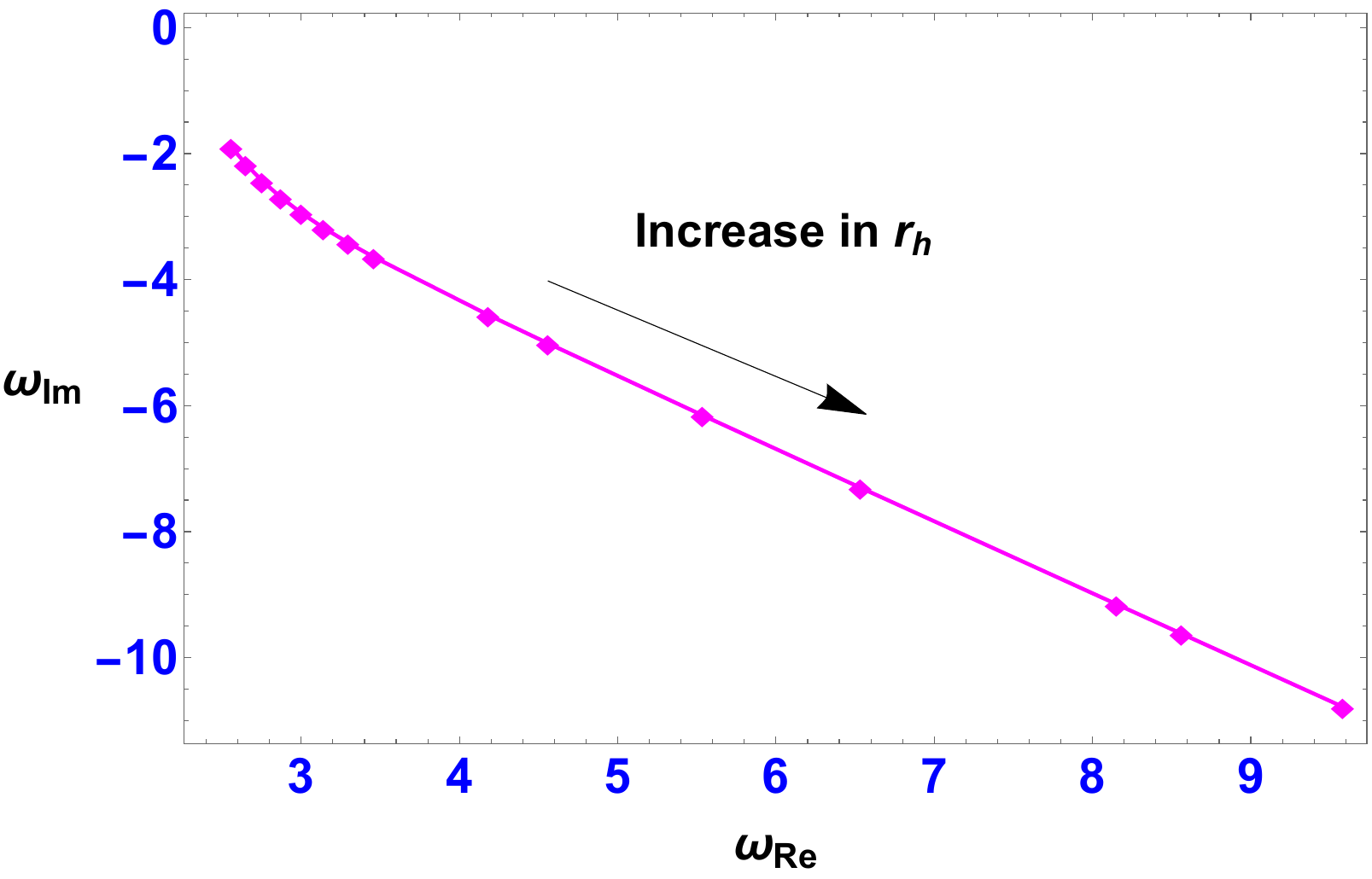}}
	\label{fig12figa}
	\hfill
	\subfigure[$a=0.1$, $q_e=0.1, q_M=0$.]{\includegraphics[width=0.3\linewidth]{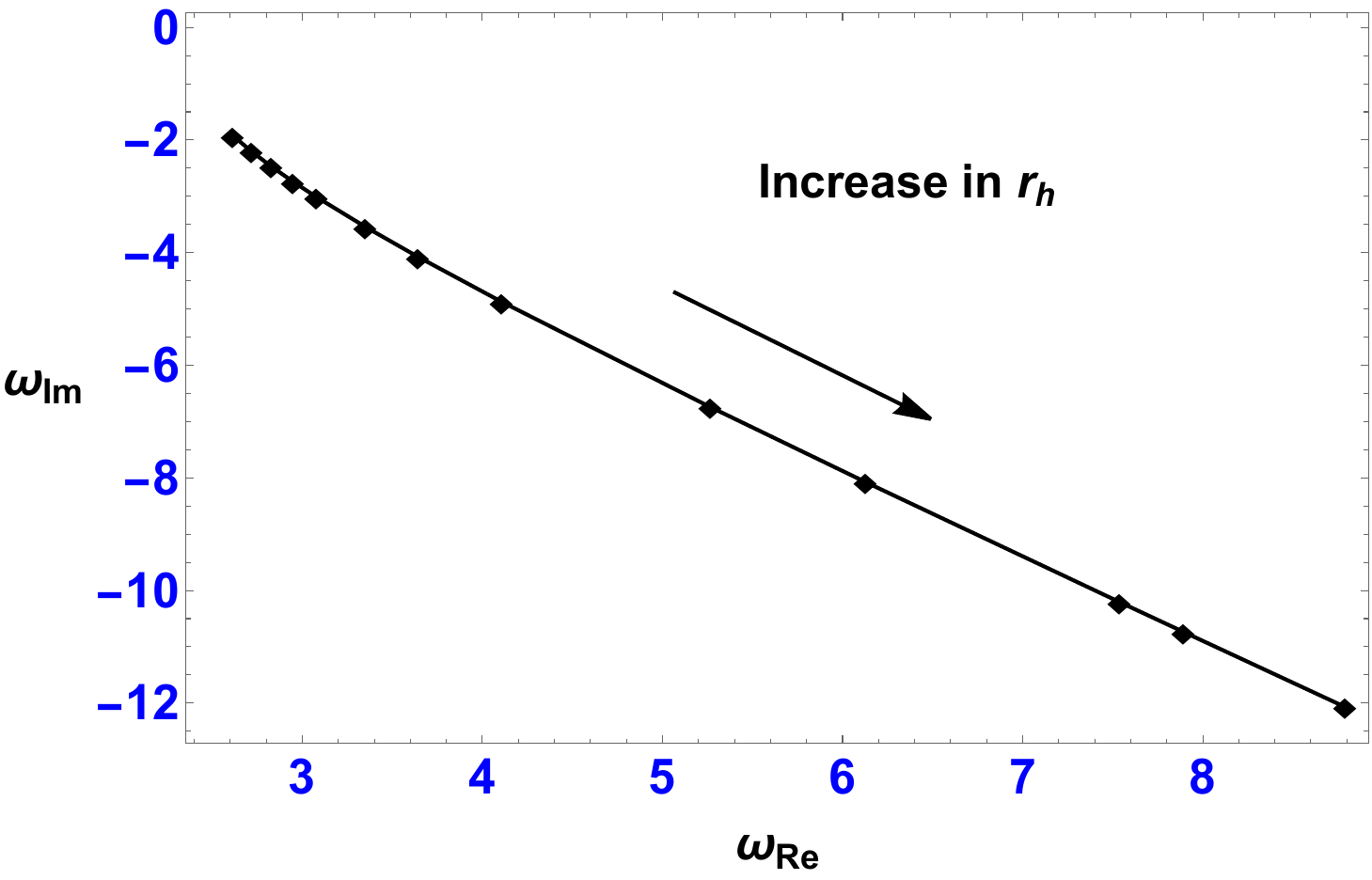}}
	\hfill
	\subfigure[$a=0.2$, $q_e=0.1, q_M=0$.]{\includegraphics[width=0.3\linewidth]{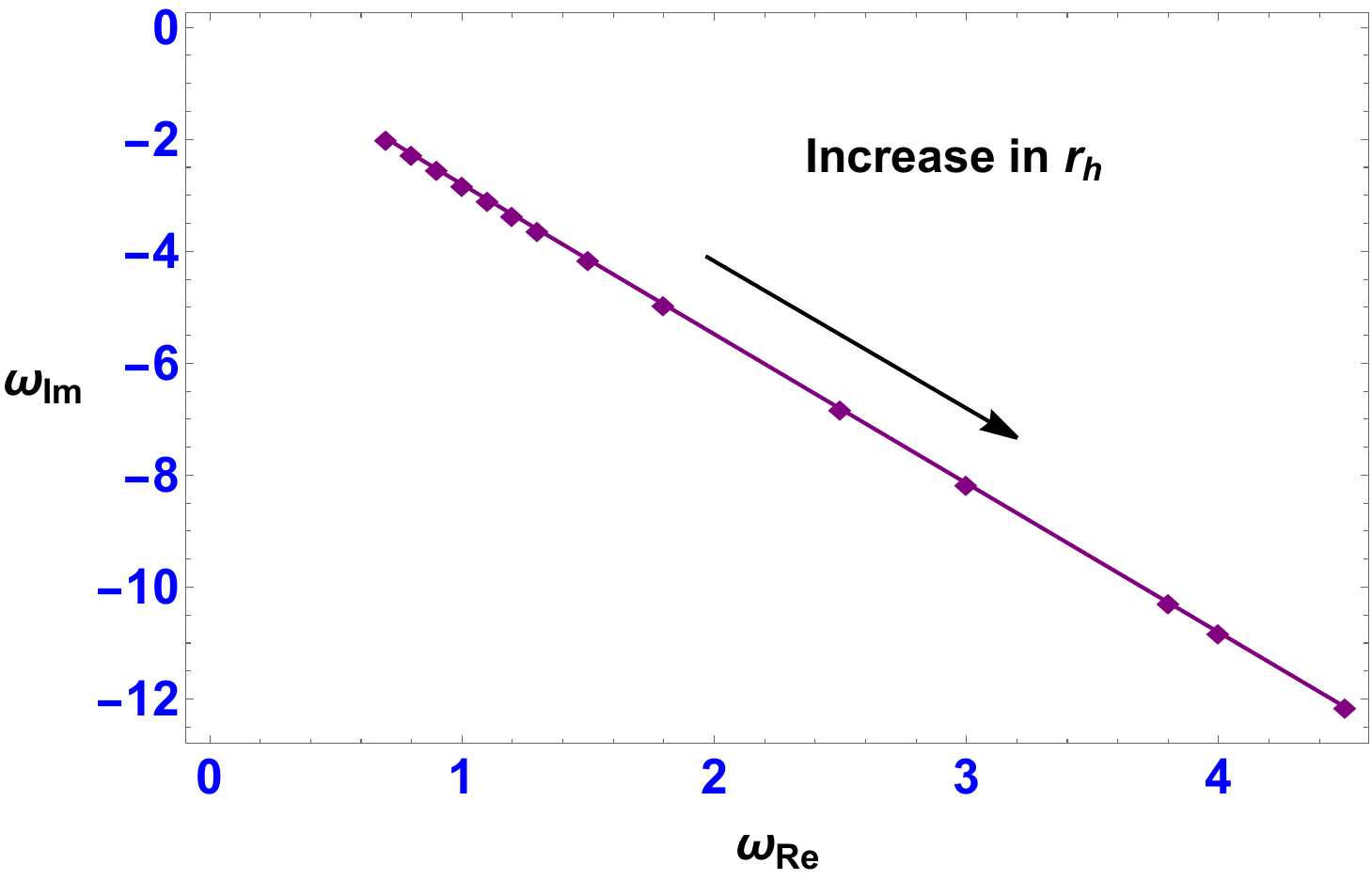}}
	
	\caption{QNM behavior in the large black hole phase in the fixed charge case.}\label{LBHP}
\end{figure}

\begin{figure}[hbtp!]
	\centering
	\subfigure[$a=0.05$, $q_e=0.1, q_M=0$.]{\includegraphics[width=0.31\linewidth]{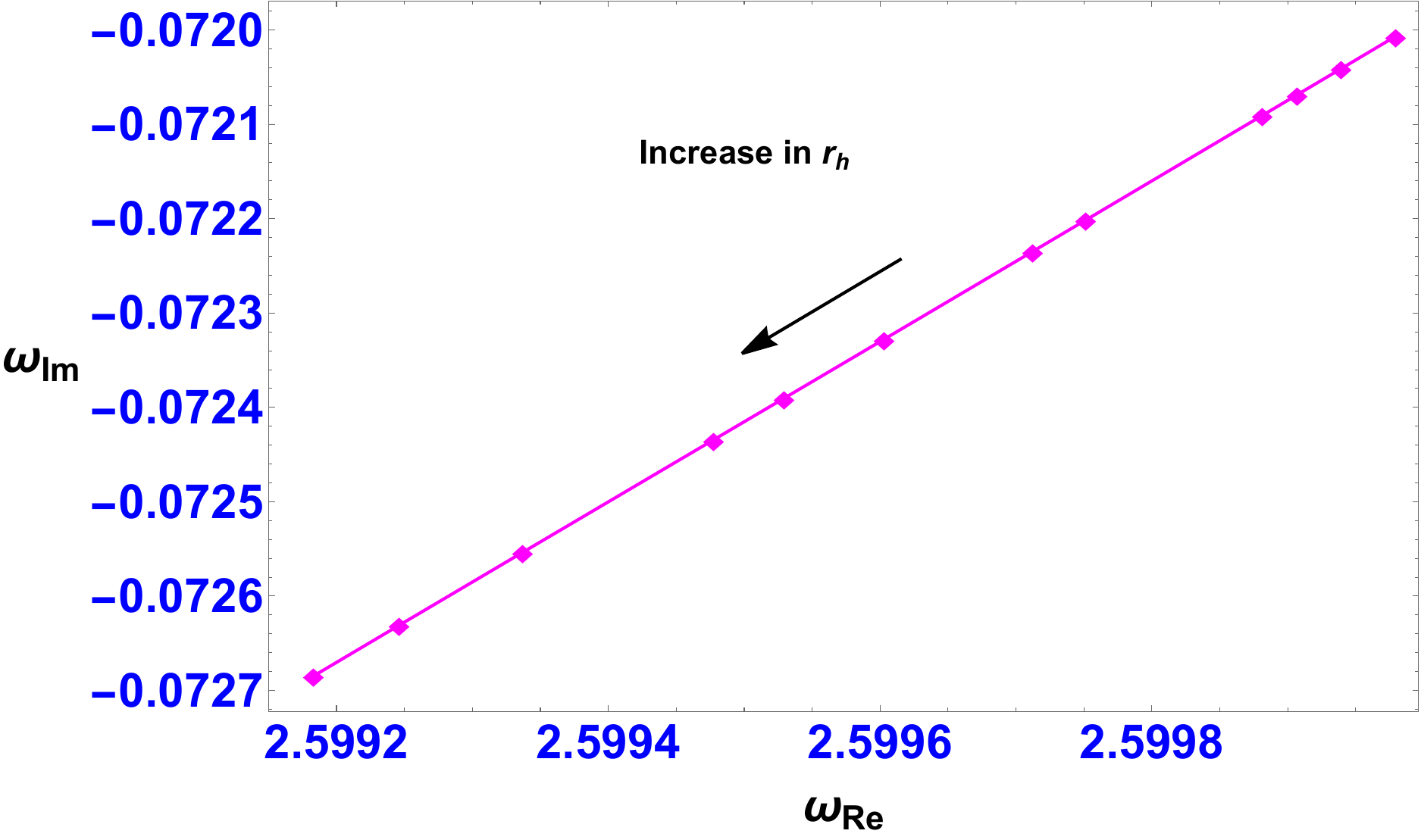}}
	\hfill
	\subfigure[$a=0.1$, $q_e=0.1, q_M=0$.]{\includegraphics[width=0.31\linewidth]{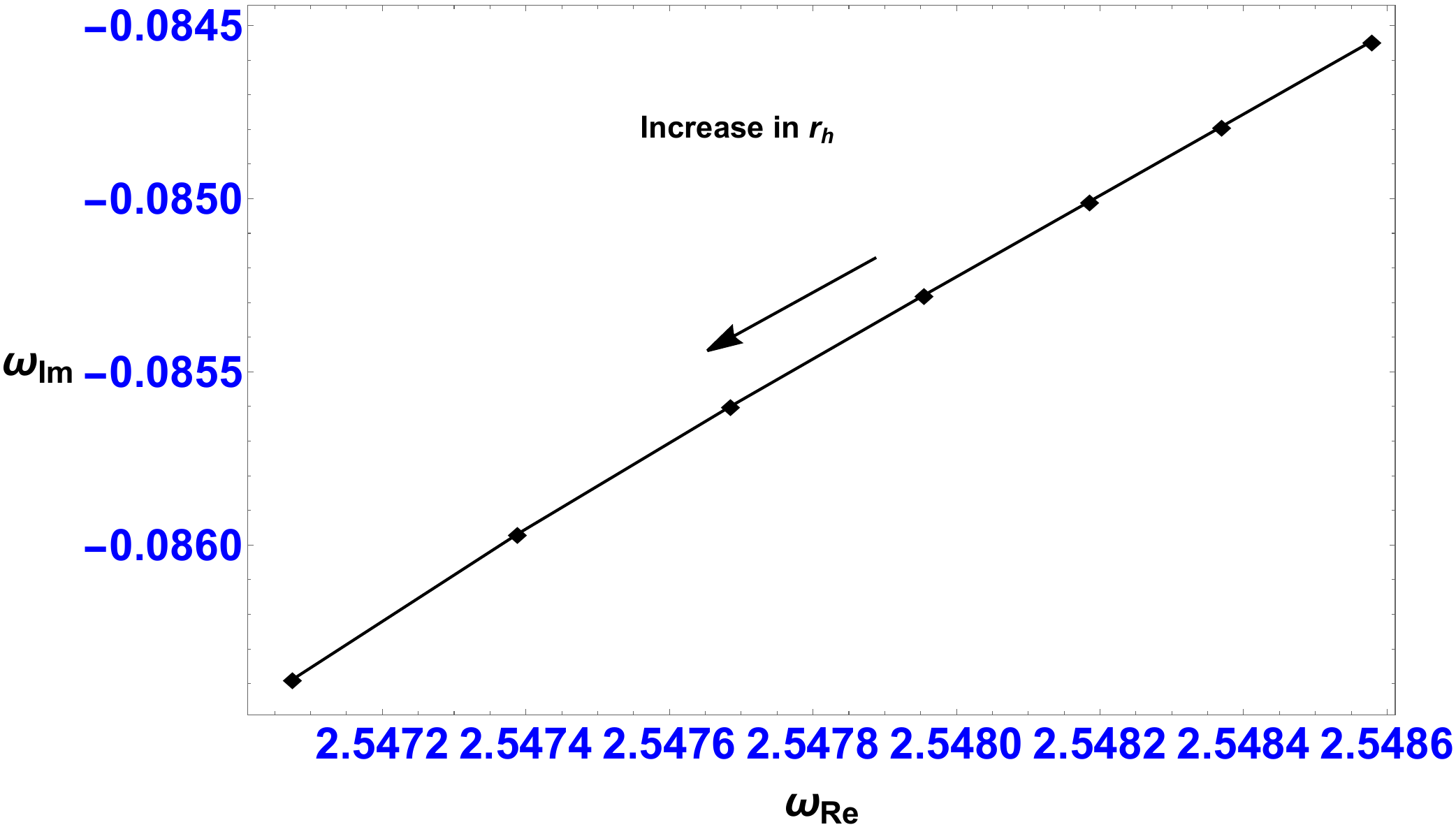}}
	\hfill
	\subfigure[$a=0.2$, $q_e=0.1, q_M=0$.]{\includegraphics[width=0.31\linewidth]{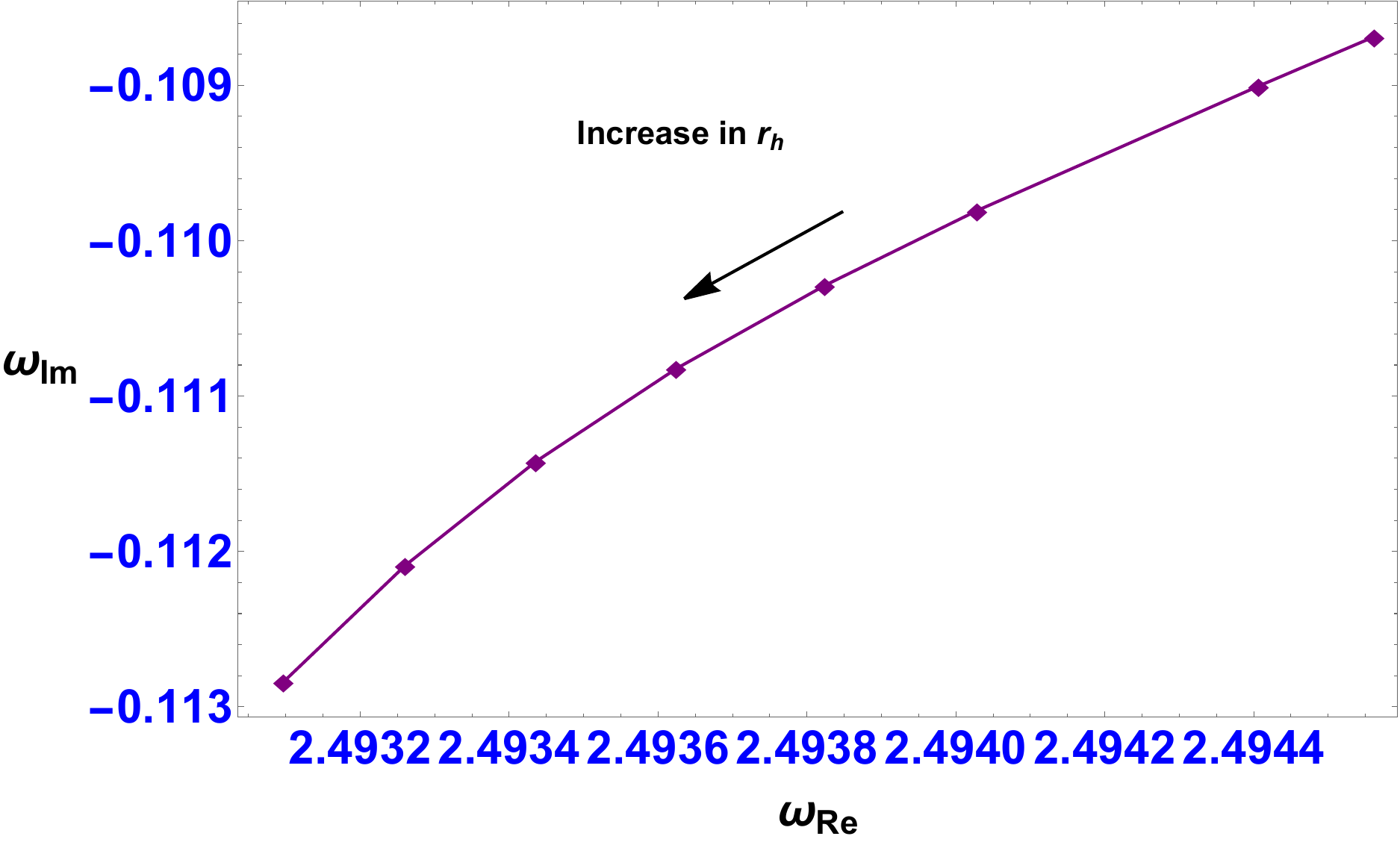}}
	
	\caption{QNM behavior in the small black hole phase in the fixed charge case.}\label{SBHP}
\end{figure}

	\begin{figure}[hbtp!]
	\centering
	\subfigure[$a=0.05$, $\mu_e=0.1, q_M=0$.]{\includegraphics[width=0.3\linewidth]{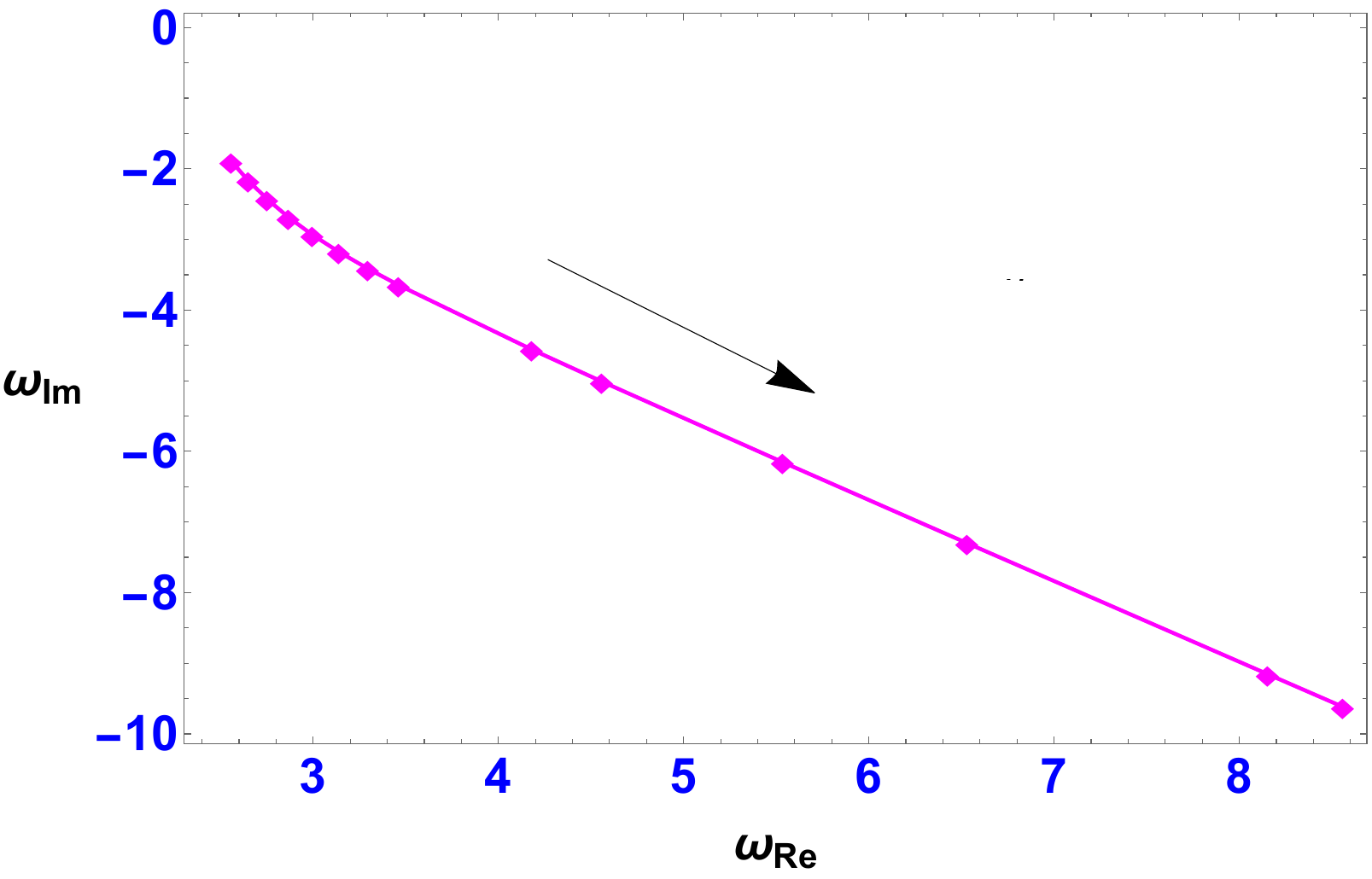}}
	\hfill
	\subfigure[$a=0.1$, $\mu_e=0.1, q_M=0$.]{\includegraphics[width=0.3\linewidth]{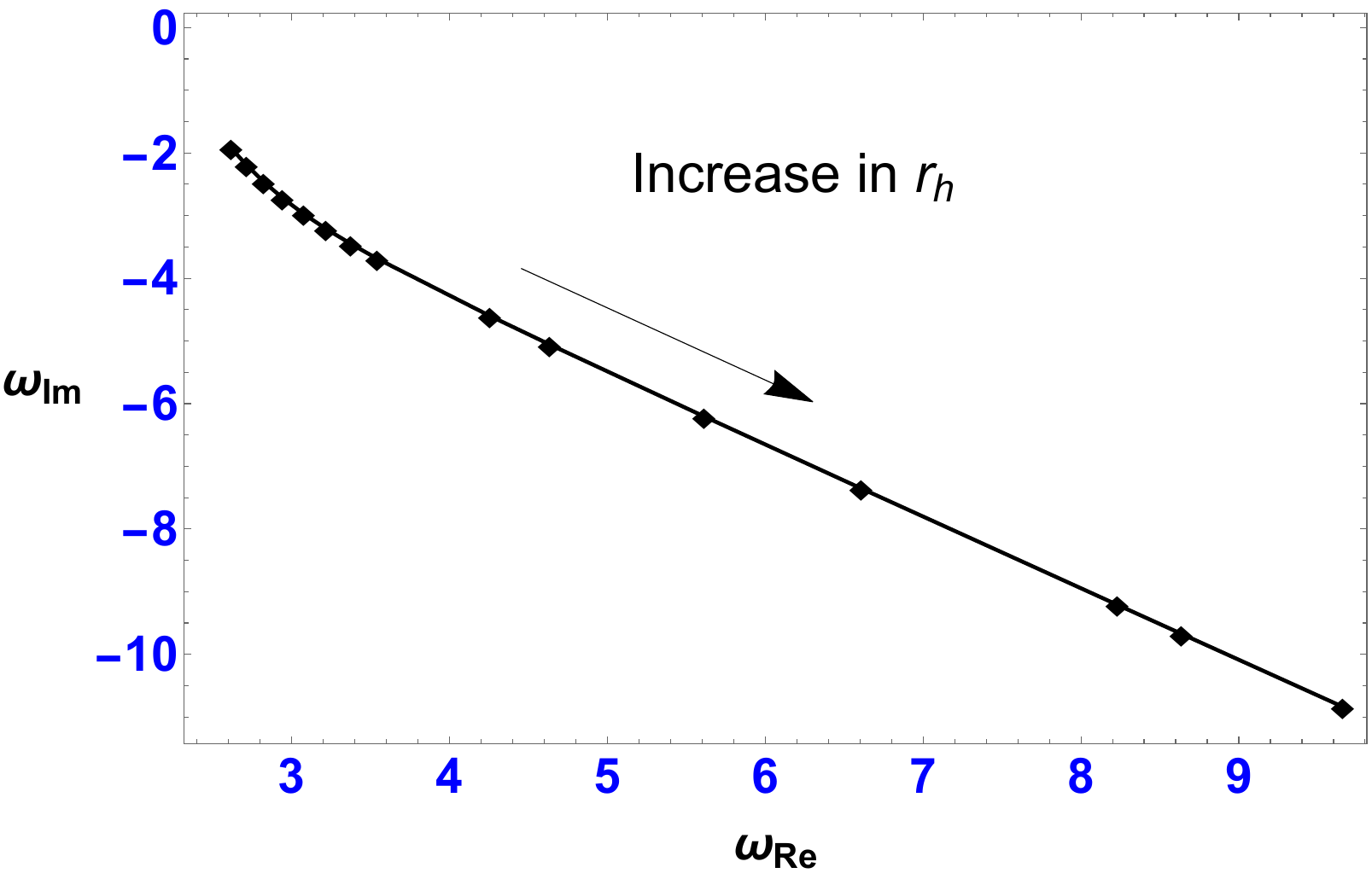}}
	\hfill
	\subfigure[$a=0.2$, $\mu_e=0.1, q_M=0$.]{\includegraphics[width=0.3\linewidth]{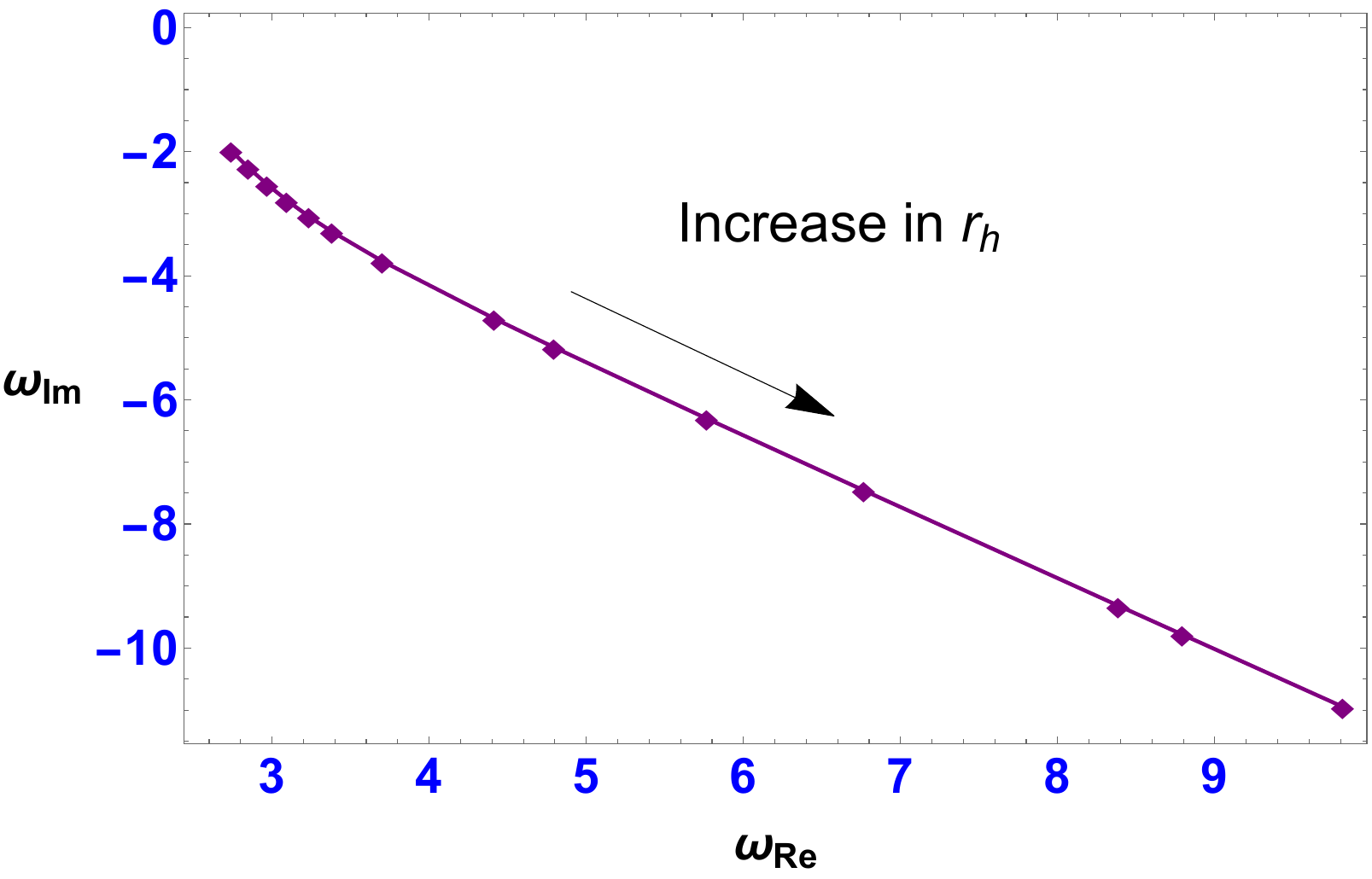}}
	
	\caption{QNM behavior in the large black hole phase in the fixed potential case.}\label{LBHP1}
\end{figure}

\begin{figure}[hbtp!]
	\centering
	\subfigure[$a=0.05$, $\mu_e=0.1, q_M=0$.]{\includegraphics[width=0.31\linewidth]{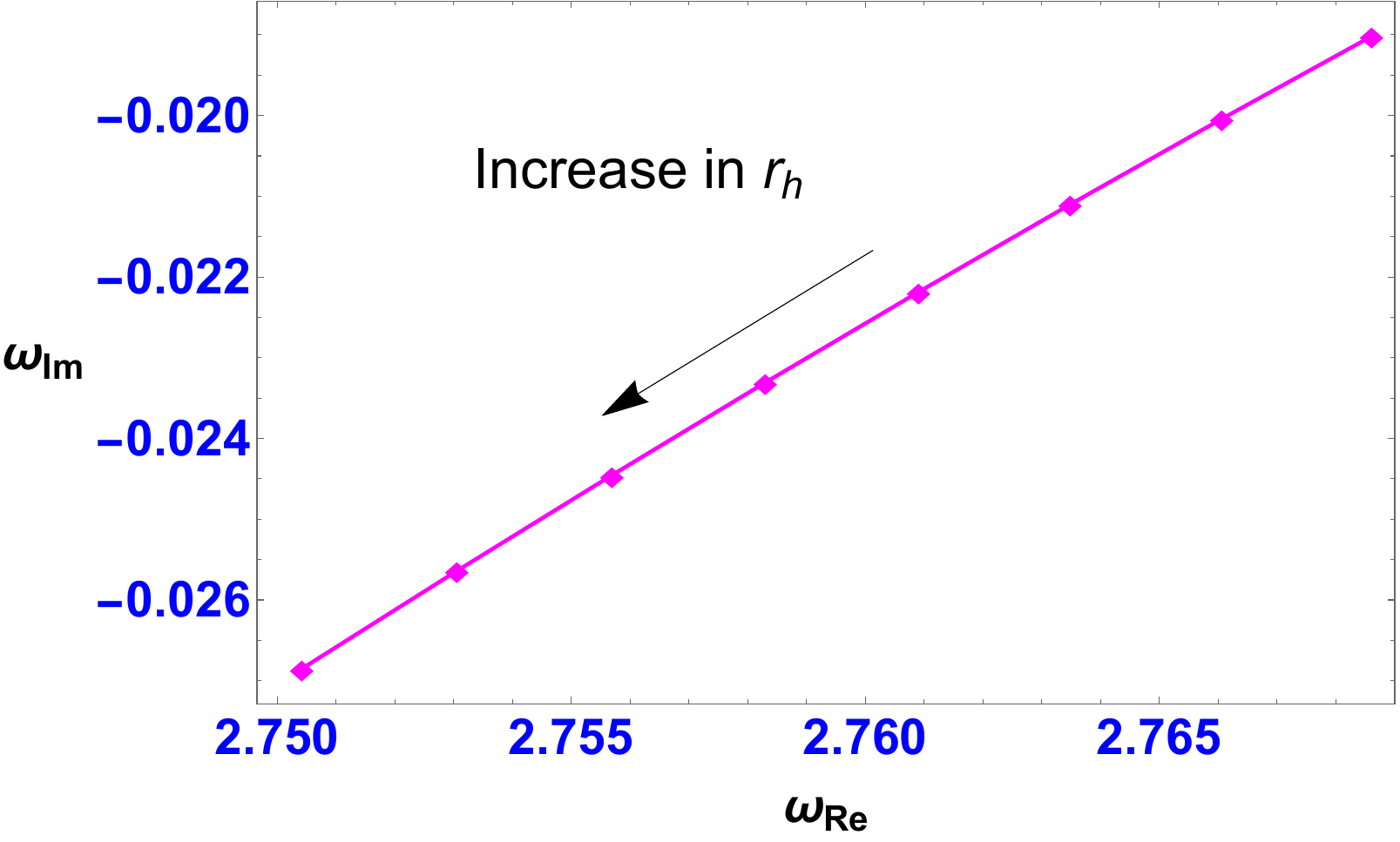}}
	\hfill
	\subfigure[$a=0.1$, $\mu_e=0.1, q_M=0$.]{\includegraphics[width=0.31\linewidth]{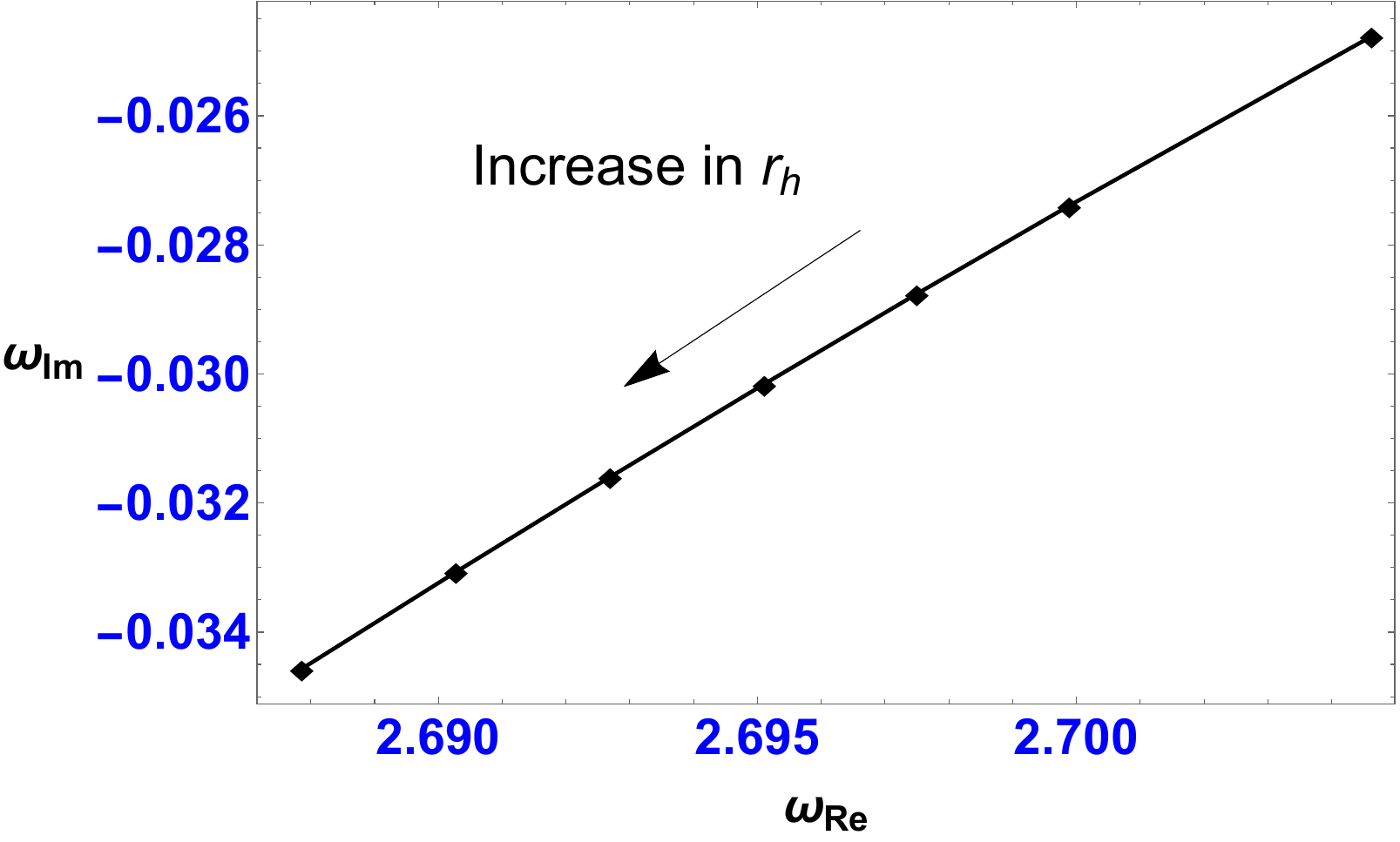}}
	\hfill
	\subfigure[$a=0.2$, $\mu_e=0.1, q_M=0$.]{\includegraphics[width=0.31\linewidth]{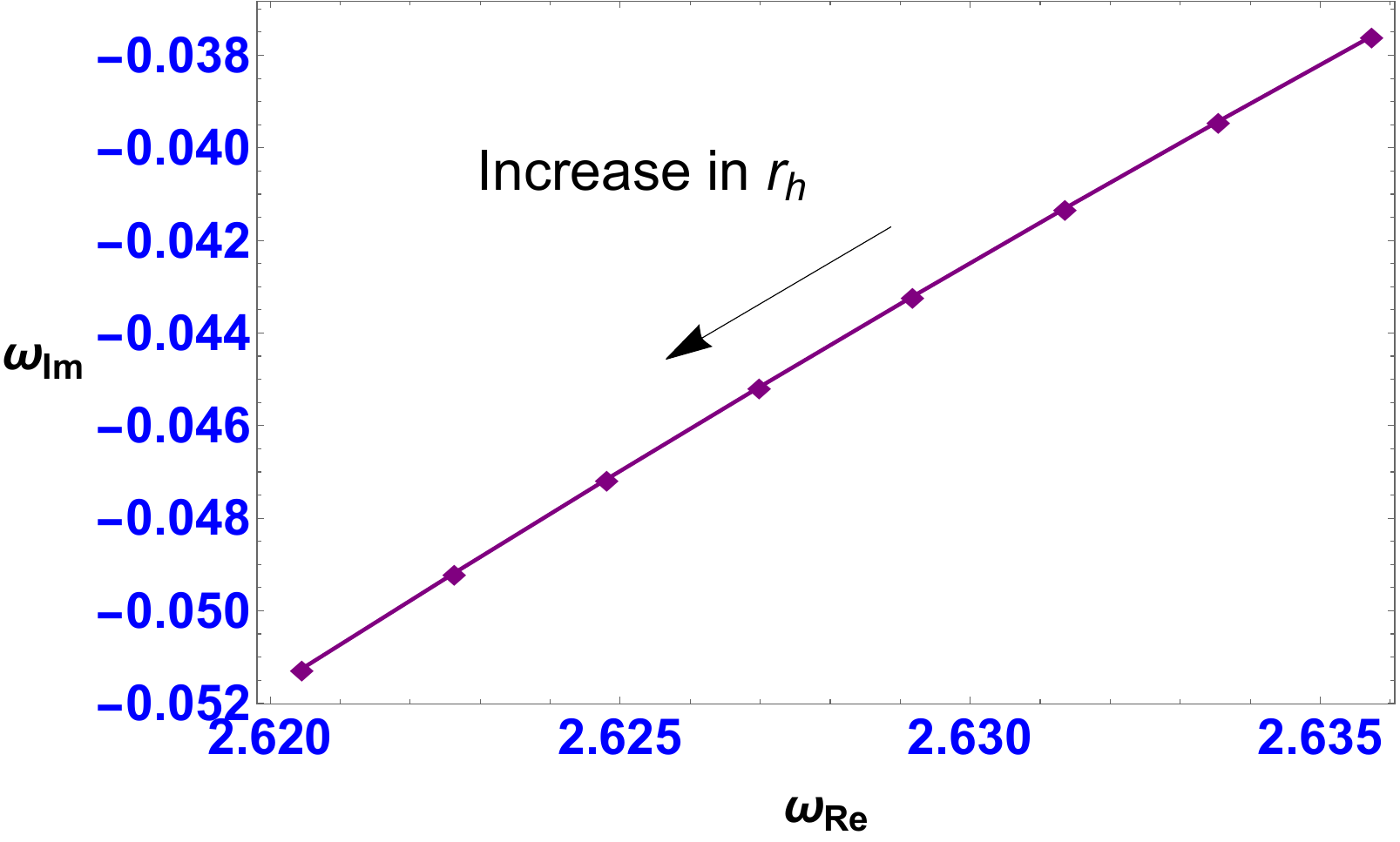}}
	
	\caption{QNM behavior in the small black hole phase in the fixed potential case.}\label{SBHP1}
\end{figure}
In Fig.~\ref{LBHP}, we have shown the behaviour of QNMs in the large black hole phase. Notice that $\omega_{Re}$ vs $\omega_{Im}$ curve traces out a negative slope as we increase the horizon radius in the large black hole phase, i.e., the magnitude of both the oscillating and decaying modes increase with the horizon radius. The same result is true for different values of $a$, as can be observed from different subfigures of Fig.~\ref{LBHP}. On the other hand, $\omega_{Re}$ vs $\omega_{Im}$ curve traces out a positive slope as we increase the horizon radius in the small black hole phase. This is shown in Fig.~\ref{SBHP}. Interestingly, the absolute value of the oscillating mode increases as the horizon radius $r_h$ increases in large hairy black hole phase. However, in the small hairy black hole phase, this value decreases with an increase in the horizon radius $r_h$. This is evident from Figs.~\ref{LBHP} and \ref{SBHP}, where an arrow inside the box indicates the direction of increase of the horizon radius. The distinct behaviour of QNM in the small and large black hole phases has been suggested as a signature of black hole phase transition \cite{liu2014signature,Mahapatra:2016dae}. We find that the
same distinct behaviour also appears in the QNM structure of small and large hairy black hole phases. Our analysis therefore supports the claim made in \cite{liu2014signature,Mahapatra:2016dae} for the QNM to be an effective probe of the black hole phase transition.

We have also analysed the QNM structure in the small/large hairy black hole phases in the fixed chemical potential ensemble. The results are shown in Figs.~\ref{LBHP1} and \ref{SBHP1}. We again find that the QNM behaves distinctly in the small and large hairy phases, suggesting that the distinct behaviour of QNM near the phases transition might be a generic feature.

\section{Conclusions}\label{section6}
In this section, we discuss the main results of our work. We explored the dynamical stability of primary scalar hair dyonic black holes, which possess both electric and magnetic charges, within the context of the Einstein-Maxwell-scalar gravity system, against the massless scalar field perturbation. The gravity solution we employ is adopted from our previous work \cite{priyadarshinee2021analytic}, where the local and global thermodynamical stability of these hairy black holes were established. Essentially, the adopted hairy dyonic black holes are not only locally stable, as they have positive specific heat, but also exhibit rich phase structure by undergoing various phase transitions.

We numerically obtained the scalar QNM values using two different methods: (i) the power series method; and (ii) the shooting method. We showed that the results from both these methods agree reasonably well. We then proceed to investigate the behaviour of QNMs for different values of hairy black hole parameters. In particular, we obtained the QNM values for different values of $\{r_h, a , q_e, q_M, \mu_e\}$ in both fixed charge and potential systems. Remarkably, in both cases, the oscillating and decaying modes displayed similar features. Importantly, for all hairy black hole parameters, the decaying mode always turned out to be negative, thereby establishing the dynamical stability of these black holes  against the scalar field perturbation. This result was also supplemented by analytic arguments. In particular, we analytically showed that $i\omega$ can not be pure real and that the imaginary part of $\omega$ has to be negative for the perturbation to satisfy the appropriate boundary conditions. Our analysis further suggested that the imaginary part of QNM increases with the hair parameter $a$, suggesting that the hairy black hole returns to the equilibrium configuration at a faster rate for larger values of $a$ as compared to the nonhairy charged and uncharged black holes. Furthermore, we conducted an analysis to observe the linear scaling of the imaginary and real parts of QNM with temperature across various values of the hairy parameter $a$.

Subsequently, we analysed the behaviour of QNMs near the small/large black hole phase transition. We discovered intriguing behaviour in the QNM profile in small and large black hole phases. In particular, in the small black hole phase below the critical temperature, the absolute value of the oscillating mode exhibits a decreasing trend with the horizon radius, whereas, in the large black hole phase above the critical temperature, the oscillating mode exhibits an increasing trend with the horizon radius. This behavior, in turn, led to different slopes for $\omega_{Re}$ -- $\omega_{Im}$ curve in small and large black hole phases. This peculiar observation indicates that QNM behaves differently in different black hole phases, even in the hairy case. Hence, our findings lend support to the assertion made in the previous literature that QNMs can serve as a dynamic tool to probe the black hole phase transition.

As a potential avenue for future research, it would be interesting to explore the QNMs for gravitational and electromagnetic perturbations. It would help us to establish the dynamical stabilities of hairy black holes against these perturbations as well. However, it is worth noting that obtaining these QNMs may pose significant numerical challenges, given that electromagnetic and gravitational perturbations arise from different parts of the gauge field and metric perturbations. We hope to comment on this issue soon.

\section*{Acknowledegment}
		The author would like to acknowledge Bhavesh Gupta for necessary discussion and Subhash Mahapatra for his contributions toward finalizing the manuscript. Also the author would like to thank Sudipti Priyadarsinee for thorough reading and identifying the essential corrections. The work of S.P is supported by Grant No. 16-6(DEC.2017)/2018 (NET/CSIR) of UGC, India.

	\bibliographystyle{ieeetr}
	\bibliography{reference}

\begin{thebibliography}{100}

\bibitem{Ruffini1971}
J.~W. R.Ruffini, ``Introducting to black hole,'' {\em Phys.Rev B}, vol.~24,
  1971.

\bibitem{Bekenstein:1972ky}
J.~Bekenstein, ``{Nonexistence of baryon number for black holes. ii},'' {\em
  Phys. Rev. D}, vol.~5, pp.~2403--2412, 1972.

\bibitem{Israel:1967wq}
W.~Israel, ``{Event horizons in static vacuum space-times},'' {\em Phys. Rev.},
  vol.~164, pp.~1776--1779, 1967.

\bibitem{Carter:1971zc}
B.~Carter, ``{Axisymmetric Black Hole Has Only Two Degrees of Freedom},'' {\em
  Phys. Rev. Lett.}, vol.~26, p.~331, 1971.

\bibitem{Israel:1967za}
W.~Israel, ``{Event horizons in static electrovac space-times},'' {\em Commun.
  Math. Phys.}, vol.~8, pp.~245--260, 1968.

\bibitem{Wald:1971iw}
R.~M. Wald, ``{Final states of gravitational collapse},'' {\em Phys. Rev.
  Lett.}, vol.~26, pp.~1653--1655, 1971.

\bibitem{Robinson:1975bv}
D.~Robinson, ``{Uniqueness of the Kerr black hole},'' {\em Phys. Rev. Lett.},
  vol.~34, pp.~905--906, 1975.

\bibitem{Teitelboim:1972qx}
C.~Teitelboim, ``{Nonmeasurability of the quantum numbers of a black hole},''
  {\em Phys. Rev. D}, vol.~5, p.~2941, 1972.

\bibitem{Mazur:1982db}
P.~Mazur, ``{PROOF OF UNIQUENESS OF THE KERR-NEWMAN BLACK HOLE SOLUTION},''
  {\em J. Phys. A}, vol.~15, p.~3173, 1982.

\bibitem{Chase}
J.~Chase, ``{Event horizons in static scalar-vacuum space-times},'' {\em
  Comm.Phys.Math.}, vol.~19, 1970.

\bibitem{bocharova1970exact}
N.~Bocharova, K.~Bronnikov, and V.~Melnikov, ``An exact solution of the system
  of einstein equations and mass-free scalar field,'' {\em Vestn. Mosk. Univ.
  Fiz. Astro}, vol.~6, p.~706, 1970.

\bibitem{Bekenstein:1974sf}
J.~Bekenstein, ``{Exact solutions of Einstein conformal scalar equations},''
  {\em Annals Phys.}, vol.~82, p.~535, 1974.

\bibitem{Bekenstein:1975ts}
J.~Bekenstein, ``{Black Holes with Scalar Charge},'' {\em Annals Phys.},
  vol.~91, p.~75, 1975.

\bibitem{Bronnikov:1978mx}
K.~Bronnikov and Y.~Kireev, ``{Instability of Black Holes with Scalar
  Charge},'' {\em Phys. Lett. A}, vol.~67, p.~95, 1978.

\bibitem{torii1999toward}
T.~Torii, K.~Maeda, and M.~Narita, ``Toward the no-scalar-hair conjecture in
  asymptotically de sitter spacetime,'' {\em Physical Review D}, vol.~59,
  no.~6, p.~064027, 1999.

\bibitem{zloshchastiev2005coexistence}
K.~G. Zloshchastiev, ``Coexistence of black holes and a long-range scalar field
  in cosmology,'' {\em Physical review letters}, vol.~94, no.~12, p.~121101,
  2005.

\bibitem{berti2013numerical}
E.~Berti, V.~Cardoso, L.~Gualtieri, M.~Horbatsch, and U.~Sperhake, ``Numerical
  simulations of single and binary black holes in scalar-tensor theories:
  circumventing the no-hair theorem,'' {\em Physical Review D}, vol.~87,
  no.~12, p.~124020, 2013.

\bibitem{herdeiro2014kerr}
C.~A. Herdeiro and E.~Radu, ``Kerr black holes with scalar hair,'' {\em
  Physical review letters}, vol.~112, no.~22, p.~221101, 2014.

\bibitem{garfinkle1991charged}
D.~Garfinkle, G.~T. Horowitz, and A.~Strominger, ``Charged black holes in
  string theory,'' {\em Physical Review D}, vol.~43, no.~10, p.~3140, 1991.

\bibitem{brodbeck1996instability}
O.~Brodbeck and N.~Straumann, ``Instability proof for einstein--yang--mills
  solitons and black holes with arbitrary gauge groups,'' {\em Journal of
  Mathematical Physics}, vol.~37, no.~3, pp.~1414--1433, 1996.

\bibitem{volkov1995number}
M.~Volkov, O.~Brodbeck, G.~Lavrelashvili, and N.~Straumann, ``The number of
  sphaleron instabilities of the bartnik-mckinnon solitons and non-abelian
  black holes,'' {\em Physics Letters B}, vol.~349, no.~4, pp.~438--442, 1995.

\bibitem{bizon1991n}
P.~Bizon and R.~M. Wald, ``The n= 1 colored black hole is unstable,'' {\em
  Physics Letters B}, vol.~267, no.~2, pp.~173--174, 1991.

\bibitem{bizon1991stability}
P.~Bizon, ``Stability of einstein yang-mills black holes,'' {\em Physics
  Letters B}, vol.~259, no.~1-2, pp.~53--57, 1991.

\bibitem{zhou1991nonlinear}
Z.-h. Zhou and N.~Straumann, ``Nonlinear perturbations of einstein-yang-mills
  solitons and non-abelian black holes,'' {\em Nuclear Physics B}, vol.~360,
  no.~1, pp.~180--196, 1991.

\bibitem{bizon1990colored}
P.~Bizon, ``Colored black holes,'' {\em Physical review letters}, vol.~64,
  no.~24, p.~2844, 1990.

\bibitem{Mahapatra:2022xea}
S.~Mahapatra and I.~Banerjee, ``{Rotating hairy black holes and thermodynamics
  from gravitational decoupling},'' {\em Phys. Dark Univ.}, vol.~39, p.~101172,
  2023.

\bibitem{Bocharova}
V.~N.~M. N.~M.~Bocharoval, K.~A.~Bronnikov, ``{An exact solution of the system
  of Einstein equations and mass-free scalar field},'' {\em Vestn.\ Mosk.\
  Univ.\ Fiz.\ Astron}, vol.~06, p.~706, 1970.

\bibitem{Bekenstein:1971hc}
J.~D. Bekenstein, ``{Nonexistence of baryon number for static black holes},''
  {\em Phys. Rev. D}, vol.~5, pp.~1239--1246, 1972.

\bibitem{Bekenstein:1995un}
J.~Bekenstein, ``{Novel ``no-scalar-hair'' theorem for black holes},'' {\em
  Phys. Rev. D}, vol.~51, no.~12, p.~6608, 1995.

\bibitem{Sudarsky:1995zg}
D.~Sudarsky, ``{A Simple proof of a no hair theorem in Einstein Higgs
  theory,},'' {\em Class. Quant. Grav.}, vol.~12, pp.~579--584, 1995.

\bibitem{Heusler:1992ss}
M.~Heusler, ``{A No hair theorem for selfgravitating nonlinear sigma models},''
  {\em J. Math. Phys.}, vol.~33, p.~3497, 1992.

\bibitem{Herdeiro:2015waa}
C.~A. Herdeiro and E.~Radu, ``{Asymptotically flat black holes with scalar
  hair: a review},'' {\em Int. J. Mod. Phys. D}, vol.~24, no.~09, p.~1542014,
  2015.

\bibitem{Hertog:2006rr}
T.~Hertog, ``{Towards a Novel no-hair Theorem for Black Holes},'' {\em Phys.
  Rev. D}, vol.~74, p.~084008, 2006.

\bibitem{Zloshchastiev:2004ny}
K.~G. Zloshchastiev, ``{On co-existence of black holes and scalar field},''
  {\em Phys. Rev. Lett.}, vol.~94, p.~121101, 2005.

\bibitem{Torii:1998ir}
T.~Torii, K.~Maeda, and M.~Narita, ``{No scalar hair conjecture in asymptotic
  de Sitter space-time},'' {\em Phys. Rev. D}, vol.~59, p.~064027, 1999.

\bibitem{Torii:2001pg}
T.~Torii, K.~Maeda, and M.~Narita, ``{Scalar hair on the black hole in
  asymptotically anti-de Sitter space-time},'' {\em Phys. Rev. D}, vol.~64,
  p.~044007, 2001.

\bibitem{Winstanley:2002jt}
E.~Winstanley, ``{On the existence of conformally coupled scalar field hair for
  black holes in (anti-de Sitter space},'' {\em Found. Phys.}, vol.~33,
  pp.~111--143, 2003.

\bibitem{Martinez:2004nb}
C.~Martinez, R.~Troncoso, and J.~Zanelli, ``{Exact black hole solution with a
  minimally coupled scalar field},'' {\em Phys. Rev. D}, vol.~70, p.~084035,
  2004.

\bibitem{Martinez:2005di}
C.~Martinez, J.~P. Staforelli, and R.~Troncoso, ``{Topological black holes
  dressed with a conformally coupled scalar field and electric charge},'' {\em
  Phys. Rev. D}, vol.~74, p.~044028, 2006.

\bibitem{Martinez:2006an}
C.~Martinez and R.~Troncoso, ``{Electrically charged black hole with scalar
  hair},'' {\em Phys. Rev. D}, vol.~74, p.~064007, 2006.

\bibitem{Hertog:2004dr}
T.~Hertog and K.~Maeda, ``{Black holes with scalar hair and asymptotics in N =
  8 supergravity},'' {\em JHEP}, vol.~07, p.~051, 2004.

\bibitem{Henneaux:2004zi}
M.~Henneaux, C.~Martinez, R.~Troncoso, and J.~Zanelli, ``{Asymptotically
  anti-de Sitter spacetimes and scalar fields with a logarithmic branch},''
  {\em Phys. Rev. D}, vol.~70, p.~044034, 2004.

\bibitem{Henneaux:2006hk}
M.~Henneaux, C.~Martinez, R.~Troncoso, and J.~Zanelli, ``{Asymptotic behavior
  and Hamiltonian analysis of anti-de Sitter gravity coupled to scalar
  fields},'' {\em Annals Phys.}, vol.~322, pp.~824--848, 2007.

\bibitem{Amsel:2006uf}
A.~J. Amsel and D.~Marolf, ``{Energy Bounds in Designer Gravity},'' {\em Phys.
  Rev. D}, vol.~74, p.~064006, 2006.
\newblock [Erratum: Phys.Rev.D 75, 029901 (2007)].

\bibitem{Mahapatra:2020wym}
S.~Mahapatra, S.~Priyadarshinee, G.~N. Reddy, and B.~Shukla, ``{Exact
  topological charged hairy black holes in AdS Space in $D$-dimensions},'' {\em
  Phys. Rev. D}, vol.~102, no.~2, p.~024042, 2020.

\bibitem{Dias:2011at}
O.~J. Dias, G.~T. Horowitz, and J.~E. Santos, ``{Black holes with only one
  Killing field},'' {\em JHEP}, vol.~07, p.~115, 2011.

\bibitem{Dias:2011tj}
O.~J. Dias, P.~Figueras, S.~Minwalla, P.~Mitra, R.~Monteiro, and J.~E. Santos,
  ``{Hairy black holes and solitons in global $AdS_5$},'' {\em JHEP}, vol.~08,
  p.~117, 2012.

\bibitem{Bhattacharyya:2010yg}
S.~Bhattacharyya, S.~Minwalla, and K.~Papadodimas, ``{Small Hairy Black Holes
  in $AdS_5 x S^5$},'' {\em JHEP}, vol.~11, p.~035, 2011.

\bibitem{Basu:2010uz}
P.~Basu, J.~Bhattacharya, S.~Bhattacharyya, R.~Loganayagam, S.~Minwalla, and
  V.~Umesh, ``{Small Hairy Black Holes in Global AdS Spacetime},'' {\em JHEP},
  vol.~10, p.~045, 2010.

\bibitem{Anabalon:2012ta}
A.~Anabalon, ``{Exact Black Holes and Universality in the Backreaction of
  non-linear Sigma Models with a potential in (A)dS4},'' {\em JHEP}, vol.~06,
  p.~127, 2012.

\bibitem{Anabalon:2012ih}
A.~Anabalon and J.~Oliva, ``{Exact Hairy Black Holes and their Modification to
  the Universal Law of Gravitation},'' {\em Phys. Rev. D}, vol.~86, p.~107501,
  2012.

\bibitem{Anabalon:2012tu}
A.~Anabalon and A.~Cisterna, ``{Asymptotically (anti) de Sitter Black Holes and
  Wormholes with a Self Interacting Scalar Field in Four Dimensions},'' {\em
  Phys. Rev. D}, vol.~85, p.~084035, 2012.

\bibitem{Kleihaus:2013tba}
B.~Kleihaus, J.~Kunz, E.~Radu, and B.~Subagyo, ``{Axially symmetric static
  scalar solitons and black holes with scalar hair},'' {\em Phys. Lett. B},
  vol.~725, p.~489, 2013.

\bibitem{Kolyvaris:2013zfa}
T.~Kolyvaris, G.~Koutsoumbas, E.~Papantonopoulos, and G.~Siopsis, ``{Phase
  Transition to a Hairy Black Hole in Asymptotically Flat Spacetime},'' {\em
  JHEP}, vol.~11, p.~133, 2013.

\bibitem{Gonzalez:2013aca}
P.~González, E.~Papantonopoulos, J.~Saavedra, and Y.~Vásquez,
  ``{Four-Dimensional Asymptotically AdS Black Holes with Scalar Hair},'' {\em
  JHEP}, vol.~12, p.~021, 2013.

\bibitem{Anabalon:2009qt}
A.~Anabalon and H.~Maeda, ``{New Charged Black Holes with Conformal Scalar
  Hair},'' {\em Phys. Rev. D}, vol.~81, p.~041501, 2010.

\bibitem{Charmousis:2009cm}
C.~Charmousis, T.~Kolyvaris, and E.~Papantonopoulos, ``{Charged C-metric with
  conformally coupled scalar field},'' {\em Class. Quant. Grav.}, vol.~26,
  p.~175012, 2009.

\bibitem{Guo:2023mda}
G.~Guo, P.~Wang, H.~Wu, and H.~Yang, ``{Scalarized Kerr-Newman Black Holes},''
  7 2023.

\bibitem{Guo:2021zed}
G.~Guo, P.~Wang, H.~Wu, and H.~Yang, ``{Scalarized
  Einstein\textendash{}Maxwell-scalar black holes in anti-de Sitter
  spacetime},'' {\em Eur. Phys. J. C}, vol.~81, no.~10, p.~864, 2021.

\bibitem{Maldacena:1997re}
J.~M. Maldacena, ``{The Large N limit of superconformal field theories and
  supergravity},'' {\em Int. J. Theor. Phys.}, vol.~38, p.~1113, 1999.

\bibitem{Gubser:2008px}
S.~S. Gubser, ``{Breaking an Abelian gauge symmetry near a black hole
  horizon},'' {\em Phys. Rev. D}, vol.~78, p.~065034, 2008.

\bibitem{Hartnoll:2008vx}
S.~A. Hartnoll, C.~P. Herzog, and G.~T. Horowitz, ``{Building a Holographic
  Superconductor},'' {\em Phys. Rev. Lett.}, vol.~101, p.~031601, 2008.

\bibitem{Hartnoll:2008kx}
S.~A. Hartnoll, C.~P. Herzog, and G.~T. Horowitz, ``{Holographic
  Superconductors},'' {\em JHEP}, vol.~12, p.~015, 2008.

\bibitem{Dey:2014voa}
A.~Dey, S.~Mahapatra, and T.~Sarkar, ``{Very General Holographic
  Superconductors and Entanglement Thermodynamics},'' {\em JHEP}, vol.~12,
  p.~135, 2014.

\bibitem{Dey:2014xxa}
A.~Dey, S.~Mahapatra, and T.~Sarkar, ``{Generalized Holographic Superconductors
  with Higher Derivative Couplings},'' {\em JHEP}, vol.~06, p.~147, 2014.

\bibitem{Mahapatra:2013vta}
S.~Mahapatra, P.~Phukon, and T.~Sarkar, ``{Generalized Superconductors and
  Holographic Optics},'' {\em JHEP}, vol.~01, p.~135, 2014.

\bibitem{Gubser:2008ny}
S.~S. Gubser and A.~Nellore, ``{Mimicking the QCD equation of state with a dual
  black hole},'' {\em Phys. Rev. D}, vol.~78, p.~086007, 2008.

\bibitem{Hawking:1982dh}
S.~W. Hawking and D.~N. Page, ``{Thermodynamics of Black Holes in anti-De
  Sitter Space},'' {\em Commun. Math. Phys.}, vol.~87, p.~577, 1983.

\bibitem{Chamblin:1999tk}
A.~Chamblin, R.~Emparan, C.~V. Johnson, and R.~C. Myers, ``{Charged AdS black
  holes and ieeetrcatastrophic holography},'' {\em Phys. Rev. D}, vol.~60,
  p.~064018, 1999.

\bibitem{Chamblin:1999hg}
A.~Chamblin, R.~Emparan, C.~V. Johnson, and R.~C. Myers, ``{Holography,
  thermodynamics and fluctuations of charged AdS black holes},'' {\em Phys.
  Rev. D}, vol.~60, p.~104026, 1999.

\bibitem{Sahay:2010wi}
A.~Sahay, T.~Sarkar, and G.~Sengupta, ``{Thermodynamic Geometry and Phase
  Transitions in Kerr-Newman-AdS Black Holes},'' {\em JHEP}, vol.~04, p.~118,
  2010.

\bibitem{Dey:2015ytd}
A.~Dey, S.~Mahapatra, and T.~Sarkar, ``{Thermodynamics and Entanglement Entropy
  with Weyl Corrections},'' {\em Phys. Rev. D}, vol.~94, no.~2, p.~026006,
  2016.

\bibitem{Witten:1998zw}
E.~Witten, ``{Anti-de Sitter space, thermal phase transition, and confinement
  in gauge theories},'' {\em Adv. Theor. Math. Phys.}, vol.~2, pp.~505--532,
  1998.

\bibitem{Dudal:2017max}
D.~Dudal and S.~Mahapatra, ``{Thermal entropy of a quark-antiquark pair above
  and below deconfinement from a dynamical holographic QCD model},'' {\em Phys.
  Rev. D}, vol.~96, no.~12, p.~126010, 2017.

\bibitem{Hartnoll:2007ai}
S.~A. Hartnoll and P.~Kovtun, ``{Hall conductivity from dyonic black holes},''
  {\em Phys. Rev. D}, vol.~76, p.~066001, 2007.

\bibitem{Dutta:2013dca}
S.~Dutta, A.~Jain, and R.~Soni, ``{Dyonic Black Hole and Holography},'' {\em
  JHEP}, vol.~12, p.~060, 2013.

\bibitem{Caldarelli:2008ze}
M.~M. Caldarelli, O.~J.~C. Dias, and D.~Klemm, ``{Dyonic AdS black holes from
  magnetohydrodynamics},'' {\em JHEP}, vol.~03, p.~025, 2009.

\bibitem{Hartnoll:2007ih}
S.~A. Hartnoll, P.~K. Kovtun, M.~Muller, and S.~Sachdev, ``{Theory of the
  Nernst effect near quantum phase transitions in condensed matter, and in
  dyonic black holes},'' {\em Phys. Rev. B}, vol.~76, p.~144502, 2007.

\bibitem{Goldstein:2010aw}
K.~Goldstein, N.~Iizuka, S.~Kachru, S.~Prakash, S.~P. Trivedi, and A.~Westphal,
  ``{Holography of Dyonic Dilaton Black Branes},'' {\em JHEP}, vol.~10, p.~027,
  2010.

\bibitem{Kundu:2012jn}
N.~Kundu, P.~Narayan, N.~Sircar, and S.~P. Trivedi, ``{Entangled Dilaton
  Dyons},'' {\em JHEP}, vol.~03, p.~155, 2013.

\bibitem{Caldarelli:2016nni}
M.~M. Caldarelli, A.~Christodoulou, I.~Papadimitriou, and K.~Skenderis,
  ``{Phases of planar AdS black holes with axionic charge},'' {\em JHEP},
  vol.~04, p.~001, 2017.

\bibitem{Donos:2015bxe}
A.~Donos, J.~P. Gauntlett, T.~Griffin, and L.~Melgar, ``{DC Conductivity of
  Magnetised Holographic Matter},'' {\em JHEP}, vol.~01, p.~113, 2016.

\bibitem{Sadeghi:2016dvc}
J.~Sadeghi, B.~Pourhassan, and M.~Rostami, ``{P-V criticality of
  logarithm-corrected dyonic charged AdS black holes},'' {\em Phys. Rev. D},
  vol.~94, no.~6, p.~064006, 2016.

\bibitem{Kruglov:2020aqm}
S.~I. Kruglov, ``{Dyonic and magnetized black holes based on nonlinear
  electrodynamics},'' {\em Eur. Phys. J. C}, vol.~80, no.~3, p.~250, 2020.

\bibitem{Kruglov:2019ybs}
S.~I. Kruglov, ``{Dyonic Black Holes with Nonlinear Logarithmic
  Electrodynamics},'' {\em Grav. Cosmol.}, vol.~25, no.~2, pp.~190--195, 2019.

\bibitem{Panahiyan:2018fpb}
S.~Panahiyan, S.~H. Hendi, and N.~Riazi, ``{$AdS_{4}$ dyonic black holes in
  gravity's rainbow},'' {\em Nucl. Phys. B}, vol.~938, pp.~388--415, 2019.

\bibitem{Hajkhalili:2018thm}
S.~Hajkhalili and A.~Sheykhi, ``{Topological dyonic dilaton black holes in AdS
  spaces},'' {\em Phys. Rev. D}, vol.~99, no.~2, p.~024028, 2019.

\bibitem{Hendi:2016uni}
S.~H. Hendi, N.~Riazi, and S.~Panahiyan, ``{Holographical aspects of dyonic
  black holes: Massive gravity generalization},'' {\em Annalen Phys.},
  vol.~530, no.~2, p.~1700211, 2018.

\bibitem{Chaturvedi:2014vpa}
P.~Chaturvedi, A.~Das, and G.~Sengupta, ``{Thermodynamic Geometry and Phase
  Transitions of Dyonic Charged AdS Black Holes},'' {\em Eur. Phys. J. C},
  vol.~77, no.~2, p.~110, 2017.

\bibitem{Kim:2015wba}
K.-Y. Kim, K.~K. Kim, Y.~Seo, and S.-J. Sin, ``{Thermoelectric Conductivities
  at Finite Magnetic Field and the Nernst Effect},'' {\em JHEP}, vol.~07,
  p.~027, 2015.

\bibitem{Amoretti:2020mkp}
A.~Amoretti, D.~K. Brattan, N.~Magnoli, and M.~Scanavino, ``{Magneto-thermal
  transport implies an incoherent Hall conductivity},'' {\em JHEP}, vol.~08,
  p.~097, 2020.

\bibitem{Bhatnagar:2017twr}
N.~Bhatnagar and S.~Siwach, ``{DC conductivity with external magnetic field in
  hyperscaling violating geometry},'' {\em Int. J. Mod. Phys. A}, vol.~33,
  no.~04, p.~1850028, 2018.

\bibitem{Lindgren:2015lia}
J.~Lindgren, I.~Papadimitriou, A.~Taliotis, and J.~Vanhoof, ``{Holographic Hall
  conductivities from dyonic backgrounds},'' {\em JHEP}, vol.~07, p.~094, 2015.

\bibitem{Zhou:2015dha}
Z.~Zhou, J.-P. Wu, and Y.~Ling, ``{DC and Hall conductivity in holographic
  massive Einstein-Maxwell-Dilaton gravity},'' {\em JHEP}, vol.~08, p.~067,
  2015.

\bibitem{Khimphun:2017mqb}
S.~Khimphun, B.-H. Lee, C.~Park, and Y.-L. Zhang, ``{Anisotropic dyonic black
  brane and its effects on holographic conductivity},'' {\em JHEP}, vol.~10,
  p.~064, 2017.

\bibitem{Bai:2020ezy}
Y.~Bai and M.~Korwar, ``{Hairy Magnetic and Dyonic Black Holes in the Standard
  Model},'' {\em JHEP}, vol.~04, p.~119, 2021.

\bibitem{Li:2016nll}
S.~Li, H.~Lu, and H.~Wei, ``{Dyonic (A)dS Black Holes in Einstein-Born-Infeld
  Theory in Diverse Dimensions},'' {\em JHEP}, vol.~07, p.~004, 2016.

\bibitem{BravoGaete:2019rci}
M.~Bravo~Gaete, S.~Gomez, and M.~Hassaine, ``{Black holes with Lambert W
  function horizons},'' {\em Eur. Phys. J. C}, vol.~79, no.~3, p.~200, 2019.

\bibitem{Cadoni:2011kv}
M.~Cadoni and P.~Pani, ``{Holography of charged dilatonic black branes at
  finite temperature},'' {\em JHEP}, vol.~04, p.~049, 2011.

\bibitem{Priyadarshinee:2021rch}
S.~Priyadarshinee, S.~Mahapatra, and I.~Banerjee, ``{Analytic topological hairy
  dyonic black holes and thermodynamics},'' {\em Phys. Rev. D}, vol.~104,
  no.~8, p.~084023, 2021.

\bibitem{Mahapatra:2018gig}
S.~Mahapatra and P.~Roy, ``{On the time dependence of holographic complexity in
  a dynamical Einstein-dilaton model},'' {\em JHEP}, vol.~11, p.~138, 2018.

\bibitem{BOHRA2020135184}
H.~Bohra, D.~Dudal, A.~Hajilou, and S.~Mahapatra, ``Anisotropic string tensions
  and inversely magnetic catalyzed deconfinement from a dynamical ads/qcd
  model,'' {\em Physics Letters B}, vol.~801, p.~135184, 2020.

\bibitem{Bohra:2020qom}
H.~Bohra, D.~Dudal, A.~Hajilou, and S.~Mahapatra, ``{Chiral transition in the
  probe approximation from an Einstein-Maxwell-dilaton gravity model},'' {\em
  Phys. Rev. D}, vol.~103, no.~8, p.~086021, 2021.

\bibitem{Dudal:2021jav}
D.~Dudal, A.~Hajilou, and S.~Mahapatra, ``{A quenched 2-flavour
  Einstein\textendash{}Maxwell\textendash{}Dilaton gauge-gravity model},'' {\em
  Eur. Phys. J. A}, vol.~57, no.~4, p.~142, 2021.

\bibitem{dudal2018interplay}
D.~Dudal and S.~Mahapatra, ``Interplay between the holographic qcd phase
  diagram and entanglement entropy,'' {\em Journal of High Energy Physics},
  vol.~2018, no.~7, pp.~1--29, 2018.

\bibitem{mahapatra2019interplay}
S.~Mahapatra, ``Interplay between the holographic qcd phase diagram and mutual
  \& n-partite information,'' {\em Journal of High Energy Physics}, vol.~2019,
  no.~4, pp.~1--37, 2019.

\bibitem{he2013phase}
S.~He, S.-Y. Wu, Y.~Yang, and P.-H. Yuan, ``Phase structure in a dynamical
  soft-wall holographic qcd model,'' {\em Journal of High Energy Physics},
  vol.~2013, no.~4, pp.~1--23, 2013.

\bibitem{Arefeva:2018hyo}
I.~Aref'eva and K.~Rannu, ``{Holographic Anisotropic Background with
  Confinement-Deconfinement Phase Transition},'' {\em JHEP}, vol.~05, p.~206,
  2018.

\bibitem{Arefeva:2020vae}
I.~Y. Aref'eva, K.~Rannu, and P.~Slepov, ``{Holographic model for heavy quarks
  in anisotropic hot dense QGP with external magnetic field},'' {\em JHEP},
  vol.~07, p.~161, 2021.

\bibitem{Arefeva:2022avn}
I.~Y. Aref'eva, A.~Ermakov, K.~Rannu, and P.~Slepov, ``{Holographic model for
  light quarks in anisotropic hot dense QGP with external magnetic field},''
  {\em Eur. Phys. J. C}, vol.~83, no.~1, p.~79, 2023.

\bibitem{Arefeva:2021mag}
I.~Y. Aref'eva, K.~Rannu, and P.~S. Slepov, ``{Anisotropic solutions for a
  holographic heavy-quark model with an external magnetic field},'' {\em Teor.
  Mat. Fiz.}, vol.~207, no.~1, pp.~44--57, 2021.

\bibitem{regge1957stability}
T.~Regge and J.~A. Wheeler, ``Stability of a schwarzschild singularity,'' {\em
  Physical Review}, vol.~108, no.~4, p.~1063, 1957.

\bibitem{zerilli1970gravitational}
F.~J. Zerilli, ``Gravitational field of a particle falling in a schwarzschild
  geometry analyzed in tensor harmonics,'' {\em Physical Review D}, vol.~2,
  no.~10, p.~2141, 1970.

\bibitem{vishveshwara1970scattering}
C.~Vishveshwara, ``Scattering of gravitational radiation by a schwarzschild
  black-hole,'' {\em Nature}, vol.~227, no.~5261, pp.~936--938, 1970.

\bibitem{PhysRevD.76.064034}
E.~Berti, V.~Cardoso, J.~A. Gonzalez, U.~Sperhake, M.~Hannam, S.~Husa, and
  B.~Br\"ugmann, ``Inspiral, merger, and ringdown of unequal mass black hole
  binaries: A multipolar analysis,'' {\em Phys. Rev. D}, vol.~76, p.~064034,
  Sep 2007.

\bibitem{horowitz2000quasinormal}
G.~T. Horowitz and V.~E. Hubeny, ``Quasinormal modes of ads black holes and the
  approach to thermal equilibrium,'' {\em Physical Review D}, vol.~62, no.~2,
  p.~024027, 2000.

\bibitem{Pretorius:2005gq}
F.~Pretorius, ``{Evolution of binary black hole spacetimes},'' {\em Phys. Rev.
  Lett.}, vol.~95, p.~121101, 2005.

\bibitem{Campanelli:2005dd}
M.~Campanelli, C.~O. Lousto, P.~Marronetti, and Y.~Zlochower, ``{Accurate
  evolutions of orbiting black-hole binaries without excision},'' {\em Phys.
  Rev. Lett.}, vol.~96, p.~111101, 2006.

\bibitem{Creighton:1999pm}
J.~D.~E. Creighton, ``{Search techniques for gravitational waves from black
  hole ringdowns},'' {\em Phys. Rev. D}, vol.~60, p.~022001, 1999.

\bibitem{Tsunesada:2005fe}
Y.~Tsunesada, D.~Tatsumi, N.~Kanda, and H.~Nakano, ``{Black-hole ringdown
  search in TAMA300: Matched filtering and event selections},'' {\em Class.
  Quant. Grav.}, vol.~22, pp.~S1129--S1138, 2005.

\bibitem{Berti:2009kk}
E.~Berti, V.~Cardoso, and A.~O. Starinets, ``{Quasinormal modes of black holes
  and black branes},'' {\em Class. Quant. Grav.}, vol.~26, p.~163001, 2009.

\bibitem{Berti:2003ud}
E.~Berti and K.~D. Kokkotas, ``{Quasinormal modes of
  Reissner-Nordstr\"om-anti-de Sitter black holes: Scalar, electromagnetic and
  gravitational perturbations},'' {\em Phys. Rev. D}, vol.~67, p.~064020, 2003.

\bibitem{Kokkotas:1999bd}
K.~D. Kokkotas and B.~G. Schmidt, ``{Quasinormal modes of stars and black
  holes},'' {\em Living Rev. Rel.}, vol.~2, p.~2, 1999.

\bibitem{Ferrari:1996vx}
V.~Ferrari, ``{The quasinormal modes of stars and black holes},'' in {\em
  {International Conference on Gravitational Waves: Sources and Detectors}},
  1996.

\bibitem{Chan:1999sc}
J.~S.~F. Chan and R.~B. Mann, ``{Scalar wave falloff in topological black hole
  backgrounds},'' {\em Phys. Rev. D}, vol.~59, p.~064025, 1999.

\bibitem{Cardoso:2001hn}
V.~Cardoso and J.~P.~S. Lemos, ``{Scalar, electromagnetic and Weyl
  perturbations of BTZ black holes: Quasinormal modes},'' {\em Phys. Rev. D},
  vol.~63, p.~124015, 2001.

\bibitem{koutsoumbas2006quasi}
G.~Koutsoumbas, S.~Musiri, E.~Papantonopoulos, and G.~Siopsis, ``Quasi-normal
  modes of electromagnetic perturbations of four-dimensional topological black
  holes with scalar hair,'' {\em Journal of High Energy Physics}, vol.~2006,
  no.~10, p.~006, 2006.

\bibitem{martinez2004exact}
C.~Martinez, R.~Troncoso, and J.~Zanelli, ``Exact black hole solution with a
  minimally coupled scalar field,'' {\em Physical Review D}, vol.~70, no.~8,
  p.~084035, 2004.

\bibitem{shen2007phase}
J.~Shen, B.~Wang, R.-K. Su, C.-Y. Lin, and R.-G. Cai, ``The phase transition
  and the quasi-normal modes of black holes,'' {\em Journal of High Energy
  Physics}, vol.~2007, no.~07, p.~037, 2007.

\bibitem{koutsoumbas2008phase}
G.~Koutsoumbas, E.~Papantonopoulos, and G.~Siopsis, ``Phase transitions in
  charged topological-ads black holes,'' {\em Journal of High Energy Physics},
  vol.~2008, no.~05, p.~107, 2008.

\bibitem{liu2014signature}
Y.~Liu, D.-C. Zou, and B.~Wang, ``Signature of the van der waals like
  small-large charged ads black hole phase transition in quasinormal modes,''
  {\em Journal of High Energy Physics}, vol.~2014, no.~9, pp.~1--20, 2014.

\bibitem{Mahapatra:2016dae}
S.~Mahapatra, ``{Thermodynamics, Phase Transition and Quasinormal modes with
  Weyl corrections},'' {\em JHEP}, vol.~04, p.~142, 2016.

\bibitem{Wei:2018aqm}
S.-W. Wei, Y.-X. Liu, and Y.-Q. Wang, ``{Probing the relationship between the
  null geodesics and thermodynamic phase transition for rotating Kerr-AdS black
  holes},'' {\em Phys. Rev. D}, vol.~99, no.~4, p.~044013, 2019.

\bibitem{Liang:2017ceh}
B.~Liang, S.-W. Wei, and Y.-X. Liu, ``{Quasinormal Modes and Van der Waals like
  phase transition of charged AdS black holes in Lorentz symmetry breaking
  massive gravity},'' {\em Int. J. Mod. Phys. D}, vol.~28, no.~09, p.~1950113,
  2019.

\bibitem{Li:2017kkj}
A.-c. Li, H.-q. Shi, and D.-f. Zeng, ``{Phase structure and quasinormal modes
  of a charged AdS dilaton black hole},'' {\em Phys. Rev. D}, vol.~97, no.~2,
  p.~026014, 2018.

\bibitem{Zou:2017juz}
D.-C. Zou, Y.~Liu, and R.-H. Yue, ``{Behavior of quasinormal modes and Van der
  Waals-like phase transition of charged AdS black holes in massive gravity},''
  {\em Eur. Phys. J. C}, vol.~77, no.~6, p.~365, 2017.

\bibitem{Chabab:2017knz}
M.~Chabab, H.~El~Moumni, S.~Iraoui, and K.~Masmar, ``{Phase Transition of
  Charged-AdS Black Holes and Quasinormal Modes : a Time Domain Analysis},''
  {\em Astrophys. Space Sci.}, vol.~362, no.~10, p.~192, 2017.

\bibitem{Chabab:2016cem}
M.~Chabab, H.~El~Moumni, S.~Iraoui, and K.~Masmar, ``{Behavior of quasinormal
  modes and high dimension RN\textendash{}AdS black hole phase transition},''
  {\em Eur. Phys. J. C}, vol.~76, no.~12, p.~676, 2016.

\bibitem{Zhang:2020khz}
M.~Zhang, C.-M. Zhang, D.-C. Zou, and R.-H. Yue, ``{Phase transition and
  Quasinormal modes for Charged black holes in 4D Einstein-Gauss-Bonnet
  gravity},'' {\em Chin. Phys. C}, vol.~45, no.~4, p.~045105, 2021.

\bibitem{Gubser:2000nd}
S.~S. Gubser, ``{Curvature singularities: The Good, the bad, and the naked},''
  {\em Adv. Theor. Math. Phys.}, vol.~4, pp.~679--745, 2000.

\bibitem{Wang:2000gsa}
B.~Wang, C.-Y. Lin, and E.~Abdalla, ``{Quasinormal modes of Reissner-Nordstrom
  anti-de Sitter black holes},'' {\em Phys. Lett. B}, vol.~481, pp.~79--88,
  2000.

\bibitem{priyadarshinee2021analytic}
S.~Priyadarshinee, S.~Mahapatra, and I.~Banerjee, ``Analytic topological hairy
  dyonic black holes and thermodynamics,'' 2021.

\end{thebibliography}
\end{document}